\documentclass[10pt]{iopart}
\usepackage{epsfig,graphicx}

\newcommand{\pprime}{{\prime\prime}}

\newcommand{\bra}{\langle}
\newcommand{\ket}{\rangle}

\newcommand{\order}{{\mathcal O}}

\newcommand{\sgn}{\textrm{sgn}}

\newcommand{\plus}{{\!+\!}}
\newcommand{\minus}{{\!-\!}}

\newcommand{\be}{\begin{equation}}
\newcommand{\ee}{\end{equation}}
\newcommand{\bd}{\begin{displaymath}}
\newcommand{\ed}{\end{displaymath}}
\newcommand{\vsp}{\vspace*{3mm}}

\newcommand{\bigroom}{\rule[-0.4cm]{0cm}{1.0cm}}

\newcommand{\R}{{\rm I\!R}}

\newcommand{\N}{{\rm I\!N}}

\newcommand{\Z}{{\mathcal Z}}

\newcommand{\bk}{\ensuremath{\mathbf{k}}}
\newcommand{\bm}{\ensuremath{\mathbf{m}}}

\newcommand{\bz}{\ensuremath{\mathbf{z}}}

\newcommand{\blambda}{{\mbox{\boldmath $\lambda$}}}

\newcommand{\bpsi}{{\mbox{\boldmath $\psi$}}}
\newcommand{\bphi}{{\mbox{\boldmath $\phi$}}}

\newcommand{\bsigma}{{\mbox{\boldmath $\sigma$}}}

\newcommand{\bGamma}{{\mbox{\boldmath $\Gamma$}}}

\newcommand{\mb}{\ensuremath{\mathbf{m}}}
\newcommand{\hsp}{\hspace*{5mm}}

\begin{document}

\title[A solvable model of the genesis of amino-acid sequences via coupled dynamics]{A solvable model of the genesis of amino-acid sequences via coupled dynamics of
folding and slow genetic variation}
\author{S Rabello$^\dag$, ACC Coolen$^{\ddag \diamond}$, CJ P\'{e}rez-Vicente$^\oplus$\\ and F Fraternali$^\diamond$}

\address{
 $\dag$ Department of Mathematics, Imperial College London, South Kensington Campus, London SW7 2AZ, U.K.\\
 $\ddag$ Department of Mathematics, King's College London, The Strand, London WC2R 2LS, U.K.\\
 $\oplus$ Departament de Fisica Fonamental, Facultat de Fisica,
 Universitat de Barcelona, 08028 Barcelona, Spain\\
 $\diamond$ Randall Division of Cell and Molecular Biophysics,
King's College London, New Hunt's House, London SE1 1UL, U.K.}

\begin{abstract}
\noindent
We study the coupled dynamics of primary and secondary structure formation (i.e. slow
genetic sequence selection and fast folding) in the context of a solvable microscopic model that includes both short-range steric
forces and and long-range polarity-driven forces. Our solution is based on the diagonalization of replicated
 transfer matrices, and leads in the thermodynamic limit
 to explicit predictions regarding phase transitions and phase diagrams at genetic equilibrium.
 The predicted phenomenology allows for natural physical interpretations, and finds satisfactory support in numerical simulations.
\end{abstract}


\ead{s.rabello@imperial.ac.uk, ton.coolen@kcl.ac.uk,
conrad@ffn.ub.es, franca.fraternali@kcl.ac.uk}


\section{Introduction}

\noindent
The constituent monomers of protein-type hetero-polymers, the amino-acids of which there exist about twenty in nature, are composed of a common backbone and a differentiating side chain, and are bound via a peptide bond. These units are connected sequentially to form a polypeptide chain. The sequence of connected amino-acids defines the so called `primary structure' of the chain.
Given the primary structure, the
 mechanical degrees of freedom of the polypeptide chain
are rotation angles at the junctions of adjacent amino-acids. They
allow proteins to fold into relatively simple repetitive local arrangements
(the `secondary structures', such as $\alpha$-helices or $\beta$-sheets)
which then combine into more complicated global arrangements in 3D
(the
 `tertiary structure'). The folding process is controlled by various combinations of
forces, such as those induced by mutual interactions between the amino-acid side chains (steric forces, Van der Waals forces),
by interactions between side-chains and
the polymer's backbone (hydrogen and sulphur bonds), and by interactions between the amino-acid side-chains and the surrounding solvent (polarity induced forces and hydrogen bonds).
 For comprehensive reviews on the physics of the interactions governing the folding of proteins see e.g. \cite{Guo,Echenique}.
Apart from `chaperone' effects (the influence of specialized proteins), it was discovered \cite{Anfinsen} that the dynamics of the folding process is for most proteins determined solely by their primary structure. Since polypeptide chains can vary in length from a few tens to tens of thousands of monomers, there is an enormous number of possible sequences. Yet only a tiny fraction of these (the actual biologically functional proteins) will represent chains that fold into a unique reproducible tertiary structure, or three-dimensional `conformation', which determines its biological function.

The protein folding problem is how to predict this conformation (the native state) of a protein, given its primary structure. It remains one of the most challenging unsolved problems in biology. Its solution would have a big impact on  medicine. The physicist's strategy in this field (as opposed to bio-informatics approaches based on simulation, see e.g. \cite{Daggett} for a recent review) is to try to understand the main physical mechanisms that drive the one-to-one correspondence between amino-acid sequence and the native state. Normally this is attempted via simple quantitative mathematical models that capture the essential phenomenology of folding and lend themselves to statistical mechanical analysis \cite{Bryg,Der,Lau} and/or are easily simulated numerically \cite{Prentiss,Chen,Yang,Akturk}.
In the language of thermodynamics and statistical mechanics,  it is
 believed that if a protein spontaneously reaches its native state at physiological conditions of temperature and pressure, its free energy landscape must possess a unique stable minimum \cite{Onuchic}.
 However, calculating free energy landscapes for biologically functional proteins is non-trivial, because of the
 frustration induced by the local steric constraints in combination with the effective interactions via polarity and hydrogen bonds, especially
   in view of the heterogeneity of the amino-acid sequences sequences. In addition we would like to understand the folding pathway that ensures a protein's fast approach to its native state in physiological conditions, by avoiding kinetic traps and minimizing the various potential
   frustration effects \cite{Pande,Faccioli}.
   Random amino-acid sequences do not fold into unique conformations, i.e. they have more complicated multi-valley free-energy landscapes,
   so one concludes that those sequences that correspond to proteins have been selected genetically on the basis of their associated free energy landscapes \cite{Dill1984,Dill1990}.

There is little consensus yet as to what is the main driving force in the folding process. Some believe the hydrophobic-hydrophilic effect
(i.e. hydrophobic side-chains try to avoid contact with the solvent, while hydrophilic side-chains seek to be in contact with it)
to be the dominant factor in secondary and tertiary structure formation \cite{Kauzmann,RoseWolfenden,Dill1984,Dill1990}, with steric constraints enforcing further microscopic specificity, and hydrogen bonds providing a locking mechanism \cite{Sippi}. Others believe the folding to be mainly driven by the formation of intra-molecular (or peptidic) hydrogen bonds on top of hydrogen bonding between side-chains and the solvent \cite{Rose}.
Most physicists' studies either resort to models similar to self-avoiding walks on regular lattices \cite{Collet,Oberdorf} (usually via graph-counting and numerical simulations), or focus on generic properties of (free) energy landscapes \cite{Abkevich,Kenzaki,Das}, or try to exploit the one-dimensional nature of the poly-peptide chains \cite{Skan,Garel1,Konkoli}. In either case,  in virtually all studies the amino-acid sequences are regarded as frozen disorder, over which appropriate averages are calculated (in statics of the free energy per monomer, in dynamics of the moment-generating dynamical functional).
This implies that the sequences at hand must be `typical' within an appropriate ensemble of sequences, which  presents us with a serious fundamental
problem.
Amino-acid sequences of proteins are far from random: they have been carefully selected during evolution on the basis of their functionality and their ability to lead to reproducible folds.
Thus one either has to define an ensemble of amino-acid sequences on the basis of the known primary sequences of real proteins that are
being collected in biological databases, which removes the possibility to carry out disorder averages in the mathematical
theory analytically, or one has to find a way to capture the essence of the observed biological sequences (as opposed to random
ones) in simple mathematical formulae. Although some analytical studies did involve non-random sequences, the sequence statistics were
usually not connected to folding quality as such \cite{Shak,Muller}.

There is an alternative strategy in the statistical mechanical
modeling of interacting many-particle systems with non-random disorder, which was followed successfully in the past for e.g.
neural networks (where the synaptic connections between neurons
represent the disorder)
\cite{Coolen1993,Penney1993,Jongen1998,Jongen2001} and for a
simple mean-field hetero-polymer model \cite{Chak}.
 Rather than
averaging over all amino-acid sequences (subject perhaps to experimentally
determined constraints), one combines the process of secondary
structure generation (folding) with a slow evolutionary process for the
amino-acid sequences (which represents the genetic selection of free energy landscapes)  and one couples these two processes
 in a biologically acceptable way. One can then try to solve for the
`slow' process upon assuming adiabatic separation of the two
time-scales, using the so-called finite-$n$ replica theory. This
results in solvable models describing structure generation in
poly-peptide chains with amino-acid sequences that are no longer
random, but selected in a manner that correlates with the folding
process, without having been required to capture the sequence statistics in a formula. It is encouraging  that we know from previous studies such as \cite{Coolen1993,Penney1993,Jongen1998,Jongen2001,Chak}
that in such models the impact of the slow genetic process is indeed generally to drive the systems away from
multi-valley energy landscapes towards single-valley ones.

 In the present paper we take the next step in this research programme.
 Whereas \cite{Chak} involved a simplified model with only
polarity-induced mean-field forces, here we develop a theory for
the coupled dynamics of (fast) folding and
 (slow) sequence selection on the basis of the more precise Hamiltonian introduced
in  \cite {Skan}, which includes also short-range steric forces
along the chain. At a technical level our problem requires the
diagonalization of replicated transfer matrices, for which
efficient methods have been developed only recently
\cite{Theo,Theo2,HSN2005,HSBB2006}.
We apply these diagonalization methods to the present model, within the ergodic (i.e. replica-symmetric, RS) ansatz,
 and show how they lead in the thermodynamic limit to closed
equations for non-trivial order parameters.
 In the context of  protein folding one expects the RS ansatz to be appropriate. In finite dimensional replica
 calculations replica symmetry is known to break down only for small values of the replica dimension $n$, i.e. at high genetic noise levels,
 whereas here our interest is mostly in the regime of low genetic noise levels. Second, given the robustness and reproducibility
 of proteins' secondary and tertiary structures one must assume these systems to operate in an ergodic regime. Third, at a mathematical level,
 our present order parameter equations
 will involve only quantities with  a single replica index, giving yet another indication that RS should hold.
 After first recovering the solutions of the order parameter equations in various known limits, we focus on the biologically most realistic regime
 of sequence selection at zero genetic noise levels. viz. $n\to\infty$,
 where we extract the non-trivial phase phenomenology and derive phase diagrams analytically.
  We find many interesting phase transitions, both continuous and discontinuous, and
 remanence effects, all of which  can be understood and explained on physical grounds.
 This is followed by a numerical analysis of the order parameter equations for nonzero genetic noise levels,
 and by tests of the theoretical predictions against numerical simulations of the coupled sequence selection and folding processes.
Within the limitations imposed by finite size and finite size effects, we find an satisfactory agreement between our theoretical predictions and the numerical simulations.

\section{Model definitions}

\subsection{The folding and sequence selection processes}

\begin{figure}[t]
\setlength{\unitlength}{1mm}
\vsp\hspace*{-15mm}
\begin{picture}(100,55)
\put( 50,  5){\epsfysize=50\unitlength\epsfbox{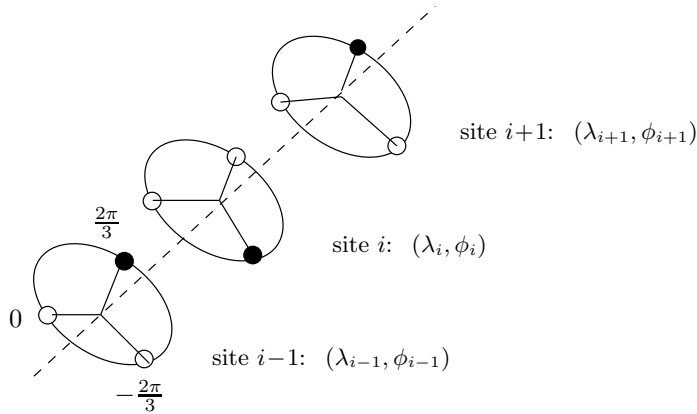}}
\put( 47, 12){$0$}
\put( 58, 25){$\frac{2\pi}{3}$}
\put( 61,  2){$-\frac{2\pi}{3}$}

\put(74,7) {\small site $i\!-\!1$: $~(\lambda_{i-1},\phi_{i-1})$}
\put(90,22) {\small site $i$: $~(\lambda_{i},\phi_{i})$}
\put(107,37) {\small site $i\!+\!1$: $~(\lambda_{i+1},\phi_{i+1})$}
\end{picture}
\vsp
\caption{Illustration of the chemical and mechanical degrees of freedom in our model.
At each site $i$ of the chain we have a discrete variable $\lambda_i$ which specifies the local amino-acid {\em type},
and a residue angle $\phi_i$ which defines its
physical {\em location} relative to the
one-dimensional polymer chain axis (the `backbone', drawn as a dashed
line). In this example the number of possible orientations of each residue is three. The black blobs
represent locations occupied by residues.
The primary structure of the polymer (its chemical composition) is thus defined by $(\lambda_1,\ldots,\lambda_N)$,
and the secondary structure by $(\phi_1,\ldots,\phi_N)$.
Both types of variables are assumed to evolve in time, although on widely separated time-scales.
}
\label{fig:definition}
\end{figure}

\noindent Our model inherits much of its initial features from
\cite{Skan}, and represents the amino-acid cores as nodes in a one
dimensional chain.
 The global conformal state of the system is defined by $N$ successive
 angles
$\bphi=(\phi_1,\ldots,\phi_N)\in \Omega^N$ of amino-acid residues,
relative to the chain's backbone. Here
$\Omega=\{0,2\pi/q,4\pi/q,\ldots,(q-1)2\pi/q\}\subset[0,2\pi)$,
where $q\in \N$. The simplified picture is that of residues being
able to rotate (with constraints, and limited to $q$ discrete
positions)  in a plane perpendicular to the chain's axis. The
primary structure (the amino-acid sequence) is written as
$\blambda=(\lambda_1,\ldots,\lambda_N)$, with
$\lambda_i\in\{1,\ldots,\Lambda\}$ denoting the residue species at
position $i$ in the chain (with $\Lambda=20$ for real proteins). See also figure \ref{fig:definition}.
In contrast to \cite{Skan}, however, the primary sequence will
here not be drawn at random, but will be generated by an
appropriate genetic selection process; this improves the
biological realism of the model, but  will change and complicate
the mathematics significantly. We will therefore only include
monomer-solvent polarity forces and steric forces, leaving out hydrogen bonds for
now. Furthermore, we refine the Hamiltonian used in \cite{Skan} to
take into account the effect of the polymer's overall polarity
balance on its ability to exhibit predominantly hydrophilic surface
residues and hydrophobic core residues; for models with
fixed primary sequences as in \cite{Skan} this would add an
irrelevant constant to the energy, but for models such as the
present where the monomer sequences evolve in time this energy
contribution will exert sequence selection pressure with
significant consequences. In many of our calculations we will also
choose $q=2$, i.e. limit the residue angles to
$\phi_i\in\{0,\pi\}$. This prevents us from having to generalize
the diagonalization methods of \cite{Theo,Theo2}, which would
probably require a separate study in itself. Thus, for a given
realization of the primary sequence $\blambda$, the folding
process is assumed to be governed by the following Hamiltonian:
\begin{eqnarray}
H_{\rm f}(\bphi|\blambda)&=& -\frac{J_p}{N}\sum_{ij}\xi(\lambda_i)
\xi(\lambda_j)~\delta_{\phi_i,\phi_j}\nonumber
\\
&&
-J_s\sum_{i}\cos[(\phi_{i+1}\!-\phi_i)-(\phi_i\!-\phi_{i-1})-a(\lambda_i)]~~~~~
\label{eq:H}
\end{eqnarray}
$\xi(\lambda)\in\R$ measures the polarity of residue $\lambda$
(with $\xi>0$ indicating hydrophobicity and $\xi_i<0$ indicating
hydrophilicity). The first term in (\ref{eq:H})  favours
conformations where hydrophobic and hydrophilic avoid identical
orientations, since this makes it easier for the polymer to find a
fold  that shields its hydrophobic residues from the solvent while
exposing its hydrophilic ones. The second term represents in a
simplified manner the effects of steric forces, characterizing
each residue $\lambda$ by a winding `distortion' angle $a(\lambda)$
for successive residue rotations.
If $a(\lambda_i)=0$, then residue $i$ will prefer to have an angle $\phi_i$ such that
torsion along the chain is homogeneous, i.e. $\phi_{i+1}\!-\phi_i=\phi_i\!-\phi_{i-1}$.
The energies $J_p>J_s>0$ control
the relative impact of each contribution. For a fixed sequence one
can define the partition function $Z_{\rm f}(\blambda)$ and the
free energy $F_{\rm f}(\blambda)$ for the equilibrium state of the
folding process at temperature $T_{\rm f}=\beta^{-1}$ (in units
where the Boltzmann constant equals $k_B=1$):
\begin{eqnarray}
\Z_{\rm f}(\blambda)&=&\sum_{\bphi}\exp[-\beta H_{\rm
f}(\bphi|\blambda)]\\ F_{\rm f}(\blambda)&=& -\beta^{-1}\log
\Z_{\rm f}(\blambda)
\end{eqnarray}
It will be convenient to characterize the relevant chemical
characteristics of amino-acids by the distribution
\begin{equation}
w(\xi,\eta)=\frac{1}{\Lambda}\sum_{\lambda=1}^\Lambda\delta[\xi-\xi(\lambda)]\delta[\eta-\cos[a(\lambda)]]
\end{equation}
 As there is no obvious structural physical/chemical link between residue
polarity and geometric (steric) properties, we assume
statistical independence, i.e. $w(\xi,\eta)=w(\xi)w(\eta)$ (this
will also induce welcome simplifications later). Typical simple
choices for $w(\xi)$ would be
$w(\xi)=\epsilon\delta(\xi)+\frac{1}{2}(1\!-\!\epsilon)
[\delta(\xi-1)+\delta(\xi+1)]$ or
$w(\xi)=\frac{1}{2}\theta[1-\xi]\theta[1+\xi]$. Note that we may
always choose the maximum polarity to be one, since alternative
values can be absorbed into the definition of the parameter $J_p$.
For $w(\eta)$, natural choices would be
$w(\eta)=\pi^{-1}\int_0^\pi\!da~\delta[\eta-\cos(a)]=\pi^{-1}[1-\arccos^2(\eta)]^{-1/2}\theta[1-\eta]\theta[1+\eta]$
or $w(\eta)=\frac{1}{2}\theta[1-\eta]\theta[1+\eta]$. Here the
allowed value range $[-1,1]$ is enforced by the physical meaning
of $\eta$.

We now follow \cite{Chak} and complement the folding process by an
adiabatically slow stochastic evolutionary selection process for
the amino-acid sequences. The assumption is that this selection
results from an interplay between the demands that (i)
 a sequence must lead to a unique and easily reproducible equilibrium conformation
 for its associated folding process,  and (ii) the resulting
structure is useful to the organism (e.g. it can act as a catalyst
of some metabolic or proteomic cellular reaction).  If one takes
the further step to quantify the quality of an equilibrium
conformation by the value of the folding free energy $F_{\rm
f}(\blambda)$ (i.e. taking `low free energy' as a proxy for `more
reproducible'), together with the direct energetic cost
$V(\blambda)$ of not having strictly hydrophilic `surface
residues' and strictly hydrophobic `core residues', and if one
assumes that biological usefulness can be measured by some utility
potential $U(\blambda)$, then the evolutionary process can be
viewed as the stochastic minimization of an effective Hamiltonian
for amino-acid sequences that takes the form
\begin{eqnarray}
H_{\rm eff}(\blambda)&=&U(\blambda)+ V(\blambda) -\beta^{-1}\log
\Z_{\rm f}(\blambda) \label{eq:Heff}
\end{eqnarray}
 If the stochastic
minimization is of the Glauber or Monte-Carlo type, the
evolutionary process will evolve itself to a Boltzmann-type
equilibrium state, namely $P_\infty(\blambda)\!\propto\!
\exp[-\tilde{\beta}H_{\rm eff}(\blambda)]$, where $\tilde{\beta}$
measures the (inverse) noise level in the genetic selection
\footnote{Another way to see why $P_\infty(\blambda)\propto
\exp[-\tilde{\beta}H_{\rm eff}(\blambda)]$ is a natural
evolutionary equilibrium state is to image having real-valued
$\blambda$, evolving according to a Langevin equation in which the
deterministic force is minus the gradient of the energy $H_{\rm
f}(\blambda)+U(\blambda)+V(\blambda)$. Given adiabatic separation of folding
and evolution time-scales, one can then integrate out the fast
variables (the conformation angles) and find the Boltzmann state
for the sequences $\blambda$ with effective Hamiltonian
(\ref{eq:Heff}). See e.g. \cite{Chak} for details.}. Our combined
model (fast folding and slow genetic sequence selection) is thus
solved in equilibrium by calculating the associated effective free energy
per monomer
\begin{eqnarray}
f_N&=& -\frac{1}{\tilde{\beta} N}\log \sum_{\blambda}
e^{-\tilde{\beta}H_{\rm eff}(\blambda)} \nonumber
\\
&=& -\frac{1}{n\beta N}\log \sum_{\blambda} [\Z_{\rm
f}(\blambda)]^n e^{-n\beta[ U(\blambda)+V(\blambda)]}
\label{eq:overallf}
\end{eqnarray}
with the noise level ratio $n=\tilde{\beta}/\beta$.
 As in
\cite{Coolen1993,Penney1993,Jongen1998,Jongen2001,Chak}, this
expression can be evaluated via the replica formalism, where $n$
is first taken to be integer and the result is subsequently
continued to non-integer values. Note that in this type of model
the replica dimension has a clear physical meaning as the ratio of
temperatures. For $n\to 0$ we recover the free energy of a system
with quenched random amino-acid sequences, for $n=1$ we have that
of  an annealed model, whereas for $n\to\infty$ the sequence
selection becomes strictly deterministic. In contrast to
previous coupled dynamics studies, however, here we have not only
mean-field forces but also short-range ones: the steric
interactions in (\ref{eq:H}). The replica calculation will
therefore be quite different.

In this paper we limit ourselves for mathematical convenience to
sequence functionality potentials of the simple form
$U(\blambda)=\sum_{i} u(\lambda_i)$. Similarly we choose the
energetic penalty $V(\blambda)$ on hydrophobic surface residues or
hydrophilic core residues to be a function only of the polarity
balance $k(\blambda)=N^{-1}\sum_{i}\xi(\lambda_i)$, putting
 $V(\blambda)=J_{g}N v(k(\blambda)-k^\star)$ with a function $v(k)$ that is minimal for $k=0$, where
 $k^\star$ represents  the `optimal' polarity balance
 that would give a protein with strictly
hydrophilic surface residues and strictly hydrophilic core
residues (which one expects to be close to zero). This form for $V(\blambda)$ would emerge naturally
 if all amino-acids were to have similar values of
$|\xi(\lambda_i)|$. The implicit assumption is that if a polarity
balance $k(\blambda)$ is energetically  favourable, i.e. close to
$k^\star$, then the protein will be able to find a fold that
realizes the desired geometric separation of core versus surface
residues. We will discuss the mathematical consequences of making
alternative choices in the discussion section.
 Since for
$N\to\infty$ chain boundary effects must vanish, we also choose
periodic boundary conditions and take $N$ even (for mathematical
reasons which will become clear later).

\subsection{Relation between model assumptions and biological reality}

\begin{figure}[t]
\vspace*{-1mm} \hspace*{27mm} \setlength{\unitlength}{0.5mm}
\begin{picture}(300,170)

   \put(0,90){\includegraphics[height=75\unitlength,width=112\unitlength]{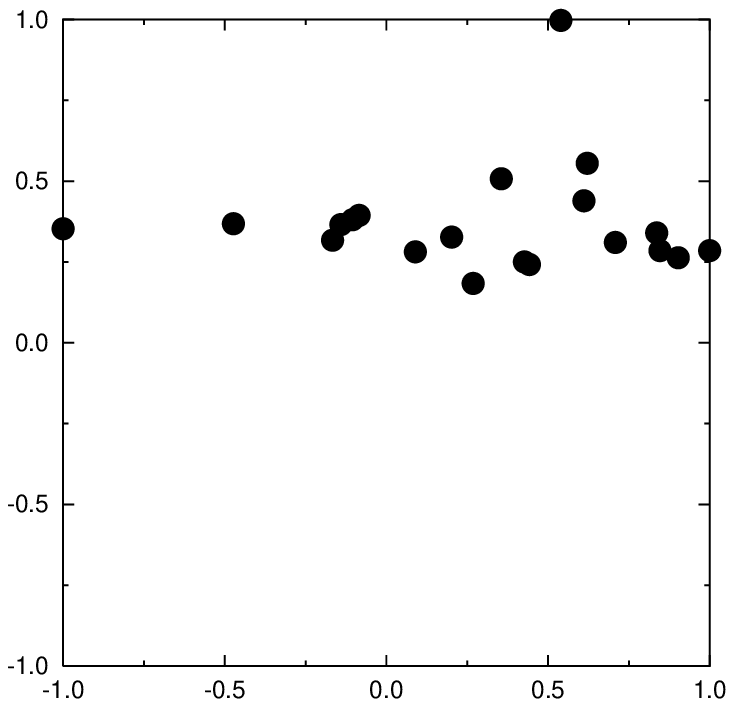}}
   \put(115,90){\includegraphics[height=75\unitlength,width=112\unitlength]{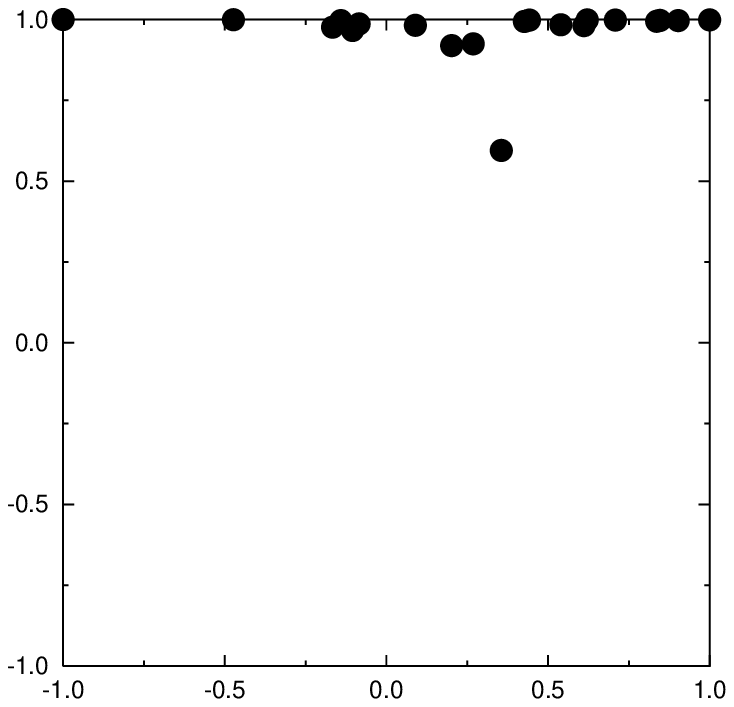}}
   \put(0,0){\includegraphics[height=75\unitlength,width=112\unitlength]{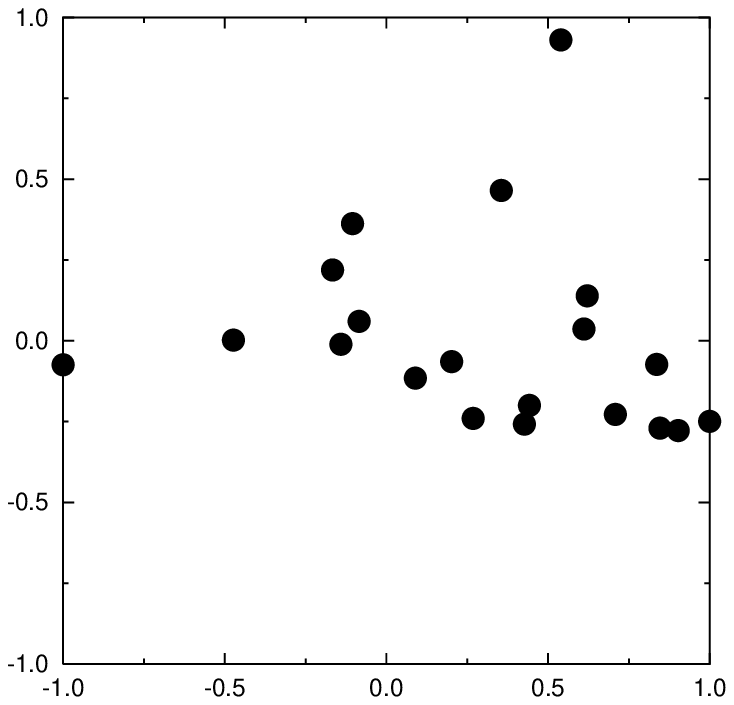}}
   \put(115,0){\includegraphics[height=75\unitlength,width=112\unitlength]{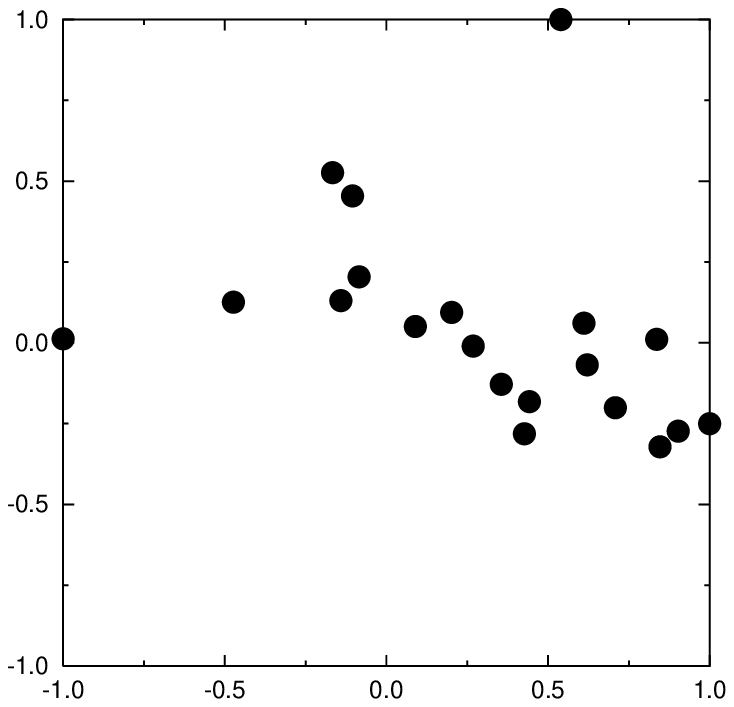}}

  \put(47,-9){$\xi$}    \put(162,-9){$\xi$}     \put(47,81){$\xi$}   \put(162,81){$\xi$}
   \put(-18,38){$\cos[\overline{\phi}]$}    \put(97,38){$\cos[\overline{\psi}]$}
   \put(-18,128){$\cos[\overline{\phi}]$}    \put(97,128){$\cos[\overline{\psi}]$}

   \put(-55,60){$\beta$-proteins:}   \put(-55,150){$\alpha$-proteins:}

\end{picture}
 \vspace*{2mm}
\caption{Diagrams showing each of the twenty amino-acids as a point in the plane, with the horizontal coordinate giving
its polarity value (taken from \cite{Eisenberg}, and normalized to the range $[-1,1]$), and with the horizontal coordinate giving the cosine
of an average conformation angle (averaged over all proteins of a given class).  left: averages calculated for conformation angle $\phi$; right: averages calculated for conformation angle $\psi$. Top row: averaging over all $\alpha$-proteins for which structures are available; bottom row: averaging over all $\beta$-proteins for which structures are available. All conformation data were extracted from \cite{protein_database1,protein_database2}.
} \label{fig:aminoacids} \vspace*{-2mm}
\end{figure}

\begin{figure}[t]
\vspace*{-1mm} \hspace*{27mm} \setlength{\unitlength}{0.5mm}
\begin{picture}(300,170)

   \put(0,90){\includegraphics[height=75\unitlength,width=112\unitlength]{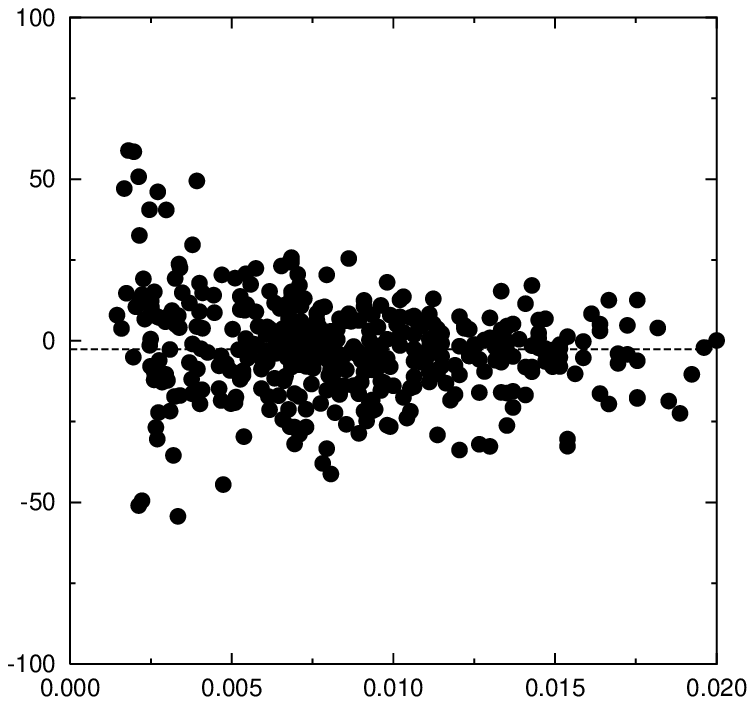}}
   \put(110,90){\includegraphics[height=75\unitlength,width=112\unitlength]{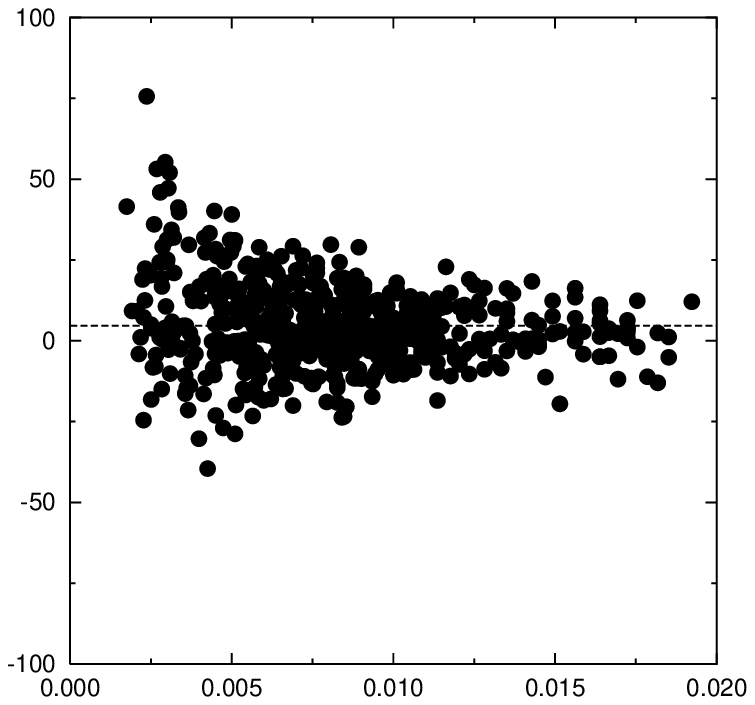}}
   \put(0,0){\includegraphics[height=75\unitlength,width=112\unitlength]{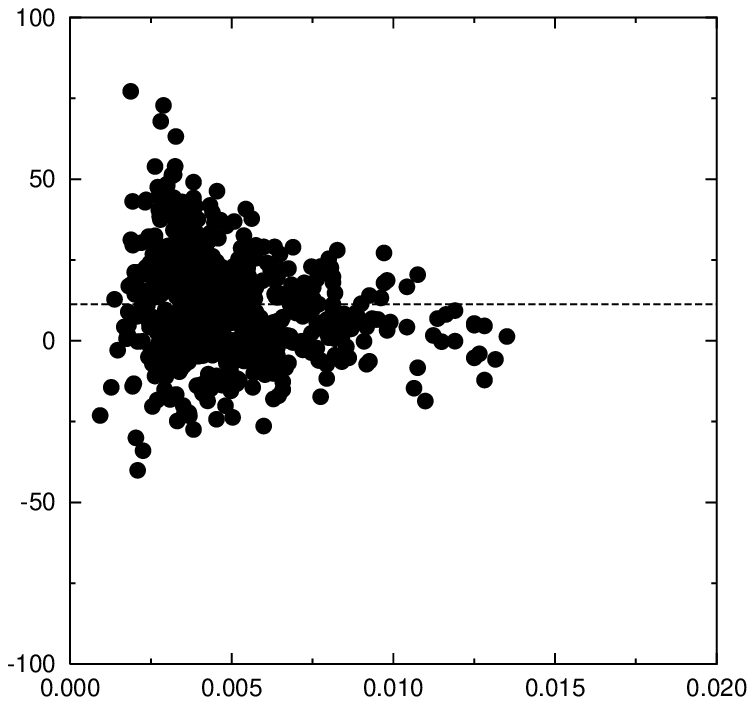}}
   \put(110,0){\includegraphics[height=75\unitlength,width=112\unitlength]{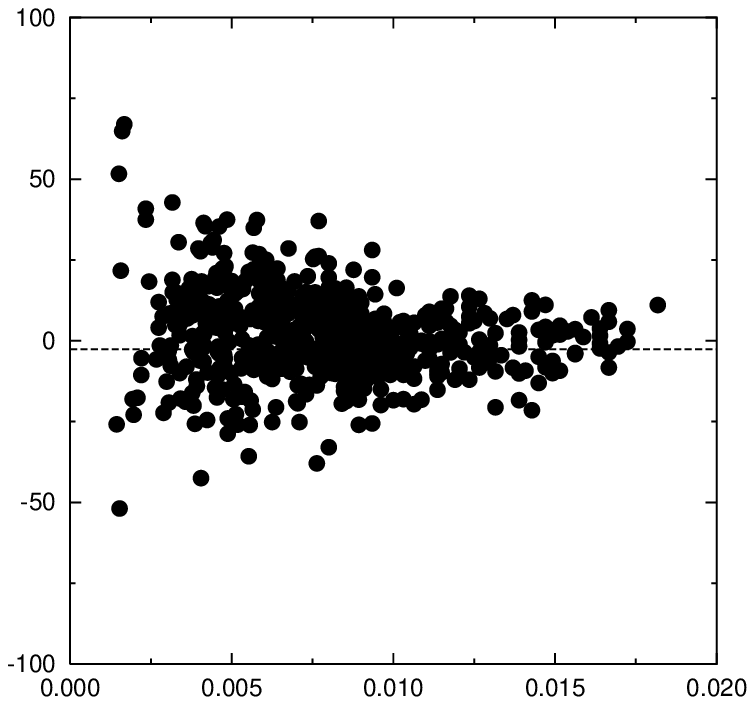}}

  \put(42,-9){\small $1/N$}    \put(152,-9){\small $1/N$}     \put(42,81){\small $1/N$}   \put(152,81){\small $1/N$}
   \put(-12,38){$k_{\rm raw}$}    \put(98,38){$k_{\rm raw}$}
   \put(-12,128){$k_{\rm raw}$}    \put(98,128){$k_{\rm raw}$}

    \put(45,153){$\alpha$-proteins} \put(155,153){$\beta$-proteins}
    \put(41,63){$\alpha\!/\!\beta$-proteins}  \put(147,63){$\alpha\!+\!\beta$-proteins}

\end{picture}
 \vspace*{2mm}
\caption{Diagrams showing all proteins in data base \cite{protein_database2}, organized into the four main protein families,
as points in the plane, with the horizontal coordinate giving their inverse size $N^{-1}$ and with the vertical coordinate giving
their {\em average} polarity value $k_{\rm raw}=N^{-1}\sum_i \xi_{i,\rm raw}$  along the chain (where the polarities $\xi_{i,\rm raw}$
of the constituent residues are taken directly from \cite{Eisenberg}, without normalization to $[-1,1]$) Dashed horizontal lines indicate the average overall polarity level found within each protein class.
} \label{fig:average_polarity} \vspace*{-2mm}
\end{figure}

Here we discuss some of the assumptions and definitions of our model
in the light of experimental evidence from real proteins.
Our choice for a single-angle representation of the mechanical degrees of freedom of a monomer
 was motivated by our desire to limit the mathematical complexity, although our methods would apply
also if we were to work with the conventional two conformation angles $(\phi,\psi)$.
In fact, there is evidence \cite{Parker} to suggest that the conventional two-angle representation is redundant,
and that only one newly defined torsion angle is needed per amino-acid to specify a protein's conformation.
If we insist on identifying the single-site degrees of freedom in our model with one of the standard conformation angles $(\phi,\psi)$,
we have to choose the one that matches our statistical assumptions best.
To do this we have calculated for individual amino-acids the average of the observed conformation angles $(\phi,\psi)$ over all occurrences of this amino-acid in the data base
of known protein structures (the SCOP data base \cite{protein_database1,protein_database2}), which resulted in the graphs of figure
 \ref{fig:aminoacids}, where we plot the cosines of the average conformation angles of all twenty amino-acids together with their polarity value (according to the Eisenberg scale, taken from \cite{Eisenberg}, and normalized linearly to the range $[-1,1]$).
 Both conformation angles $(\phi,\psi)$
 give averages that have cosines of both signs, both are biased towards positive values; however, the bias is more extreme in the case of $\psi$.
 Since there is no such bias in our theory, the most suitable conformation angle to correspond to the orientation degrees of freedom in our model appears to be $\phi$.
In the same figure we can also see that there is no obvious correlation between polarity characteristics and steric characteristics.
In our model this is assumed to be a property of the amino-acids, and we will find in our analysis that neither the primary structure generation nor the secondary structure generation introduces any such correlations.
Finally, let us turn to the postulated preferred average polarity of any amino-acid chain (which was used in our phenomenological Hamiltonian),
 purely on the basis of the energetic need to shield
hydrophobic residues from the solvent and to expose hydrophilic ones. There is certainly evidence for the link between the average polarity of
a sequence and the surface-exposure pattern of the associated protein structure \cite{Moelbert}.
If we plot all those proteins for which primary structure data are available as points in a plane, with the inverse size $1/N$ as horizontal coordinate and the average polarity as vertical coordinate, we obtain figure \ref{fig:average_polarity}. This figure supports strongly the existence of a an energetically preferred average polarity $k^\star$, with a value close to zero in rescaled polarity units $\xi\in[-1,1]$.

\section{Replica analysis of the model}

\noindent For integer $n$ one can write the $n$-th power of the folding partition function
$\Z_{\rm f}(\blambda)$ in (\ref{eq:overallf}) in terms of $n$
replicas of the original system, to be labeled by $\alpha=1\ldots
n$. If the sum over the sequences $\blambda$ is carried out before
the sum over conformations, one finds an effective theory in which
the $n$ replicas are coupled:
\begin{eqnarray}
f_N&\!=\!& -\frac{1}{n\beta N}\log \sum_{\bphi^1\ldots \bphi^n}
e^{- \beta {\cal H}(\bphi^1,\ldots,\bphi^n)}
\label{eq:replicated_f}
\\
{\cal H}(\ldots)&\!=\!& -\!\frac{1}{\beta}\log \sum_{\blambda}
e^{-\beta \sum_{\alpha}H_{\rm f}(\bphi^\alpha|\blambda)-n\beta[
U(\blambda)+V(\blambda)]}~~~ \label{eq:calH}
\end{eqnarray}
For $\beta\to 0$ (infinite temperature) we have $\beta {\cal
H}(\ldots)\to -N \log \Lambda$ and the free energy retains only
entropic terms, viz. $\lim_{\beta\to 0}(\beta f_N)=-\log q
-n^{-1}\log  \Lambda$.
 Upon
using (\ref{eq:H}), and inserting $\sum_\phi
\delta_{\phi,\phi_i^\alpha}$ into the polarity term of the folding
energy, we can work out the effective Hamiltonian (\ref{eq:calH}). If we introduce appropriate integrals over
$\delta$-functions (written in integral representation) to isolate
the quantities
$N^{-1}\sum_i\xi(\lambda_i)\delta_{\phi,\phi_i^\alpha}$, viz.
\begin{equation}
1=\int\!\frac{dz_{\alpha\phi}
d\hat{z}_{\alpha\phi}}{2\pi}~e^{i\hat{z}_{\alpha\phi}[z_{\alpha\phi}-N^{-1}\sum_i\xi(\lambda_i)\delta_{\phi,\phi_i^\alpha}]}
\end{equation}
 we can carry out the
sum over $\blambda$ in (\ref{eq:calH}) and find,  with the
abbreviation $\bz=\{z_{\alpha\phi}\}$,
\begin{eqnarray}
\hspace*{-20mm}
-\frac{\beta}{N} {\cal H}(\ldots)&=& \frac{1}{N}\log
\sum_{\blambda} e^{-n\beta[\sum_i u(\lambda_i)+N J_g
v(k(\blambda)-k^\star)] + \frac{\beta
J_p}{N}\sum_\phi\sum_\alpha[\sum_{i}\xi(\lambda_i)
\delta_{\phi,\phi_i^\alpha}]^2}
\nonumber
\\[-1mm]
\hspace*{-20mm}
&&\hspace*{30mm}\times
e^{\beta
J_s\sum_{i\alpha}\cos[\phi^\alpha_{i+1}+\phi^\alpha_{i-1}-2\phi^\alpha_i-a(\lambda_i)]
} \nonumber
\\[2mm]
\hspace*{-20mm}
&=& \frac{1}{N}\log \int\!\!\frac{d\bz d\hat{\bz}}{(2\pi/\beta
N)^{qn}}e^{\beta N[i\sum_{\alpha\phi}\hat{z}_{\alpha\phi}
z_{\alpha\phi}+ J_p \bz^2- n  J_g
v(\frac{1}{n}\sum_{\alpha\phi}z_{\alpha\phi}-k^\star)]} \nonumber
\\
\hspace*{-20mm}
&&\hspace*{-15mm} \times \prod_i\left\{ \sum_{\lambda}e^{-n\beta
u(\lambda)-i\beta\xi(\lambda)
\sum_{\alpha\phi}\hat{z}_{\alpha\phi}\delta_{\phi,\phi_i^\alpha}
+\beta
J_s\sum_{\alpha}\cos[\phi^\alpha_{i+1}+\phi^\alpha_{i-1}-2\phi^\alpha_i-a(\lambda)]
} \right\}~~~~
\end{eqnarray}
Inserting this into (\ref{eq:replicated_f}) leads to an expression
for the asymptotic free energy per monomer
$f=\lim_{N\to\infty}f_N$ that can be evaluated by steepest
descent. Upon eliminating the conjugate integration variables
$\{\hat{z}_{\alpha\phi}\}$ by variation of $\{z_{\alpha\phi}\}$,
giving $ i\hat{z}_{\alpha\phi} =
 J_g v^\prime(\frac{1}{n}\sum_{\alpha\phi}z_{\alpha\phi}-k^\star)- 2J_p
 z_{\alpha\phi}$,
 and upon defining the replicated single-site vectors
$\bphi_i=(\phi_i^1,\ldots,\phi_i^n)$ the result takes the form
$f={\rm extr}_{\bz} \varphi_n(\bz)$ with
\begin{eqnarray}
\hspace*{-15mm}
n\varphi_n(\bz)&=& J_p\sum_{\alpha\phi}z_{\alpha\phi}^2 +
nJ_g\Big[v(\frac{1}{n}\!\sum_{\alpha\phi}\!z_{\alpha\phi}\!-\!k^\star)
-(\frac{1}{n}\!\sum_{\alpha\phi}\!z_{\alpha\phi})v^\prime(\frac{1}{n}\!\sum_{\alpha\phi}\!z_{\alpha\phi}\!-\!k^\star)\Big]
 \nonumber
\\
\hspace*{-15mm}
&& \hspace*{10mm}
-\frac{1}{\beta }\log \Lambda -\lim_{N\to\infty}\frac{1}{\beta N}\log
\!\sum_{\bphi^1\ldots \bphi^n}\!\prod_i\!
M[\bphi_{i-1},\bphi_i,\bphi_{i+1}|\bz]
 \label{eq:full_phi}~~~~
\\
\hspace*{-15mm}&&\hspace*{-18mm}
 M[\bphi_{i-1},\bphi_i,\bphi_{i+1}|\bz]=
\frac{1}{\Lambda} \sum_{\lambda=1}^{\Lambda}e^{\beta\xi(\lambda)
\sum_{\alpha}[2J_p z_{\alpha\phi^\alpha_i} -J_g
v^\prime(\frac{1}{n}\sum_{\alpha\phi}z_{\alpha\phi}-k^\star)]}
\nonumber
\\
\hspace*{-15mm}&& \hspace*{20mm}\times e^{\beta
J_s\sum_{\alpha}\cos[\phi^\alpha_{i+1}+\phi^\alpha_{i-1}-2\phi^\alpha_i-a(\lambda)]-n\beta
u(\lambda) }
 \label{eq:full_M}
\end{eqnarray}
 We recognize in
(\ref{eq:full_phi},\ref{eq:full_M}) a replicated
transfer matrix product embedded within a mean-field calculation,
and conclude that this model is therefore in principle solvable.
The only amino-acid characteristics that affect the folding
process are its polarity $\xi(\lambda)$ and steric angle
$a(\lambda)$, so we will from now on choose the single site
functionality potential to have the form $u(\lambda)=\mu
\xi(\lambda)+\nu\cos[a(\lambda)]$ (where $\mu$ and $\nu$ are
control parameters).

\subsection{The case $q=2$}

\noindent Our calculations become significantly simpler and more
transparent for $q=2$. Here, after a uniform basis rotation, the
allowed residue angles are $\phi_i\in\{-\pi/2,\pi/2\}$, which can
be written in terms of Ising spin variables $\sigma_i\in\{-1,1\}$
as $\phi_i=\sigma_i\pi/2$. We transform the $2n$ remaining
replicated order parameters, which can be written as $z_{\alpha
\pm}$, into new order parameters $m_\alpha=z_{\alpha +}-z_{\alpha
-}$ and $k_\alpha=z_{\alpha +}+z_{\alpha -}$.
 Our equations will now involve the replicated
spin variables $\bsigma_i=(\sigma_i^1,\ldots,\sigma_i^n)$, and the
cosine term in the exponent of the transfer matrix simplifies to
$\sigma_{i+1}^\alpha\sigma_{i-1}^\alpha \cos[a(\lambda)]$. With
the short-hands $\mb=(m_1,\ldots,m_n)$ and $\bk=(k_1,\ldots,k_n)$,
$\bra g(\xi)\ket_\xi=\int\!d\xi~w(\xi)g(\xi)$, and $\bra
g(\eta)\ket_\eta=\int\!d\eta~w(\eta)g(\eta)$, our previous
 expressions (\ref{eq:full_phi},\ref{eq:full_M}) take the form
\begin{eqnarray}
\hspace*{-22mm}
n\varphi_n(\mb,\bk)&=& \frac{1}{2}J_p(
\bk^2\!\!+\!\mb^2)
+ nJ_g
\Big[v(\frac{1}{n}\!\sum_{\alpha}\!k_{\alpha}\!-\!k^\star)
-(\frac{1}{n}\!\sum_{\alpha}\!k_{\alpha})v^\prime(\frac{1}{n}\!\sum_{\alpha}\!k_{\alpha}\!-\!k^\star)\Big]
\nonumber \\
 \hspace*{-22mm} &&
 -\frac{1}{\beta }\log \Lambda
-\lim_{N\to\infty}\!\frac{1}{\beta N}\log\!
\!\sum_{\bsigma_{\!1}\ldots\bsigma_{\!N}}\!\prod_i\!
M[\bsigma_{i-1},\bsigma_i,\bsigma_{i+1}|\mb,\bk]
  \label{eq:q2phi}~~~~
\\
\hspace*{-22mm} &&\hspace*{-20mm}
 M[\bsigma_{i-1},\bsigma_i,\bsigma_{i+1}|\mb,\bk]=
\bra e^{ \beta\xi \big[J_p\sum_{\alpha}(k_\alpha +
m_{\alpha}\sigma_i^\alpha)-n\mu
 -nJ_g
v^\prime(\frac{1}{n}\sum_{\alpha}\!k_{\alpha }-k^\star)]\big]}
\ket_\xi
\nonumber
\\
\hspace*{-22mm}&& \hspace*{25mm} \times
\bra e^{\beta \eta
[J_s\bsigma_{i+1}\cdot\bsigma_{i-1}-n\nu] }\ket_\eta
  \label{eq:q2M}
\end{eqnarray}
 The disconnection
inside $M[\ldots|\ldots]$ of the factor involving $\bsigma_i$ from
that involving $\bsigma_{i+1}\cdot\bsigma_{i-1}$ allows us to
rewrite $\varphi_n(\mb,\bpsi)$ into a more convenient form, with a
new replicated transfer matrix $\bGamma(\mb,\bk)$ that involves
only the two sites $i-1$ and $i+1$:
\begin{equation}
\prod _i
 M[\bsigma_{i-1},\bsigma_i,\bsigma_{i+1}|\mb,\bk]=\prod _i
 \Gamma_{\bsigma_{i-1},\bsigma_{i+1}}(\mb,\bk)
 \label{eq:decomposition}
 \end{equation}
 where
\begin{eqnarray}\hspace*{-20mm}
 \Gamma_{\bsigma\bsigma^\prime}(\mb,\bk)
 &=& \big\bra
e^{\beta\eta[J_s\bsigma\cdot\bsigma^\prime\!-n\nu]}
\big\ket_{\eta}
\big\bra e^{\beta\xi[ J_p (\sum_\alpha
\!k_\alpha+\mb\cdot\bsigma)-n\mu-nJ_g
v^\prime(\frac{1}{n}\sum_{\alpha}\!k_{\alpha
}-k^\star)]}\big\ket_{\xi}
 \label{eq:Gamma}
\end{eqnarray}
 Since $N$ is even and we have periodic boundaries,
even sites thereby disconnect from odd sites. The trace in
$\varphi_n$ is now in leading order for large $N$ expressed in the
usual manner in terms of the largest eigenvalue $\lambda(\mb,\bk)$
of the matrix $\bGamma(\mb,\bk)$:
\begin{eqnarray}
&&\hspace*{-15mm} \lim_{N\to\infty}\frac{1}{N}\log
\sum_{\bsigma_1\ldots \bsigma_N} \prod _i
 M[\ldots|\ldots]\nonumber
 \\
 &=& \lim_{N\to\infty}\frac{2}{N}\log {\rm
 Tr}[\bGamma^{N/2}(\mb,\bk)] =\log \lambda(\mb,\bk)
 \end{eqnarray}
 In fact, the specific dependence of $\varphi_n(\mb,\bk)$ on $\bk$ via (\ref{eq:q2M}) is such that all its
 saddle-points will have $\bk=k(1,\ldots,1)$. This reduces the number
 of order parameters from $2n$ to $n+1$. We now have $f={\rm extr}_{\mb,k}
 \varphi_n(\mb,k)$,with
 \begin{eqnarray}
 &&
 \varphi_n(\mb,\bk)= \frac{1}{2}J_p(\frac{\mb^2\!\!}{n}\!+\!k^2)+ J_g
\big[v(k\!-\!k^\star)\!-\!kv^\prime(k\!-\!k^\star)\big]-\frac{1}{n\beta }\log \Lambda\nonumber
\\ &&\hspace*{25mm}
 -\frac{1}{n\beta }\log\lambda(\mb,k)
\label{eq:q2phi_new}
\\
&&\hspace*{-5mm}
 \Gamma_{\bsigma\bsigma^\prime}(\mb,k)
 =
 \big\bra
e^{\beta\eta[J_s\bsigma\cdot\bsigma^\prime\!-n\nu]}
\big\ket_{\eta}
\big\bra e^{n\beta\xi[ J_p (k+n^{-1}\mb\cdot\bsigma)-\mu-J_g
v^\prime(k-k^\star)]}\big\ket_{\xi}
 \label{eq:Gamma2}
\end{eqnarray}
Our problem has been reduced to the diagonalization of the
$2^n\times 2^n$ replicated transfer matrix (\ref{eq:Gamma2}). This
 matrix can be simplified to a form analyzed in
\cite{Theo,Theo2,HSN2005,HSBB2006} upon making the so-called
replica symmetric (RS) ansatz, which is equivalent to assuming
ergodicity. Since the order parameters in the present model have
at most one replica index, one expects RS to be exact at all
temperatures. Now one has $m_\alpha=m$ for all $\alpha$, which
simplifies our solution to $f={\rm extr}_{m,k} \varphi_{\rm
RS}(m,k)$ in which
 \begin{eqnarray}
\varphi_n(m,k)&=& \frac{1}{2}J_p(m^2+k^2)+ J_g
\big[v(k\!-\!k^\star)\!-\!kv^\prime(k\!-\!k^\star)\big]\nonumber
\\ &&- \frac{\log\Lambda+\log \lambda_{\rm RS}(m,k)}{\beta n}
 ~~~~\label{eq:q2phi_RS}
\end{eqnarray}
where $\lambda_{\rm RS}(m,k)$ is the largest eigenvalue
of
\begin{eqnarray}
\hspace*{-10mm}
 \Gamma^{\rm RS}_{\bsigma\bsigma^\prime}(m,k)
 &=&
 \big\bra
e^{\beta\eta[J_s\bsigma\cdot\bsigma^\prime\!-n\nu]}
\big\ket_{\eta}
\big\bra e^{n\beta\xi[ J_p (k+
\frac{m}{n}\sum_\alpha\sigma_\alpha)-\mu-J_g
v^\prime(k-k^\star)]}\big\ket_{\xi}
\label{eq:GammaRS}
\end{eqnarray}
Working out the saddle-point equations for $\{m,k\}$ from
(\ref{eq:q2phi_RS}) leads us to
\begin{eqnarray}
m&=& \frac{1}{\beta n J_p}\frac{\partial}{\partial m}\log
\lambda_{\rm RS}(m,k)~~~~ \label{eq:eqn_for_m}
\\
k&=& \frac{1}{\beta n[J_p\!-\!J_g
v^{\prime\prime\!}(k\!-\!k^\star)] }\frac{\partial}{\partial
k}\log \lambda_{\rm RS}(m,k)~~~~
 \label{eq:eqn_for_k}
\end{eqnarray}
An alternative (but equivalent) form for our order parameter
equations that does not require differentiation of $\lambda_{\rm
RS}(m,k)$ is obtained if we extremize $\varphi_n(m,k)$ at the
stage where it is still expressed in terms of a trace of powers of
the matrix $\Gamma_{\rm RS}(m,k)$, viz.
\begin{eqnarray}
m&=& \frac{2}{\beta n
J_p}\lim_{N\to\infty}\frac{\partial}{\partial m}\log ~{\rm
Tr}[\Gamma^{N/2}_{\rm RS}(m,k)]\nonumber
\\
&=&\frac{1}{\beta n J_p}\lim_{N\to\infty}\frac{{\rm
Tr}[\frac{\partial}{\partial m}\Gamma_{\rm
RS}(m,k).\Gamma^{N/2}_{\rm RS}(m,k)]}{{\rm Tr}[\Gamma^{N/2}_{\rm
RS}(m,k)]}
\\
k&=& \frac{2}{\beta
n}\frac{\lim_{N\to\infty}\frac{\partial}{\partial k}\log ~{\rm
Tr}[\Gamma^{N/2}_{\rm RS}(m,k)]}{J_p\!-\!J_g
v^{\prime\prime\!}(k\!-\!k^\star) } \nonumber
\\
&=& \frac{1}{\beta n }\lim_{N\to\infty}\frac{{\rm
Tr}[\frac{\partial}{\partial k}\Gamma_{\rm
RS}(m,k).\Gamma^{N/2}_{\rm RS}(m,k)]}{[J_p\!-\!J_g
v^{\prime\prime\!}(k\!-\!k^\star)]{\rm Tr}[\Gamma^{N/2}_{\rm
RS}(m,k)]}~~~
\end{eqnarray}
Upon working out the partial derivatives of $\Gamma_{\rm
RS}(m,k)$, and upon writing the left- and right eigenvectors of
$\Gamma_{\rm RS}(m,k)$ corresponding to the largest eigenvalue as
$\{u^{\rm L}_{\bsigma}\}$ and $\{u^{\rm R}_{\bsigma}\}$, the
limit $N\to\infty$ can be taken. To avoid unwieldy equations we
 drop the explicit mentioning of the arguments
$(m,k)$ for quantities such as $\lambda_{\rm RS}$,
$\{u^{\rm L}_{\bsigma}\}$ or $\{u^{\rm R}_{\bsigma}\}$ from now on;
the formulae should make this dependence clear. Using the replica
permutation invariance of RS equations, the result can be written
as
\begin{eqnarray}
m&=& \frac{\sum_{\bsigma\bsigma^\prime}u^{\rm L}_{\bsigma}
\sigma_1 Y_{\bsigma\bsigma^\prime}u^{\rm R}_{\bsigma^\prime}}{
\lambda_{\rm RS}\sum_{\bsigma}u^{\rm L}_{\bsigma}u^{\rm
R}_{\bsigma}} \label{eq:m_in_Y}
\\
k&=& \frac{\sum_{\bsigma\bsigma^\prime}u^{\rm L}_{\bsigma}
Y_{\bsigma\bsigma^\prime}u^{\rm R}_{\bsigma^\prime}}{ \lambda_{\rm
RS}\sum_{\bsigma}u^{\rm L}_{\bsigma}u^{\rm R}_{\bsigma}}
\label{eq:k_in_Y}
\end{eqnarray}
where
\begin{eqnarray}
 Y_{\bsigma\bsigma^\prime}
 &=&
 \big\bra
e^{\beta\eta[J_s\bsigma\cdot\bsigma^\prime\!-n\nu]}
\big\ket_{\eta}   \big\bra
\xi e^{n\beta\xi[ J_p (k+
\frac{m}{n}\sum_\alpha\sigma_\alpha)-\mu-J_g
v^\prime(k-k^\star)]}\big\ket_{\xi}
\label{eq:Y}
\end{eqnarray}
Finally,  the physical meaning of the order parameters $m$ and
$k$, expressed in terms of the original variables
$\{\sigma_i,\lambda_i\}$ and averages over the equilibrated
coupled relaxation processes, is found to be (see \ref{app:meaning}):
\begin{eqnarray}
m&=& \lim_{N\to\infty}\frac{1}{N}\sum_i \bra\bra
\xi(\lambda_i)\sigma_i\ket\ket \label{eq:meaning_m}
\\
k&=& \lim_{N\to\infty}\frac{1}{N}\sum_i \bra\bra
\xi(\lambda_i)\ket\ket \label{eq:meaning_k}
\end{eqnarray}
(with double brackets $\bra\bra\ldots\ket\ket$ denoting
equilibrium averages over both the fast secondary structure
formation process and the slow sequence selection process). Within
 the present model we may interpret $m=0$, where the
equilibrium amino-acid residue orientations are uncorrelated with
amino-acid species, as describing a `swollen' state where secondary
structure fails to develop (although, as we will find, for
$m=0$ there could still be phase transitions in terms of the
amino-acid statistics, as measured by the order parameter
$k$). States with $m\neq 0$ would exhibit secondary, and by construction (via the polarity term in the folding Hamiltonian)
also tertiary structure, so should be described as `collapsed' states.

\subsection{Solution of the replicated eigenvalue problem}

\noindent It was argued in \cite{Theo} that the left and right
eigenvectors $\{u^{\rm L}_{\bsigma}\}$ and $\{u^{\rm
R}_{\bsigma}\}$ corresponding to the largest eigenvalue of
matrices of the class (\ref{eq:GammaRS}) are of the following
form:
\begin{eqnarray}
u^{\rm R}_{\bsigma}&=& \int\!dx~\Phi(x) e^{\beta x\sum_\alpha
\sigma_\alpha} \label{eq:right_ev}\\
 u^{\rm L}_{\bsigma}&=& \int\!dy~\Psi(y)
e^{\beta y\sum_\alpha \sigma_\alpha} \label{eq:left_ev}
\end{eqnarray}
Inserting (\ref{eq:right_ev},\ref{eq:left_ev}) into the right/left eigenvalue equations $\sum_{\bsigma^\prime} \Gamma^{\rm
RS}_{\bsigma\bsigma^\prime}u^{\rm R}_{\bsigma^\prime}=\lambda_{\rm
RS}u^{\rm R}_{\bsigma}$ and $\sum_{\bsigma^\prime} \Gamma^{\rm
RS}_{\bsigma^\prime\bsigma}u^{\rm L}_{\bsigma^\prime}=\lambda_{\rm
RS}u^{\rm L}_{\bsigma}$, followed by use of the identity $g(\pm
1)=\exp[\beta (B\pm A)]$ with
\begin{eqnarray}
&&
A= \frac{1}{2\beta}\log[g(1)/g(-1)],~~~~~~ B=
\frac{1}{2\beta}\log[g(1)g(-1)]
\end{eqnarray}
 leads us
to a re-formulation of our eigenvalue problems in terms of
integral operators, where the role of $n$ has changed from
controlling the dimension of the problem (limited to integer values) to that of a simple
parameter that can be continued to the real line:
\begin{eqnarray}
\lambda_{\rm RS}\Phi(x)  &=& \int\!dx^\prime
\Lambda_{\Phi}(x,x^\prime) \Phi(x^\prime)~~~ \label{eq:Phi_eqn}
\\
\lambda_{\rm RS}\Psi(x)  &=& \int\!dx^\prime
\Lambda_{\Psi}(x,x^\prime) \Psi(x^\prime)~~~ \label{eq:Psi_eqn}
\end{eqnarray}
With help of the short-hands
\begin{eqnarray}
A(x,y) &=&\frac{1}{\beta}\tanh^{-1}[\tanh(\beta x)\tanh(\beta y)]
\label{eq:Afunc}
\\
B(x,y)&=&\frac{1}{2\beta}\log[4 \cosh[\beta(x\!+\!y)]
\cosh[\beta(x\!-\!y)]]~~~~ \label{eq:Bfunc}
\end{eqnarray}
one can write the kernels in (\ref{eq:Phi_eqn},\ref{eq:Psi_eqn})
(of which again we seek the largest eigenvalue) as
\begin{eqnarray}
\hspace*{-20mm}
\Lambda_{\Phi}(x,x^\prime) &=&\Big\bra\!\Big\bra \delta\big[x\!-\! \xi
J_p m  \!-\! A(x^\prime\!,\eta J_s)\big]
 e^{n\beta [B(x^\prime\!,\eta
J_s)+\xi( J_p k  -\mu- J_g v^{\prime\!}(k-k^\star)) - \nu\eta]}
\Big\ket\!\Big\ket_{\xi,\eta} \nonumber
\\
\hspace*{-20mm}
\label{eq:Phi_kernel}
\\
\hspace*{-20mm}
\Lambda_{\Psi}(x,x^\prime) &=& \Big\bra\!\Big\bra \delta\big[x\!-\!
A(x^\prime\!+\!\xi J_p m,\eta J_s)\big]
 e^{n\beta [B(x^\prime\!+\xi J_p m,\eta
J_s)+\xi( J_p k  -\mu- J_g v^{\prime\!}(k-k^\star))- \nu\eta]}
\Big\ket\!\Big\ket_{\xi,\eta}\nonumber
\hspace*{-10mm}
\\
\hspace*{-20mm}
 \label{eq:Psi_kernel}
\end{eqnarray}
 Both kernels $\Lambda_{\Phi}(x,x^\prime)$
and $\Lambda_{\Psi}(x,x^\prime)$ take only non-negative values,
so the eigenvalue problems (\ref{eq:Phi_eqn},\ref{eq:Psi_eqn})
support solutions where $\Phi(x)\geq 0$ and $\Psi(x)\geq 0$ for
all $x\in\R$. We may then normalize these functions according to
$\int\!dx~\Phi(x)=\int\!dx~\Psi(x)=1$ and interpret both, in view
of (\ref{eq:right_ev},\ref{eq:left_ev}), as field
distributions. A consequence of
this normalization convention is that we obtain two relatively
simple (and equivalent) expressions for the eigenvalue
$\lambda_{\rm RS}$ upon integration of
(\ref{eq:Phi_eqn},\ref{eq:Psi_eqn}) over $x$:
\begin{eqnarray}
\lambda_{\rm RS}  &\!=\!& \!\int\!dx~\Phi(x)\Big\bra\!\!\Big\bra
 e^{n\beta [B(x,\eta
J_s)+\xi( J_p k  -\mu- J_g v^{\prime\!}(k-k^\star))- \nu\eta]}
\Big\ket\!\!\Big\ket_{\!\xi,\eta} ~~ \label{eq:lambdaPhi}
\\
\lambda_{\rm RS} &\!=\!& \!\int\!dx~\Psi(x) \Big\bra\!\!\Big\bra
 e^{n\beta [B(x+\xi J_p m,\eta
J_s)+\xi( J_p k  -\mu- J_g v^{\prime\!}(k-k^\star))- \nu\eta]}
\Big\ket\!\!\Big\ket_{\!\xi,\eta} ~~ \label{eq:lambdaPsi}
\end{eqnarray}
 Given the normalized solutions $\Psi(x)$ and $\Psi(x)$ of
(\ref{eq:Phi_eqn},\ref{eq:Psi_eqn}) with the largest eigenvalue,
which will generally have to be obtained by numerical iteration, we can work out the remaining
contributions to our order parameter equations
(\ref{eq:m_in_Y},\ref{eq:k_in_Y}), such as
\begin{eqnarray}
\hspace*{-20mm}
\sum_{\bsigma}u^{\rm L}_{\bsigma}u^{\rm R}_{\bsigma} &=& 2^n
\!\int\!dxdx^\prime \Phi(x^\prime)\Psi(x)\cosh^n[\beta (x+
x^\prime)] \label{eq:inp1}
\\
\hspace*{-20mm}
\sum_{\bsigma\bsigma^\prime}u^{\rm L}_{\bsigma}
Y_{\bsigma\bsigma^\prime}u^{\rm R}_{\bsigma^\prime} &=&
2^n\!\int\!dx dx^\prime \Phi(x^\prime)\Psi(x) \Big\bra\!\Big\bra
\xi e^{n\beta[B(x^\prime\!,\eta J_s)+\xi( J_p k  -\mu- J_g
v^{\prime\!}(k-k^\star))-\eta\nu]}\nonumber
\\
\hspace*{-20mm}
&&\times \cosh^n[\beta (x+\xi J_p m+A(x^\prime\!,\eta
J_s))]\Big\ket\!\Big\ket_{\xi,\eta}~~~~ \label{eq:inp2}
\\
\hspace*{-20mm}
\sum_{\bsigma\bsigma^\prime}u^{\rm L}_{\bsigma} \sigma_1
Y_{\bsigma\bsigma^\prime}u^{\rm R}_{\bsigma^\prime}&=&
2^n\!\int\!dx dx^\prime \Phi(x^\prime)\Psi(x)
 \Big\bra\!\Big\bra \xi
\tanh[\beta (x\!+\!\xi J_p m\!+\!A(x^\prime\!,\eta J_s))]
\nonumber
\\
\hspace*{-20mm}&&
\times e^{n\beta[B(x^\prime\!,\eta J_s)+\xi( J_p k  -\mu- J_g
v^{\prime\!}(k-k^\star))-\eta\nu]}\nonumber
\\
\hspace*{-20mm}
&&\times \cosh^n[\beta (x+\xi J_p m+A(x^\prime\!,\eta
J_s))]\Big\ket\!\Big\ket_{\xi,\eta} \label{eq:inp3}
\end{eqnarray}

\subsection{Simplified form of the theory}

\noindent Equations
(\ref{eq:m_in_Y},\ref{eq:k_in_Y},\ref{eq:Phi_eqn},\ref{eq:Psi_eqn})
(where we need the eigenfunctions with the largest eigenvalue)
together with the supporting expressions
(\ref{eq:Phi_kernel},\ref{eq:Psi_kernel},\ref{eq:lambdaPhi},\ref{eq:lambdaPsi},\ref{eq:inp1},\ref{eq:inp2},\ref{eq:inp3})
constitute a closed set of equations for the RS order
parameters $\{m,k,\Phi(x),\Psi(x)\}$ of our model. We now simplify this set further. First we define the following
polarity probability density:
\begin{eqnarray}
p(\xi) &=& \frac{w(\xi)
 e^{n\beta \xi( J_p k  -\mu- J_g
v^{\prime\!}(k-k^\star))}} { \int\!d\xi^\prime
~w(\xi^\prime)e^{n\beta \xi^\prime( J_p k -\mu- J_g
v^{\prime\!}(k-k^\star))}  }~~~ \label{eq:p_xi}
\end{eqnarray}
It represents the amino-acid statistics that would
have been observed in the absence of the fast process (see
 \ref{app:meaning}). If we normalize the eigenfunctions
$\{\Phi(x),\Psi(x)\}$ according to
$\int\!dx~\Phi(x)=\int\!dx~\Psi(x)=1$ we find that they are to be solved from
\begin{eqnarray}
\hspace*{-20mm}
\Phi(x) &=&\int\!d\xi~p(\xi) \left\{\frac{\int\!dx^\prime
\Phi(x^\prime)\int\!d\eta~w(\eta) \delta\big[x\!-\! \xi J_p m \! -\!
A(x^\prime\!,\eta J_s)\big]
 e^{n\beta [B(x^\prime\!,\eta
J_s)- \nu\eta]}}
 {\int\!dx^\prime \Phi(x^\prime)
 \int\!d\eta~w(\eta)
 e^{n\beta [B(x^\prime\!,\eta
J_s)- \nu\eta]}}\right\}
\nonumber
\\
\hspace*{-20mm}&& \label{eq:normalized_eve_phi}
\\
\hspace*{-20mm}
\Psi(x)&=& \frac{\int\! dx^\prime \int\!d\xi~p(\xi)
\Psi(x^\prime\!-\xi J_p m) \int\!d\eta~w(\eta)\delta\big[x-
A(x^\prime\!,\eta J_s)\big]
 e^{n\beta [B(x^\prime\!,\eta
J_s)- \nu\eta]}} { \int\!dx^\prime \int\!d\xi~p(\xi)
\Psi(x^\prime\!-\xi J_p m)\int\!d\eta~w(\eta)
 e^{n\beta [B(x^\prime\!,\eta
J_s)- \nu\eta]}}\nonumber
\\
\hspace*{-20mm}&& \label{eq:normalized_eve_psi}
\end{eqnarray}
 The variables $x$ in $\Phi(x)$ and $\Psi(x)$ have
the dimension (in spin language) of fields, so $\Phi(x)$ and
$\Psi(x)$ must represent field distributions. In fact they are
connected in a very explicit way: they can be expressed in terms
of each other via
\begin{eqnarray}
\Psi(x) &=& \frac{\int\!dx^\prime \Phi(x^\prime)
\int\!d\eta~w(\eta)\delta\big[x- A(x^\prime\!,\eta J_s)\big]
 e^{n\beta [B(x^\prime\!,\eta
J_s)- \nu\eta]}} {\int\!dx^\prime
\Phi(x^\prime)\int\!d\eta~w(\eta) e^{n\beta [B(x^\prime\!,\eta
J_s)- \nu\eta]}} \label{eq:Psi_in_Phi}
\\
\Phi(x)&=& \int\!d\xi~p(\xi)\Psi(x-J_pm\xi) \label{eq:Phi_in_Psi}
\end{eqnarray}
 One proves these statements by substituting
(\ref{eq:Psi_in_Phi}) into (\ref{eq:normalized_eve_psi}) and (\ref{eq:Phi_in_Psi}) into
(\ref{eq:normalized_eve_phi}), which shows in either case that
both sides of the respective equation are identical. The remaining
 eigenvalue problem (\ref{eq:normalized_eve_psi}) is still
nontrivial, but some properties of its solution(s) can be
established easily. First, it follows from $|\tanh(\beta
A(x^\prime,\eta J_s))|=|\tanh(\beta x^\prime)\tanh(\beta \eta
J_s)|\leq \tanh(\beta |\eta| J_s)$ that any solution $\Psi(x)$
must have $\Psi(x)=0$ for $|x|>J_s\max_{\eta,w(\eta)>0}|\eta|$.
Second, as soon as $J_s>0$ and $J_p m\neq 0$ there cannot be
 solutions of the trivial form $\Psi(x)=\delta(x-x^\star)$
 for finite $n$. This is clear upon inserting
$\Psi(x^\prime)=\delta(x^\prime-x^\star)$ into the right-hand side
of (\ref{eq:normalized_eve_psi}): for $J_s>0$ and $J_pm\neq 0$
there will {\em always} be multiple values of $A(x^\star,\eta
J_s)$ (since $\eta$ and $\xi$ take multiple nonzero values), so it
is impossible for the right-hand side of
(\ref{eq:normalized_eve_psi}) to produce a $\delta$-function.

We can now eliminate the distribution $\Phi(x)$ and its eigenvalue
problem from our theory, and reduce our order parameter equations
to a set involving $\{m,k,\Psi(x)\}$ only. The function $\Psi$ is
still to be solved from the eigenvalue equation
(\ref{eq:normalized_eve_psi}), whereas our two scalar order
parameter equations can now be made to take the transparent form
\begin{eqnarray}
m&=&   \int\!d\xi  dh~W(h,\xi)~ \xi\tanh[\beta h]
\label{eq:final_m_with_fields}
\\
k&=&  \int\!d\xi dh~W(h,\xi)~ \xi \label{eq:final_k_with_fields}
\end{eqnarray}
with the joint equilibrium distribution $W(h,\xi)$ of
local effective fields and polarities:
\begin{eqnarray}
W(h,\xi)&=& \frac{p(\xi)\cosh^n[\beta h]\int\!dx~ \Psi(x)
\Psi(h\!-\!x\!-\! J_p m\xi)
  } {\int\!d\xi^\prime dh^\prime p(\xi^\prime)\cosh^n[\beta h^\prime] \int\!dx~ \Psi(x)\Psi(h^\prime\!-x\!-\!J_pm
\xi^\prime)} \label{eq:effective_fields}
\end{eqnarray}
 Upon calculating the equilibrium distribution
\begin{equation}
\pi(\xi,\eta)= \lim_{N\to\infty}\frac{1}{N}\sum_i
\bra\bra\delta\big[\xi-
\xi(\lambda_i)\big]\delta\big[\eta-\cos[a(\lambda_i)]\big]\ket\ket
\end{equation}
(see \ref{app:meaning})  one finds that
$\pi(\xi,\eta)=\pi(\xi)\pi(\eta)$, and that
$\pi(\xi)=\int\!dh~W(h,\xi)$.
The equilibrium distributions $\pi(\xi)$ and $\pi(\eta)$ will generally
differ from the prior distributions $w(\xi)$ and $w(\eta)$ that
would be found upon simply drawing amino-acids at each site
randomly and independently. However, the factorization
$\pi(\xi,\eta)=\pi(\xi)\pi(\eta)$ tells us that, although it
impacts on amino-acid statistics,  in the present model the
sequence selection process does not induce correlations between
polarity and steric angles.\vsp

Given a solution of equations
(\ref{eq:normalized_eve_psi},\ref{eq:final_m_with_fields},\ref{eq:final_k_with_fields})
we can evaluate whether it is the physical one (i.e. the
one with the lowest free energy) by calculating
(\ref{eq:q2phi_RS}), which now takes the simple form
 \begin{eqnarray}
\varphi&=&  \frac{1}{2}J_p(m^2+k^2)+ J_g
\big[v(k\!-\!k^\star)\!-\!kv^\prime(k\!-\!k^\star)\big] -
\frac{\log\Lambda}{\beta n} \nonumber
\\
&&
 - \frac{1}{\beta n}\log
\int\!d\xi~w(\xi)
 e^{n\beta \xi( J_p k  -\mu- J_g
v^{\prime\!}(k-k^\star))}
  \label{eq:final_fRS}
\\
&& - \frac{1}{\beta n}\log \int\!dx
d\xi~p(\xi)\Psi(x\!-\!J_pm\xi)\int\! d\eta~w(\eta)
 e^{n\beta [B(x,\eta
J_s)- \nu\eta]}
\nonumber
\end{eqnarray}

\section{Solution of order parameter equations for special cases}

\subsection{The state without secondary structure}
\label{sec:mzero}

\noindent Our equations always allow for solutions
with $m=0$, describing states where no secondary structure
develops. To see this we first note that now $\Psi(x)=\Phi(x)$ for
any $x$, that (\ref{eq:Afunc},\ref{eq:Bfunc}) obey
$A(-x,y)=-A(x,y)$ and $B(-x,y)=B(x,y)$, and that
\begin{eqnarray}
\hspace*{-20mm}
\Lambda_{\Psi}(x,x^\prime|0,k) &=&\bra \delta\big[x \!-\!
A(x^\prime\!,\eta J_s)\big]
 e^{n\beta [B(x^\prime\!,\eta
J_s)- \nu\eta]} \ket_{\eta} \bra
 e^{n\beta \xi( J_p k  -\mu-J_g v^{\prime\!}(k-k^\star))}\ket_{\xi}
\end{eqnarray}
 Due to the above symmetries of $A(x,y)$ and $B(x,y)$, one has $\Lambda_{\Psi}(x,x^\prime)
=\Lambda_{\Psi}(-x,-x^\prime)$, so $\Lambda_{\Psi}$ commutes with
the parity operator. Its eigenfunctions are therefore
either symmetric or anti-symmetric. The anti-symmetric
eigenfunctions are ruled out by the requirement $\Psi(x)\geq 0$,
so we conclude that $\Psi(x)$
 must be symmetric in $x$, and that therefore
$W(-h,\xi)=W(h,\xi)$. From this it follows, via the saddle-point
equation for $m$, that $m=0$ indeed solves our equations for any
choice of the control parameters.

The distribution $\Psi(x)$ is for $m=0$ to be solved from
\begin{eqnarray}
\Psi(x)&=&
\frac{\int\!dx^\prime \Psi(x^\prime)
\int\!d\eta~w(\eta)\delta\big[x- A(x^\prime\!,\eta J_s)\big]
 e^{n\beta [B(x^\prime\!,\eta
J_s)- \nu\eta]}} {\int\!dx^\prime
\Psi(x^\prime)\int\!d\eta~w(\eta) e^{n\beta [B(x^\prime\!,\eta
J_s)- \nu\eta]}} \label{eq:final_Phi_m0}
\end{eqnarray}
This equation has the trivial solution
$\Psi(x)=\delta(x)$, which is in fact unique. To prove uniqueness
we use $|\tanh[\beta A(x^\prime,\eta J_s)]|=|\tanh(\beta
x^\prime)|\tanh(\beta J_s)$. Since $\Psi(x)=0$ for $|x|>J_s$ we
can define the largest interval $[-u,u]\subseteq [-J_s,J_s]$ such
that $\Psi(x)=0$ for $x\notin[-u,u]$. Inside the numerator of
(\ref{eq:final_Phi_m0}) we now know that any nonzero contribution
to the integral must have $|x^\prime|\leq u$, so $|\tanh[\beta
A(x^\prime,\eta J_s)]|\leq |\tanh(\beta u)|\tanh(\beta J_s)$.
Hence equation (\ref{eq:final_Phi_m0}) tells us that if
$\Psi(x)\neq 0$ then $|\tanh(\beta x)|\leq \tanh(\beta
J_s)|\tanh(\beta u)|$, but now one must also have $|\tanh(\beta
u)|\leq \tanh(\beta J_s)|\tanh(\beta u)|$. Clearly the only $u$
that satisfies the latter inequality is $u=0$, which completes the
proof that $\Psi(x)=\delta(x)$.

Furthermore, upon inserting $\Psi(x)=\delta(x)$ equation
(\ref{eq:effective_fields}) tells us that
 $W(h,\xi)=p(\xi)\delta(h)$, with $p(\xi)$ given by
(\ref{eq:p_xi}). This makes sense, since the contribution to the
fields that depends on the polarity does so via the mean-field
forces, which are absent for $m=0$, whereas in the absence of
long-range folding forces the remaining one-dimensional chain
cannot order (hence all effective fields are zero). The
saddle-point equation for $k$ can be simplified to
\begin{eqnarray}
k&=& \frac{\int\!d\xi~\xi ~w(\xi) e^{n\beta\xi[J_p k-\mu-J_g
v^{\prime\!}(k-k^\star)]}}{\int\!d\xi~w(\xi)
 e^{n\beta\xi[ J_p k  -\mu-J_g v^{\prime\!}(k-k^\star)]}}
 \label{eq:mzero_keqn}
\end{eqnarray}
This equation shows that even for $m=0$ (i.e. no secondary
structure) there is still an effect of the coupling
between sequence selection and residue orientation: there will still be an effective preference
for homogeneous sequences, due to the increased potential for
energy gain (via $J_p$) if monomers are of the same type, which is
however counter-acted by the energy cost of polarity homogeneity
as controlled by $J_g$.

 Finally, using the above results as well as $B(0,\eta J_s)=\beta^{-1}\log[2\cosh(\beta \nu J_s)]$,
  the free energy of the $m=0$ state is seen to take the
value
\begin{eqnarray}
\hspace*{-20mm}
\varphi&=&  \frac{1}{2}J_pk^2+ J_g
\big[v(k\!-\!k^\star)\!-\!kv^\prime(k\!-\!k^\star)\big]
- \frac{1}{\beta n}\log
\int\!d\xi~w(\xi)
 e^{n\beta \xi[ J_p k  -\mu- J_g
v^{\prime\!}(k-k^\star)]}
 \nonumber
\\
\hspace*{-20mm}
&&
 -\frac{\log 2}{\beta}
-
\frac{\log\Lambda}{\beta n}
  - \frac{1}{\beta n}\log\int\!\! d\eta~w(\eta)
 e^{n[\log\cosh(\beta \nu J_s)- \beta \nu\eta]}~~~~
 \label{eq:final_fm0}
\end{eqnarray}

\subsection{Analytical solution in verifiable limits}

\subsubsection{Infinite temperature.}

\noindent In the infinite temperature limit $\beta\to 0$ one has
$A(x,y)=\beta xy+\order(\beta^3)$ and $B(x,y)=\beta^{-1}\log 2
+\frac{1}{2}\beta(x^2+y^2)+\order(\beta^3)$. From this we can
immediately extract the following solution of our saddle-point
equations
(\ref{eq:normalized_eve_psi},\ref{eq:final_m_with_fields},\ref{eq:final_k_with_fields}):
\begin{eqnarray}
&& \lim_{\beta\to 0}\Psi(x)=\delta(x),~~~~~~\lim_{\beta\to 0} W(\xi,h)=w(\xi)\delta(h) \\
 && \lim_{\beta\to
0}m=0~~~~~\lim_{\beta\to 0}k=\int\!d\xi~\xi ~w(\xi)
\end{eqnarray}
The corresponding value for the free energy (\ref{eq:final_fRS})
is \begin{equation} \lim_{\beta\to 0} \beta f=-n^{-1}\log\Lambda
-\log 2
\end{equation}
This is just the $\beta\to 0$ limit of the $m=0$ (swollen) state.  We recognize the free energy
reducing to the entropic contributions from the angular
($-\beta^{-1}\log 2$) and from the sequence ($-(\beta n)^{-1}\log
\Lambda$) degrees of freedom, and the average polarity $k$
reduces to that of the amino-acid pool. All this is easily
understood on physical grounds.

\subsubsection{Random sequences: $n\to 0$.}

\noindent According to $n=\tilde{\beta}/\beta$
this limit describes the case where monomer sequences are
selected fully randomly, independent of the functionality
potential or the secondary structure they would  generate.
Our equations must for $n\to 0$ therefore reproduce the theory
developed for random hetero-polymers in \cite{Skan}, provided  we
set the hydrogen bond coupling in \cite{Skan} to zero. Here we
find for $n\to 0$ that our equations indeed simplify considerably.
As expected we obtain $p(\xi)=w(\xi)$,
 since sequences are selected randomly from
the amino-acid pool, and hence $k=\bra \xi\ket_\xi$.
We are then left with the following eigenvalue problem for $\Psi(x)$:
\begin{eqnarray}
\Psi(x)&=& \!\int\!dy~\Psi(y) \bra\!\bra\delta\big[x-
A(y\!+\!J_pm\xi,\eta J_s)\big]\ket\!\ket_{\xi,\eta} ~~~
\label{eq:Psi_n0}
\end{eqnarray}
with $\Phi(x)= \bra\Psi(x-J_pm\xi)\ket_\xi$. But now the order
parameter $m$ (which measures the degree of orientation
specificity along the chain of hydrophobic versus hydrophilic
residues) is to be solved from
\begin{eqnarray}
m&=&   \bra \xi \int\!dh~W(h|\xi)\tanh[\beta h]\ket_\xi \\
W(h|\xi)&=& \int\!dx~ \Psi(x) \Psi(h\!-\!x\!-\! J_p m\xi)
\end{eqnarray}
Equivalently, upon using (\ref{eq:Psi_n0}):
\begin{eqnarray}
m&=& \int\!dx dx^\prime \Phi(x^\prime)\Psi(x)
\bra\bra \xi \tanh[\beta (x\!+\!\xi J_p m\!+\!A(x^\prime\!,\eta
J_s))] \ket\ket_{\xi,\eta}
\end{eqnarray}
The corresponding free energy per monomer is
 \begin{eqnarray}
\lim_{n\to 0}\left(f \!+\! \frac{\log\Lambda}{\beta n}\right)&=&
\frac{1}{2}J_p\Big(m^2\!-\bra\xi\ket_\xi^2\Big) +  \mu\bra\xi\ket+
\nu \bra\eta\ket_\eta  +J_g v(\bra\xi\ket_\xi\!-\!k^\star)
 \nonumber
\\
&&-   \int\!dx~\Phi(x)\bra B(x,\eta J_s)\ket_\eta
 \label{eq:final_fRS_n0}
\end{eqnarray}
The theory for random sequences in \cite{Skan} (based on random field techniques
rather than the replica  formalism) involved as its main order
parameter the distribution $P_\infty(\bk|\beta J_pm)$ of three
ratios $\bk=(k_1,k_2,k_3)$ of conditioned partition functions
(condition on the values of the last two spins of the chain). The
link between our present equations and those in \cite{Skan} is
made via the identification
\begin{eqnarray}
\hspace*{-10mm}
P_\infty(\bk|\beta J_pm)&=&
\delta(k_3\!-\!k_1)\int\!dxdy~\Phi(x)\Psi(y)\delta(k_1\!-\!e^{2\beta y})\delta(k_2\!-\!e^{2\beta (x-y)})~~~~
\label{eq:Pinfty}
\end{eqnarray}
Using $\Phi(x)= \bra\Psi(x-J_pm\xi)\ket_\xi$, equation
(\ref{eq:Psi_n0}) and the relation $A(x,y)=(2\beta)^{-1}
\log[\cosh(\beta x+\beta y)/\cosh(\beta x-\beta y)]$ one proves
that (\ref{eq:Pinfty}) obeys
\begin{eqnarray}
\hspace*{-10mm}
P_\infty(\bk|\beta J_pm)&=&\int\!d\bk^\prime
P_\infty(\bk^\prime|\beta J_pm)
\Big\bra\!\Big\bra \delta\left[\bk-\left(\!\begin{array}{c}
{\cal F}_1(\bk^\prime|\beta J_s\eta)\\
 {\cal F}_2(\bk^\prime|\beta J_s\eta,\beta J_p m\xi)\\ {\cal F}_3(\bk^\prime|\beta J_s\eta)
 \end{array}
 \!\!\right)\right]\Big\ket\!\Big\ket_{\xi,\eta}
 \nonumber
 \end{eqnarray}
 with
 \begin{eqnarray}
{\cal F}_1(\bk|x)&=& \frac{e^{x}k_1 k_2+e^{-x}}{e^{-x}k_1 k_2+e^x},~~~~~~
 {\cal F}_2(\bk|x,y)= \frac{e^{-x}k_1 k_2+e^{x}}{e^{x}k_1
 k_2+e^{-x}}~k_3 e^{2y}
 \\ {\cal F}_3(\bk|x) &=&  \frac{e^{x}k_2 k_3+e^{-x}}{e^{-x}k_2 k_3+e^x}
 \end{eqnarray}
 which is indeed the limit $J_{Hb}\to 0$ of the
 equation derived for $q=2$ in \cite{Skan}.

\subsubsection{Mean field limit.}

\noindent A second limit which can be verified using earlier work
is that where $J_s\to 0$ and $J_g\to 0$, describing the coupled
dynamics of sequence selection and secondary structure generation
in heteropolymers with (one type of) polarity energies only, the
simpler case studied in \cite{Chak}. In this limit the model
contains only mean-field forces, and no longer involves transfer
matrices. Using the identities $A(x,0)=0$ and
$B(x,0)=\beta^{-1}\log[2 \cosh(\beta x) ]$ one
extracts from (\ref{eq:normalized_eve_psi}) that
$\Psi(x)=\delta(x)$, and so
\begin{eqnarray}
m&=&   \int\!d\xi  dh~W(h,\xi)~ \xi\tanh[\beta h]
\\
k&=&  \int\!d\xi dh~W(h,\xi)~ \xi
\\
W(h,\xi)&=& \frac{\cosh^n[\beta h]p(\xi) \delta[h- \xi J_p m] }{
\int\!dh^\prime \cosh^n[\beta h^\prime]\int\!d\xi^\prime
p(\xi^\prime) \delta[h^\prime- \xi^\prime J_p m] } ~~~~~
\end{eqnarray}
As could have been expected, the equations for the scalar order
parameters $(m,k)$ already close onto themselves. In explicit form
they are
\begin{eqnarray}
m&=& \frac{
 \bra \xi \tanh(\beta
\xi J_p m) e^{n\beta\xi(J_p k-\mu)}\cosh^n(\beta \xi J_p
m)\ket_{\xi}}{ \bra e^{n\beta \xi( J_p k  -\mu)} \cosh^n(\beta \xi
J_p m)\ket_\xi }~~~~ \label{eq:m_meanfield}
\\
k&=& \frac{ \bra \xi
 e^{n\beta\xi(J_p k-\mu)}\cosh^n(\beta \xi
J_p m)\ket_{\xi} }{\bra e^{n\beta \xi( J_p k  -\mu)} \cosh^n(\beta
\xi J_p m) \ket_\xi}~~~~ \label{eq:k_meanfield}
\end{eqnarray}
The amino-acid statistics in equilibrium are given by
\begin{eqnarray}
p(\xi) &=& \frac{w(\xi)
 e^{n\beta \xi( J_p k  -\mu)}} { \int\!d\xi^\prime w(\xi^\prime)e^{n\beta \xi^\prime( J_p k
-\mu)}  },~~~~~~
p(\eta) = \frac{ w(\eta)e^{- n\beta\nu\eta} }
{\int\!d\eta^\prime w(\eta^\prime) e^{-n\beta \nu\eta^\prime}}
\end{eqnarray}
Finally, using  $\Phi(x)= \int\!d\xi~p(\xi)\delta[x- \xi J_p m]$
we may work out the value of the free energy per monomer for
$J_s=0$:
 \begin{eqnarray}
\lim_{J_s\to 0}f&=& \frac{1}{2}J_p(m^2+k^2)-
\frac{\log\Lambda}{\beta n}-\frac{\log 2}{\beta}
- \frac{1}{\beta n}\log
\int\!d\eta~w(\eta)
 e^{-n\beta \nu\eta}
 \nonumber
\\
&&- \frac{1}{\beta n}\log \int\!d\xi~w(\xi) \cosh^n(\beta \xi
J_pm)
 e^{n\beta \xi( J_p k  -\mu)}
 \label{eq:f_meanfield}
\end{eqnarray}
The specific model studied in \cite{Chak} had $\xi\in\{-1,1\}$
 and
$w(\xi)=\frac{1}{2}(\delta_{\xi,1}+\delta_{\xi,-1})$, i.e. no
varying degrees of hydrophobicity or hydrophilicity and a
polarity-unbiased amino-acid pool. Steric effects did not come
into play in \cite{Chak} (monomers were characterized only
by their polarity), so we may here simply take $\nu=0$. These
choices simplify our two remaining order parameter equations
(\ref{eq:m_meanfield},\ref{eq:k_meanfield})
to
\begin{equation}
m=  \tanh(\beta
 J_p m), ~~~~~~
k= \tanh[n\beta(J_p k-\mu)]
\end{equation}
 These equations are indeed identical to
those of \cite{Chak}, given $q=2$. Similarly,
for $w(\xi)=\frac{1}{2}(\delta_{\xi,1}+\delta_{\xi,-1})$ and
$\nu=0$ the free energy per monomer (\ref{eq:f_meanfield}) now
simplifies to
 \begin{eqnarray}
\lim_{J_s\to 0}f&=& \frac{1}{2}J_p m^2-
\frac{\log(\Lambda/2)}{\beta n}- \frac{1}{\beta }\log[2
\cosh(\beta J_pm)] \nonumber
\\
&&+\frac{1}{2}J_p k^2 - \frac{1}{\beta n}\log[2 \cosh(n\beta (J_p
k \!-\!\mu))]~~~
\end{eqnarray}
Apart from the excess entropy $-\log(\Lambda/2)/\beta n$ due to
the extra chemical degrees of freedom of our present monomers
compared to the ones in \cite{Chak} (and modulo a trivial typo in
\cite{Chak}) this is indeed the free energy expression found in
\cite{Chak} for $q=2$.

\section{Transitions and phase diagrams for deterministic sequence selection}

\noindent We now turn to nontrivial regimes where analytical
solution is still possible, but where our model does not map onto
any existing model in literature. The biologically most relevant regime is that of low or even absent
genetic noise levels, viz. $n\to\infty$.  We still have to select a form for the polarity balance
potential. Since $v(k-k^\star)$ must be minimal at
$k=k^\star$ and increase monotonically with $|k-k^\star|$, we
choose a simple quadratic form $v(u)=\frac{1}{2}u^2$. Thus
from now on we will have $v^\prime(u)=u$ and
$v^{\prime\prime}(u)=1$.

 Since
$n=\tilde{\beta}/\beta$, the limit $n\to\infty$ corresponds to the
case where monomer sequences are selected fully deterministically,
such as to minimize the effective Hamiltonian (\ref{eq:Heff}).
Here, in view of many exponents in our equations growing with $n$,
we may evaluate virtually all integrations by steepest descent.
 With a modest amount of foresight we define the
canonical polarity balance $k_0$ as
\begin{equation}
k_0=\frac{k^\star-\mu/J_g}{1-J_p/J_g}
\end{equation}
Clearly $\lim_{J_g\to\infty}k_0=k^\star$, and $\lim_{J_g\to
0}k_0=\mu/J_p$. To keep our analysis as transparent as possible we
will not consider pathological parameter coincidences but restrict
our discussion to the generic scenario where $J_g\neq J_s$,
$\nu\neq 0$, and $k_0\in(-1,1)$; the system behaviour in the
pathological cases can always be understood as specific limits
and/or degeneracies of the more generic solutions. Here the order
parameters $\{m,k,\Psi(x)\}$ are to be found by analyzing the
solutions for $n\to\infty$ of the following equations, where the
complications are mainly in the subtle dependence of the
distribution $\Psi(x)$ on $n$:
\begin{eqnarray}
\hspace*{-23mm}
\Psi(x)&=& \frac{\int\! dx^\prime \int\!d\xi~p(\xi) \Psi(x^\prime)
\int\!d\eta~w(\eta)\delta\big[x\!-\! A(x^\prime\!+\!J_pm\xi,\eta
J_s)\big]
 e^{n\beta [B(x^\prime\!+J_pm\xi,\eta
J_s)- \nu\eta]}} { \int\!dx^\prime \int\!d\xi~p(\xi)
\Psi(x^\prime)\int\!d\eta~w(\eta)
 e^{n\beta [B(x^\prime\!+J_pm\xi,\eta
J_s)- \nu\eta]}}\nonumber
\\
\hspace*{-23mm}&& \label{eq:ninfty_psi}
\\
\hspace*{-23mm}
m &=&  \frac{\int\!d\xi~p(\xi)\xi \int\!dxdy~ \Psi(x)
\Psi(y)\tanh[\beta(J_p m\xi\!+\!x\!+\!y)]\cosh^n[\beta(J_p m\xi\!+\!x\!+\!y)]
  } {\int\!d\xi~p(\xi) \int\!dxdy~ \Psi(x)
\Psi(y)\cosh^n[\beta(J_p m\xi+x+y)]
 }
 \label{eq:ninfty_m}
\\
\hspace*{-23mm}
 k&=&  \frac{\int\!d\xi~p(\xi)\xi \int\!dxdy~ \Psi(x)
\Psi(y)\cosh^n[\beta(J_p m\xi+x+y)]
  } {\int\!d\xi~p(\xi) \int\!dxdy~ \Psi(x)
\Psi(y)\cosh^n[\beta(J_p m\xi+x+y)]
 }
 \label{eq:ninfty_k}
 \end{eqnarray}
 with the abbreviations
 \begin{eqnarray}
 p(\xi)
&=& \frac{w(\xi)
 e^{n\beta \xi(J_p-J_g)(k-k_0)}} { \int\!d\xi^\prime ~w(\xi^\prime)e^{n\beta \xi^\prime(J_p-J_g)(k -k_0)}  }
\\
k_0&=&(k^\star-\mu/J_g)/(1-J_p/J_g)
\end{eqnarray}
Once the above equations have been solved for $n\to\infty$, the
associated values of the free energy per monomer subsequently
follows upon taking the $n\to\infty$ limit in
(\ref{eq:final_fRS}).

\subsection{The two simple cases  $J_s=0$ and $J_pm=0$}

\noindent In both these special cases our problem simplifies
significantly due to $\Psi(x)=\delta(x)$ (a property which has
been established earlier). If we define
\begin{eqnarray}
L(\xi)&=& \frac{1}{\beta}\log\cosh(\beta J_p m\xi)+
\xi(J_p\!-\!J_g)(k\!-\!k_0) ~~~~
\end{eqnarray}
we see that our remaining equations for $m$ and $k$ reduce to a
simple form in which for $n\to\infty$ the integration over $\xi$
is dominated by the maximum of $L(\xi)$, subject to the constraint
$\xi\in[-1,1]$ imposed by the measure $w(\xi)$:
 \begin{eqnarray}
 m &=&\lim_{n\to\infty}
\frac{\int\!d\xi~w(\xi)\xi\tanh(\beta J_p m\xi)e^{n\beta L(\xi)}
  } {\int\!d\xi~w(\xi)e^{n\beta L(\xi)}}
\\
 k&=& \lim_{n\to\infty} \frac{\int\!d\xi~w(\xi)\xi e^{n\beta L(\xi)}
  } {\int\!d\xi~w(\xi)e^{n\beta L(\xi)}
 }
 \end{eqnarray}
 If $J_pm\neq 0$ then
$L(\xi)$ is maximal either for $\xi=\sgn[(J_p-J_g)(k-k_0)]$ (if
$\lim_{n\to\infty}k\neq k_0$), or for $\xi=\pm 1$ (if
$\lim_{n\to\infty}k=k_0$). In either case one has
$\xi\in\{-1,1\}$, so we always find for $n\to\infty$ the simple
Curie-Weiss law
 $m = \tanh(\beta J_p m)$ which describes a transition to secondary structure at $T=J_p$.
 Our equation for $k$, on the other
 hand, will produce for $n\to\infty$ only solutions  of
 \begin{eqnarray}
 k&=& \sgn[(J_p-J_g)(k-k_0)]
 \end{eqnarray}
(this includes the case $k=k_0$). Graphical inspection of this
equation shows immediately that for $J_p<J_g$ the only solution is
$k=k_0$, whereas for $J_p>J_g$ we have the additional solutions
$k=\pm 1$. If $J_pm=0$ then $L(\xi)$ is either maximal for
$\xi=\sgn[(J_p-J_g)(k-k_0)]$ (if $\lim_{n\to\infty}k\neq k_0$), or
it is a constant on $\xi\in[-1,1]$ (if $\lim_{n\to\infty}k=k_0$).
Here one has $\xi\in\{-1,1\}$ only if $k\neq k_0$.

Working out the free energy per monomer (\ref{eq:final_fRS})
gives, using $B(x,y)=B(|x|,|y|)$ and the property that
$B(|x|,|y|)$ increases monotonically with both $|x|$ and $|y|$,
 \begin{eqnarray}
 \hspace*{-15mm}
\varphi&=&  \frac{1}{2}J_p m^2 +\frac{1}{2}J_g k^{\star 2}
+\frac{1}{2}(J_p-J_g) k^2
 \nonumber
\\
\hspace*{-15mm}
&& - \lim_{n\to\infty}\frac{1}{\beta
n}\log\int\!d\eta~w(\eta) \int\!d\xi~w(\xi)
 e^{n\beta[ \xi(J_p-J_g)(k-k_0)+ B(J_pm\xi,\eta
J_s)- \nu\eta]} \nonumber
\\
\hspace*{-15mm}
&=& \frac{1}{2}J_p m^2 +\frac{1}{2}J_g k^{\star 2}
+\frac{1}{2}(J_p-J_g) k^2
 \nonumber
\\
\hspace*{-15mm}
&& - \max_{\xi,\eta\in[-1,1]} \Big\{
  \xi(J_p\!-\!J_g)(k\!-\!k_0)+ B(J_pm\xi,\eta
J_s)\!-\! \nu\eta\Big\} \nonumber
\\
\hspace*{-15mm}
&=&
 \frac{1}{2}J_p m^2\!- B(J_p|m|,
J_s) +\frac{1}{2}J_g k^{\star 2}\!-|\nu|
+\frac{1}{2}(J_p\!-\!J_g) k^2\!-
  |J_p\!-\!J_g| |k\!-\!k_0|
\end{eqnarray}
where the maximum corresponds to $\eta=-\sgn(\nu)$ and
$\xi=\sgn[(J_p-J_g)(k-k_0)]$. The last line reveals that in
cases where we have multiple solutions, viz. $J_p>J_g$, the
solution $k=k_0$ is always a local maximum of $\varphi$ and $k=\pm
1$ are always local minima. Of the latter two, the lowest
free energy is found for $k=-\sgn(k_0)$ (this is therefore the
state that it not only locally stable but also thermodynamically
stable). Therefore
\begin{eqnarray}
&&
J_p>J_g:~~ k=\pm 1,~~~~~~
J_p<J_g:~~ k=k_0
\end{eqnarray}
 This implies a discontinuous phase transition at $J_p=J_g$,
 where we go from $k=\pm 1$ (homogeneous
polarity sequences) to $k=k_0\in(-1,1)$, where the sequence
becomes inhomogeneous in polarity.

If we calculate the distribution $W(\xi,h)$ for the above
solutions we always find $W(\xi,h)=\pi(\xi)\delta(h)$, but with
potentially different polarity statistics. For the $k=\pm 1$
states one has $\pi(\xi)=\delta(\xi-k)$. For the $k=k_0$ solution,
however, we need to look beyond the leading order and write
$k=k_0+n^{-1}k_1+\order(n^{-2})$. Here we find for $n\to\infty$:
\begin{eqnarray}
\pi(\xi)= \frac{w(\xi) e^{\beta \xi(J_p-J_g)k_1}}
{\int\!d\xi^\prime~w(\xi^\prime)e^{\beta \xi^\prime(J_p-J_g)k_1}}
\label{eq:specialcase_ninfty}
\end{eqnarray}
with $k_1$ to be solved from
\begin{eqnarray}
k_0= \frac{\int\!d\xi~w(\xi)\xi~ e^{\beta \xi(J_p-J_g)k_1}}
{\int\!d\xi~w(\xi)e^{\beta \xi(J_p-J_g)k_1}}
\label{eq:specialcase_k1}
\end{eqnarray}
 This
concludes our solution for the simple cases $J_s=0$ and $J_pm=0$.
From now on we consider the case where $J_s>0$ and $J_pm\neq 0$.

\subsection{Summary of the $n\to\infty$ theory}

\noindent
The full analysis of our order parameter equations in the limit $n\to\infty$ via saddle-point analysis, for arbitrary $(J_s,J_p)$,
turns out to be nontrivial; details of this calculation would interrupt the flow of the paper and have therefore been delegated to
\ref{app:saddlepoint_deterministic}. The end result, however, is surprisingly simple.
We can summarize the final equations for our order
parameters $(k,m)$ describing the systems states as identified in
the limit $n\to\infty$ in the following compact way:
\begin{eqnarray}
J_g\!>\!J_p\!:&~~~~ k=k_0,~~~~~& m=0~~{\rm or}~~F_{\beta J_p}(m)=-\tanh(\beta J_s)
\\[1mm]
J_p\!>\!J_g\!:&~~~~ k=\pm 1,~~~~~& m=0~~{\rm or}~~F_{\beta J_p}(m)=\sgn(\nu)\tanh(\beta
J_s)
\end{eqnarray}
in which the function $F_x(m)$ is defined as
\begin{eqnarray}
 F_{x}(m)&=&\frac{\tanh[ \frac{1}{2}xm-\frac{1}{2}\tanh^{-1}(m) ]}{\tanh[\frac{1}{2}xm+\frac{1}{2}\tanh^{-1}(m)]}
\label{eq:DefineF}
\end{eqnarray}
Only the solutions with $k=k_0$ as obtained for $J_g>J_p$
correspond hetero-polymers with inhomogenous polarity along the
chain, i.e. to systems of the protein type. The solutions  with
$k=\pm 1$ (with $k=-\sgn(k_0)$ being also
thermodynamically stable) describe a situation where the sequence
selection results in polymers with homogeneous polarity.
For $J_p>J_g$ we have two further conditions
(\ref{eq:Homopolb2},\ref{eq:Homopolb3}); these are always
satisfied for $m=0$, but may be violated by saddle-points for
which $|m|$ is too large. We observe that for $\nu<0$ the
homogeneous  polarity states and the inhomogeneous polarity states
exhibit fully identical levels of secondary structure (as measured
by $m$), for any combination of $\beta J_p$ and $\beta J_s$. Here
is it therefore also easy to show by comparing the two free energy
expressions (\ref{eq:f_homopol}) and (\ref{eq:f_inhomopol}) that
for $J_p>J_g$ the free energy per monomer of the $k=\pm 1$ state
is lower than that of the state $k=k_0$, whereas for $J_g>J_p$ the
free energy of the $k=k_0$ state is lower.  For $\nu>0$, however,
the two states no longer have identical values of $m$, with that
of the $k=\pm 1$ state being lower; here the system finds it
increasingly difficult to combine homogeneous polarity sequences
with secondary structure.

\begin{figure}[t]
\vspace*{5mm} \hspace*{45mm}\setlength{\unitlength}{0.85mm}
\begin{picture}(80,60)
\put(5,5){\epsfysize=60\unitlength\epsfbox{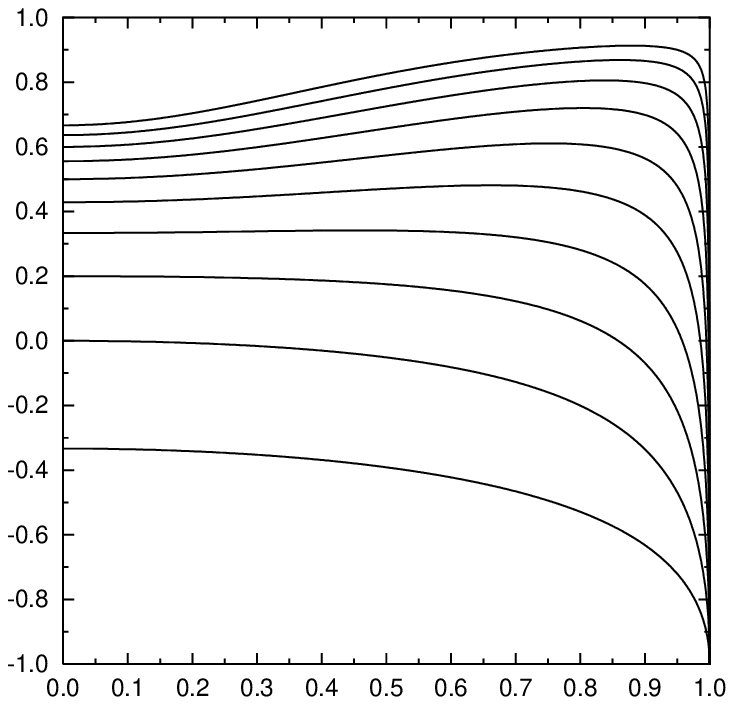}}
\put(38,-2){$m$}
 \put(-6,34){$F_x(m)$}
\end{picture}
\vspace*{3mm} \caption{The function $F_x(m)$ for
$x\in\{\frac{1}{2},1,\frac{3}{2},2,\frac{5}{2},3,\frac{7}{2},4,\frac{9}{2},5\}$
(from bottom to top). Solving the equation $F_x(m)=y$ for $m$ can
give at most one positive solution if $x<\sqrt{3}$, where $[d^2
F_x(m)/dm^2]|_{m=0}<0$. It may have two positive solutions if
$x>\sqrt{3}$, where $[d^2 F_x(m)/dm^2]|_{m=0}>0$, provided one
also has $y>F_x(0)= (x-1)/(x+1)$. For sufficiently large $y$ the
equation $F_x(m)=y$ will no longer have any solutions. }
\label{fig:FxM}
\end{figure}

Let us inspect  the bifurcation phenomenology for the order parameter $m$.
Note that $F_{0}(m)=-1$ for all $m\in[-1,1]$, and that
$F_\infty(m)=1$ for all $m\in[-1,1]$. For $x>0$ the function
$F_x(m)$ is symmetric in $m$, with $F_{x}(\pm 1)=-1$ and with
\begin{eqnarray}
 F_{x}(m)&=& \frac{x-1}{x+1}-m^2
 \frac{x(3-x^2)}{3(x+1)^2}+\order(m^4)~~~~
 \end{eqnarray}
(see also figure \ref{fig:FxM}). In view of the symmetry
$F_x(-m)=F_x(m)$, we conclude that (depending on the values of
$(x,y)$), the equation $F_x(m)=y$ has either zero, two ($\pm
m^\star$), or four ($\pm m^\star,\pm m_0$) nontrivial solutions in
$m$.

\begin{figure}[t]
\vspace*{4mm} \hspace*{43mm}\setlength{\unitlength}{0.66mm}
\begin{picture}(160,60)
\put(0,0){\epsfysize=58\unitlength\epsfbox{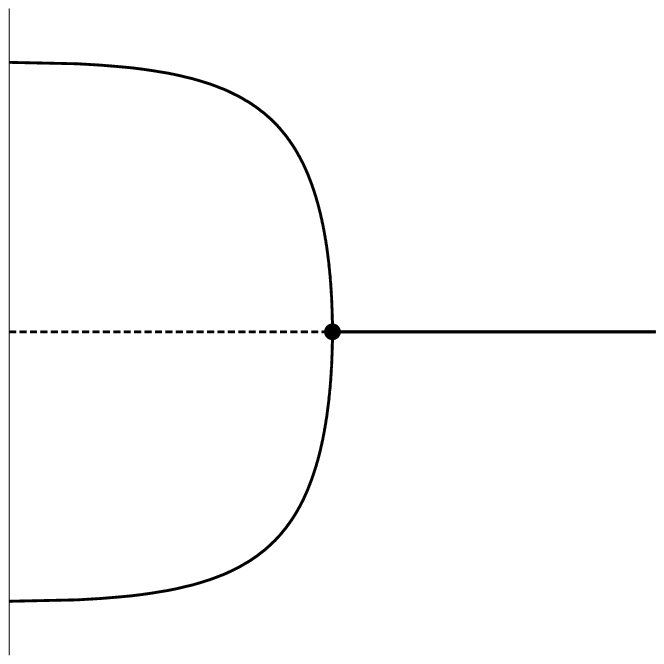}}
\put(51,24){$y$} \put(20,57) {$x<\sqrt{3}$}
 \put(-5,40){$m$} \put(-3,50){$1$} \put(-6,5){$-1$}
 \put(63,0){\epsfysize=58\unitlength\epsfbox{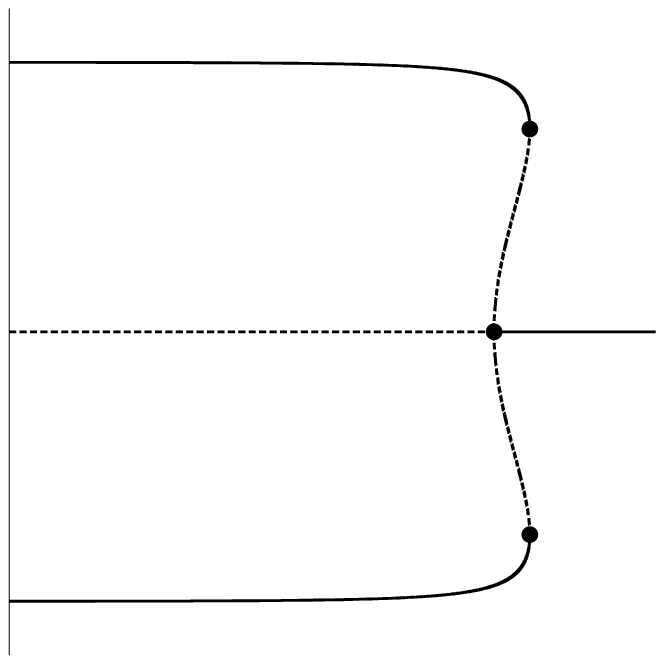}}
\put(114,24){$y$} \put(83,57) {$x>\sqrt{3}$}
 \put(58,40){$m$} \put(60,50){$1$} \put(57,5){$-1$}
\end{picture}
\vspace*{-3mm} \caption{ The bifurcation scenarios for the
solutions $m$ of the equation $F_{x}(m)=y$, with $x=\beta J_p$ and
with $y={\rm sgn}(\nu)\tanh(\beta J_s)\in[-1,1]$ for $J_p>J_g$ and
$y=-\tanh(\beta J_s)\in[-1,0]$ for $J_g>J_p$. Solid lines
correspond to stable solutions (local minima of the free energy),
whereas dashed lines correspond to unstable ones. The trivial solution $m=0$
changes stability at $\beta J_s=\frac{1}{2}{\rm
sgn}(\nu)\log(\beta J_p)$ for $J_p>J_g$ and at $\beta
J_s=\frac{1}{2}\log(\beta J_p)$ for $J_g>J_p$.}
\label{fig:Msolutions}
\end{figure}

 In the $(x,y)$ plane, where $x=\beta J_p$ and $y=\tanh(\sigma\beta J_s)$ with $\sigma=\pm 1$
 (so $\sigma=\sgn(\nu)$ for $J_p>J_g$ and $\sigma=-1$ for $J_g>J_p$),
 the bifurcation scenarios for our saddle-point
equation $F_x(m)=y$ can now be summarized as:
\begin{eqnarray*}
x<\sqrt{3}: &~~~& {\rm continuous~
transition~at}~y_c=(x\!-\!1)/(x\!+\!1)\\
 && y<y_c:~m\in \{0,\pm m^\star(x)\}\\
 && y>y_c:~m=0
\\[2mm]
x>\sqrt{3}:&~~~& {\rm
continuous~transition~at}~y_c=(x\!-\!1)/(x\!+\!1)\\
 && y<y_c:~m\in\{0,\pm m^\star(x)\}\\
 && y>y_c:~m\in\{0,\pm m_0(x),\pm m^\star(x)\}
 \\[2mm]
 && {\rm discontinuous~transition~at}~y^\prime_c >(x\!-\!1)/(x\!+\!1)\\
 && y<y_c^\prime:~m\{0,\pm m_0(x),\pm m^\star(x)\}\\
 && y>y_c^\prime:~m=0
\end{eqnarray*}
The result is shown in figure \ref{fig:Msolutions}.

\subsection{Phases, transition lines, and phase diagrams}

\noindent We can characterize the phases of our system for
$n\to\infty$ in terms of the values for the order parameters
$(k,m)$, where $k$ provides information on the primary structure
(the average polarity) and $m$ provided information on the
secondary structure (the extent of order in the side-chain
orientations). The system is found to exhibit five phases:
\begin{description}
\item[~~~HS] (`homogenous \& swollen'):
 \hsp $k=\pm 1$, $m=0$\\
 \hsp primary structure but no secondary structure,\\ \hsp selected sequences are homogeneous in polarity
 \item[~~~HC] (`homogenous \& collapsed'):
 \hsp $k=\pm 1$, $m\neq 0$
 \\
 \hsp both primary and secondary structure,\\ \hsp selected sequences are homogeneous in polarity
 \item[~~~HM] (`homogenous \& mixed'): \hsp $k=\pm 1$, coexistence of
 $m=0$ and $m\neq 0$
\\
 \hsp primary structure, with secondary structure controlled by remanence, \\
 \hsp sequences are homogeneous in polarity
 \item[~~~IS] (`inhomogenous \& swollen'):
 \hsp $k=k_0$, $m=0$\\
  \hsp primary structure but no secondary structure,\\ \hsp selected sequences are inhomogeneous in polarity
 \item[~~~IC] (`inhomogenous \& collapsed'):
 \hsp $k=k_0$, $m\neq 0$
 \\
  \hsp both primary and secondary structure,\\ \hsp selected sequences are inhomogeneous in polarity
 \end{description}
 There is no random (paramagnetic) phase
 $m=k=0$. This is a consequence of the $n\to\infty$ limit: since the noise in the genetic selection (viz. mutations)
 is removed, there is at least always a primary structure
 developing as measured by $k\neq 0$.
 \vsp

 Similarly,
 we can summarize the transitions we have by now identified:
\begin{itemize}
\item HS$\to$IS and HC$\to$IC:
 discontinuous transitions, at
 \begin{equation}
J_g=J_p
\end{equation}
 The  HS$\to$IS line is found in the regime of
small values of $J_p$. The  HC$\to$IC line is found for large
values of $J_p$. Along the latter line,  if $\nu<0$ only $k$ is
changed at the transition, if
 $\nu>0$ both $k$ and $m$ are changed.

 \item HS$\to$HC, IS$\to$IC, and HC$\to$HM: continuous transitions,
 at
\begin{eqnarray}
\beta J_s&=& \left\{\begin{array}{lll} \frac{1}{2}{\rm
sgn}(\nu)\log(\beta J_p)&& {\rm if}~~J_p>J_g\\[1mm]
-\frac{1}{2}\log(\beta J_p)&& {\rm if}~~J_g>J_p
\end{array}\right.
\label{eq:cont_bif_ninfty}
\end{eqnarray}
The HC$\to$HM line exists only when $J_p>J_g$ and $\nu>0$ (where
the coexistence phase HM is found).

\item HS$\to$HM: discontinuous transition, to be solved from
the coupled equations
\begin{eqnarray}
\hspace*{-10mm}
&& \frac{\tanh[ \frac{1}{2}\beta J_p m-\frac{1}{2}\tanh^{-1}(m)
]}{\tanh[\frac{1}{2}\beta J_p
m+\frac{1}{2}\tanh^{-1}(m)]}=\tanh(\beta J_p)
\\
\hspace*{-10mm}
&& \frac{1\!-\!\tanh^2[ \frac{1}{2}\beta
J_pm\!-\!\frac{1}{2}\tanh^{-1}(m)
]}{1\!-\!\tanh^2[\frac{1}{2}\beta
J_pm\!+\!\frac{1}{2}\tanh^{-1}(m)]} ~\frac{\beta
J_p(1\!-\!m^2)\!-\!1}{\beta J_p(1\!-\!m^2)\!+\!1} =\tanh(\beta J_s)
\end{eqnarray}
where the second equation is obtained from combining $F_{\beta
J_p}(m)=\tanh(\beta J_s)$ with $\frac{\partial }{\partial
m}F_{\beta J_p}(m)=0$. This line starts at the triple point
$(\beta J_p,\beta J_s)=(\sqrt{3},\frac{1}{4}\log 3)$ in the
$(\beta J_p,\beta J_s)$ plane, and rises continually for $\beta
J_p>\sqrt{3}$. It emerges only for $J_p>J_g$ and $\nu>0$ (where
the coexistence phase HM is found).
\end{itemize}
At the continuous transition (\ref{eq:cont_bif_ninfty}) the $m\neq
0$ state always takes over the stability from the trivial one.
This can be seen upon expanding the two free energy expressions
(\ref{eq:f_homopol},\ref{eq:f_inhomopol}) for small $m$. For both
expressions this gives
\begin{eqnarray}
 && \beta(\varphi-\varphi_{m=0})=
\frac{1}{8}m^2(\beta J_p\!+\!1)^2\Big\{
 \tanh^2(\beta
 J_s)-\Big(\frac{\beta J_p\!-\!1}{\beta J_p\!+\! 1}\Big)^2
\Big\}+\order(m^3)\nonumber
\end{eqnarray}
\begin{figure}[t]
\vspace*{-5mm} \hspace*{45mm}\setlength{\unitlength}{1.0mm}
\begin{picture}(80,70)
\put(5,5){\epsfysize=60\unitlength\epsfbox{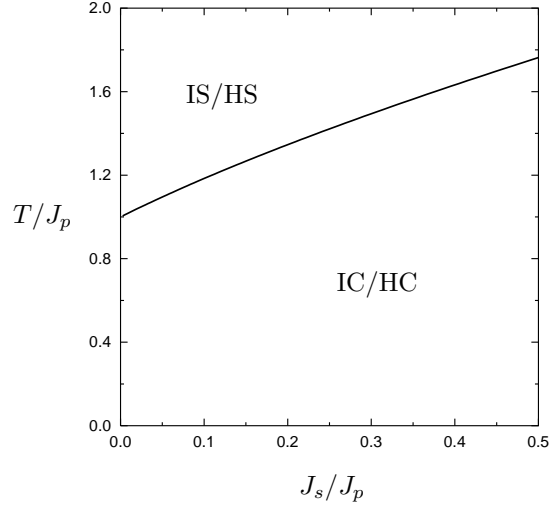}}
\put(35,-2){$J_s/J_p$} \put(-3,34){$T/J_p$} \put(40,25){IC/HC}
\put(20,50){IS/HS}
\end{picture}
\vspace*{2mm} \caption{Phase diagram cross-section for
$n\to\infty$ (deterministic sequence selection) for the cases
where either $J_g>J_p$ (protein-like inhomogeneous polarity sequences, $k=k_0$)
or where $J_g<J_p$ (homogeneous polarity sequences, $k=\pm 1$) but
with $\nu<0$.  Solid line: transition marking the continuous
 bifurcation of collapsed ($m\neq 0$) states, although for $J_g>J_p$ this transition is discontinuous in the polarity statistics.
 Phases are defined and described in the main text.
 } \label{fig:phasediag_ninfty_nocoex}
\end{figure}
\begin{figure}[t]
\vspace*{-5mm} \hspace*{45mm}\setlength{\unitlength}{1.0mm}
\begin{picture}(80,70)
\put(5,5){\epsfysize=60\unitlength\epsfbox{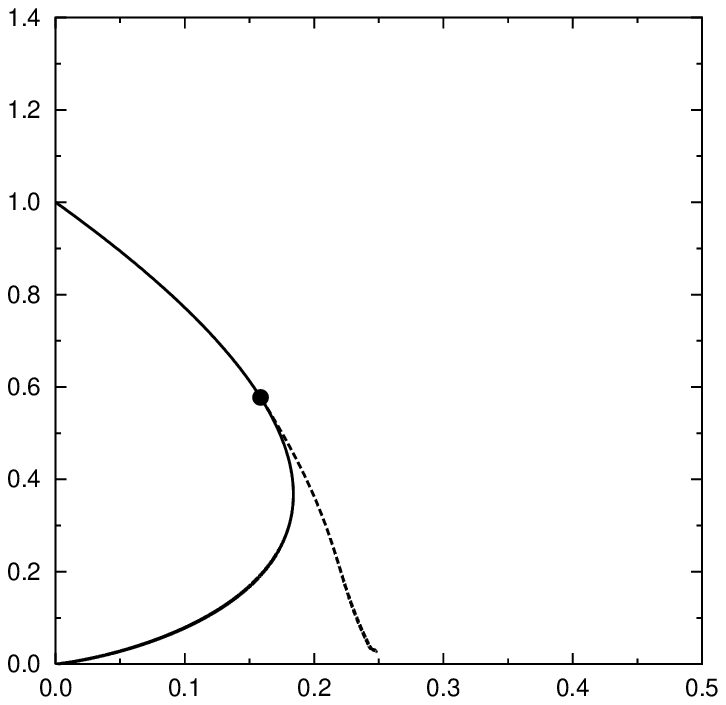}}
\put(35,-2){$J_s/J_p$} \put(-3,34){$T/J_p$} \put(17,25){HC}
\put(29,10){HM} \put(45,42){HS}
\end{picture}
\vspace*{2mm} \caption{Phase diagram cross-section for
$n\to\infty$ (deterministic sequence selection) for the case where
$J_p>J_g$ and $\nu>0$ (homogeneous polarity sequences, unlike proteins). Here  the
system is
 unable to minimize steric and polar energies
 simultaneously.
 Solid line: the continuous
 transitions between swollen ($m=0$) and collapsed ($m\neq 0$) solutions. Dashed: the discontinuous transition.
 Phases
 are defined and described in the main text.
 } \label{fig:phasediag_ninfty_coex}
\end{figure}
Although both are co-located and are
continuous in the fundamental order parameters $(m,k)$, there is
an important  difference between the HS$\to$HC and the IS$\to$IC
transitions, which involves the behaviour of the polarity
distribution $\pi(\xi)$. As one crosses from HS into HC,
$\pi(\xi)$ remains unchanged, taking the value
$\pi(\xi)=\delta(\xi-k)$ in both states. In contrast, we know from
 (\ref{eq:specialcase_ninfty}) that the IS state has a continuous
polarity distribution $\pi(\xi)=\int\!dh~W(\xi,h)$ whereas the IC
state has the binary distribution
$\pi(\xi)=\frac{1}{2}(1\!+\!k_0)\delta(\xi\!-\!1)+\frac{1}{2}(1\!-\!k_0)\delta(\xi\!+\!1)$.
Thus, the transition IS$\to$IC is in fact {\em discontinuous}, in
spite of it involving no jump in the order parameter $m$ itself.

 Upon translating our results into the original
control parameters $\beta J_p$ and $\beta J_s$ one obtains the
phase diagram cross-sections shown in figures \ref{fig:phasediag_ninfty_nocoex}
and \ref{fig:phasediag_ninfty_coex}. The phase where compact
($m\neq 0$) and swollen ($m=0$) states coexist will be
characterized by strong remanence effects. The thermodynamic
transition line (calculated by selecting the solution with the
lowest free energy) coincides with the second order transition for
$\beta J_p<\sqrt{3}$, and will be found inside the coexistence
region for $\beta J_p>\sqrt{3}$.

Without noise (i.e. random mutations) in the sequence
selection process, $\tilde{\beta}=\infty$, we can summarize the
behaviour of the system as follows. For $J_p>J_g$ it
always finds itself in states where any infinitesimal functional
advantage of either the hydrophilic or the hydrophobic monomers
leads to amino-acid sequences that are, unlike proteins, fully homogeneous in their
polarity. The phenomenology described by the remaining equations
for $m$ and the resulting phase diagram
reflect the interplay between the tendencies of the
polarity-homogeneous system to have similarly oriented amino-acid
residues (induced by the long-range forces) and low steric
energies (induced by the short-range forces). The system behaves
as an Ising chain with random short-range bonds and uniform long
range bonds. In those cases where the amino-acids are
forced by steric effects to have non-identical side chain
orientations (i.e. for $\nu>0$) there is a complex competition
between long range and short-range order, which leads
to low values of $|m|$ and strong remanence effects, in sharp contrast to the
situation in mean field models \cite{Chak}. In contrast,
for $\nu<0$ both the long range and the short range forces promote
similar side chain orientations; the absence of frustration is
responsible for the absence of remanence effects and for having
large $|m|$ (strong secondary structure). For $J_g>J_p$ it is no
longer energetically advantageous to select chains with uniform
polarity, and here we find the protein-like states. The polarity inhomogeneity of the sequence reduces
dramatically the energetic impact of the long-range forces
compared to the case $k\pm 1$, and this decouples the
strength $|m|$ of the secondary structure from any preference for
aligning or anti-aligning short-range forces, as controlled by $\nu$.

\section{Transitions and phase diagrams for non-deterministic sequence selection}

In this section we extract solutions, transition
lines and phase diagrams from our order parameter equations for
non-deterministic selection of primary sequences, viz. finite $n$.
Full analytical
solution of our equations is generally ruled out, so we restrict
ourselves to the study of instabilities and to
collecting further information on phases by solving our equations
numerically. As in the previous section we restrict ourselves to
simple parameter choices, in particular we
 take $v(u)=\frac{1}{2}u^2$ and $k_0\in(-1,1)$.

\subsection{Continuous transitions away from $m=0$}

\noindent We first derive exact conditions marking continuous
phase transitions
away from the state $m=0$ without secondary structure as defined
and studied earlier, for arbitrary $n$. For $m=0$ one has
$\Psi(x)=\Phi(x)=\delta(x)$, and $k$ is to be solved from
(\ref{eq:mzero_keqn}).
We make in our order parameter equations
(\ref{eq:p_xi},\ref{eq:normalized_eve_psi},\ref{eq:final_m_with_fields},\ref{eq:final_k_with_fields},\ref{eq:effective_fields}) the substitutions
$m\!\to\!\Delta m$, $k\!\to\! k\!+\!\Delta k$, and
$\Psi(x)\!\to\!\delta(x)\!+\!\Delta\Psi(x)$. We next expand these equations
in $\{\Delta m,\Delta k,\Delta\Psi(x)\}$ and locate their linear
instabilities. In doing so we may use $k=\int\!d\xi~p(\xi)\xi$,
which holds for $m=0$. In practice it turns out somewhat easier to
involve also the auxiliary distribution $\Phi(x)$, and replace
(\ref{eq:normalized_eve_psi}) by the pair
(\ref{eq:Psi_in_Phi},\ref{eq:Phi_in_Psi}). First, substitution in and expansion of equations
(\ref{eq:p_xi}) and (\ref{eq:Phi_in_Psi})
gives
\begin{eqnarray}
 \Delta p(\xi)
&=& n\beta (J_p\!-\!J_g)(\xi\!-\!k) p(\xi)\Delta k
+\order(\Delta^2)~~\\
 \Delta\Phi(x)&=&
\Delta\Psi(x) -J_p k \delta^\prime(x)\Delta m  + \order(\Delta^2)
\end{eqnarray}
These results are then substituted into (\ref{eq:Psi_in_Phi}), which leads
to an equation for $\Delta\Psi(x)$:
\begin{eqnarray}
\hspace*{-25mm}
\Delta\Psi(x) &=& \\
\hspace*{-25mm}&&\hspace*{-15mm}
\frac{\int\!dx^\prime [ \Delta\Psi(x^\prime)
\!-\!J_p k\Delta m \delta^\prime(x^\prime) ] \int\!d\eta~w(\eta)\big\{
\delta[x\!-\! A(x^\prime\!,\eta J_s)]\!-\!\delta(x)\big\}
 e^{n\beta [B(x^\prime\!,\eta
J_s)- \nu\eta]}} { \int\!d\eta~w(\eta) e^{n\beta [B(0,\eta J_s)-
\nu\eta]}}+\order(\Delta^2)\nonumber
\end{eqnarray}
 We next separate $\Delta\Psi(x)$ into its symmetric and
anti-symmetric parts,
$\Delta\Psi(x)=\Delta\Psi_S(x)+\Delta\Psi_A(x)$, giving up to order $\Delta$:
\begin{eqnarray}
\hspace*{-25mm}
\Delta\Psi_S(x) &=& \frac{\int\!dx^\prime\Delta\Psi_S(x^\prime)
\int\!d\eta~w(\eta)\big\{ \delta[x\!-\! A(x^\prime\!,\eta
J_s)]\!-\!\delta(x)\big\}
 e^{n\beta [B(x^\prime\!,\eta
J_s)- \nu\eta]}} { \int\!d\eta~w(\eta) e^{n\beta [B(0,\eta J_s)-
\nu\eta]}}
 \label{eq:PsiS}
\\
\hspace*{-25mm}
\Delta\Psi_A(x) &=& \frac{\int\!dx^\prime [ \Delta\Psi_A(x^\prime)
\!-\!J_p k\Delta m \delta^\prime(x^\prime) ] \int\!d\eta~w(\eta)
\delta[x\!-\! A(x^\prime\!,\eta J_s)]
 e^{n\beta [B(x^\prime\!,\eta
J_s)- \nu\eta]}} { \int\!d\eta~w(\eta) e^{n\beta [B(0,\eta J_s)-
\nu\eta]}}\nonumber
\\
\hspace*{-25mm}&& \label{eq:PsiA}
\end{eqnarray}
 The symmetric and anti-symmetric
parts obey independent equations, and only the anti-symmetric
part $\Psi_A(x)$ is coupled to the bifurcation of $m\neq 0$. Apparently, any nonzero solution of equation
(\ref{eq:PsiS}) describes transitions from one $m=0$ state to
another, whereas equation (\ref{eq:PsiA})
controls the bifurcations away from $m=0$.

In order to expand equations
(\ref{eq:final_m_with_fields},\ref{eq:final_k_with_fields}) for
the scalar order parameters we need to vary the distribution
$W(\xi,h)$ defined in (\ref{eq:effective_fields}), which we first
rewrite as
\begin{eqnarray}
W(\xi,h)&=& \frac{p(\xi)\cosh^n(\beta
h)\int\!dxdy~ \Psi(x) \Psi(y)\delta(h\!-\!J_p m\xi\!-\!x\!-\!y)}
 {\int\!d\xi^\prime p(\xi^\prime) \int\!dxdy~ \Psi(x)
\Psi(y)\cosh^n[\beta(J_p m\xi^\prime\!+\!x\!+\!y)]} \nonumber
\end{eqnarray}
 Upon varying this equation around the $m=0$ state we then find
\begin{eqnarray}
\hspace*{-15mm}
\Delta W(\xi,h)&=& \Delta p(\xi) \delta(h) + 2p(\xi) \cosh^n(\beta
h) \Delta\Psi(h)
 -J_p \Delta m ~\xi p(\xi)
\cosh^n(\beta h) \delta^\prime(h) \nonumber\\
\hspace*{-15mm}
 &&
 -2 p(\xi)\delta(h)\!\int\!dy \cosh^n(\beta y)
 \Delta\Psi_S(y)+\order(\Delta^2)
 \nonumber
 \\
 \hspace*{-15mm}
 &=& p(\xi)\Big\{ n\beta (J_p\!-\!J_g)(\xi\!-\!k)\delta(h)
 +  \cosh^n(\beta
h)[ 2\Delta\Psi(h) -J_p \xi\Delta m~\delta^\prime(h)]\nonumber
\\
\hspace*{-15mm} &&  -2
\delta(h)\int\!dy~\cosh^n(\beta y)
 \Delta\Psi_S(y)\Big\}+\order(\Delta^2)
\end{eqnarray}
 Insertion into
(\ref{eq:final_m_with_fields},\ref{eq:final_k_with_fields}) then
gives, using $\int\! dh~\tanh(\beta h)
\cosh^n(\beta h) \delta^\prime(h)=-\beta$:
\begin{eqnarray}
\hspace*{-20mm} \Delta m&=& 2k \!\int\! dh~\tanh(\beta h)
\cosh^n(\beta h) \Delta\Psi_A(h) +\beta J_p \Delta
m\!\int\!d\xi~p(\xi)\xi^2\! +\order(\Delta^2) ~~~~\label{eq:general_bif_m}
 \\
 \hspace*{-20mm}
 \Delta k&=&  n\beta
(J_p\!-\!J_g)\Big[\!\int\!d\xi~\xi^2p(\xi)\!-\!k^2\Big]\Delta k
+\order(\Delta^2)~~~~~ \label{eq:general_bif_k}
\end{eqnarray}
As expected, the perturbations $\Delta m$ couple only to the
anti-symmetric part of $\Delta\Psi(x)$; the $m\neq 0$ bifurcations
are the instabilities of the coupled pair
(\ref{eq:PsiA},\ref{eq:general_bif_m}). Furthermore, equation
(\ref{eq:general_bif_k}) for $\Delta k$ does not depend on
the symmetric part of $\Delta\Psi(x)$, so we may for the purpose
of studying continuous transitions away from the $m=0$ state
regard $\delta \Psi(x)$ as strictly anti-symmetric and extract
instabilities involving $k$ only from (\ref{eq:general_bif_k}).

\begin{figure}[t]
\vspace*{-5mm} \hspace*{20mm}\setlength{\unitlength}{1.0mm}
\begin{picture}(220,70)
\put(15,5){\epsfysize=60\unitlength\epsfbox{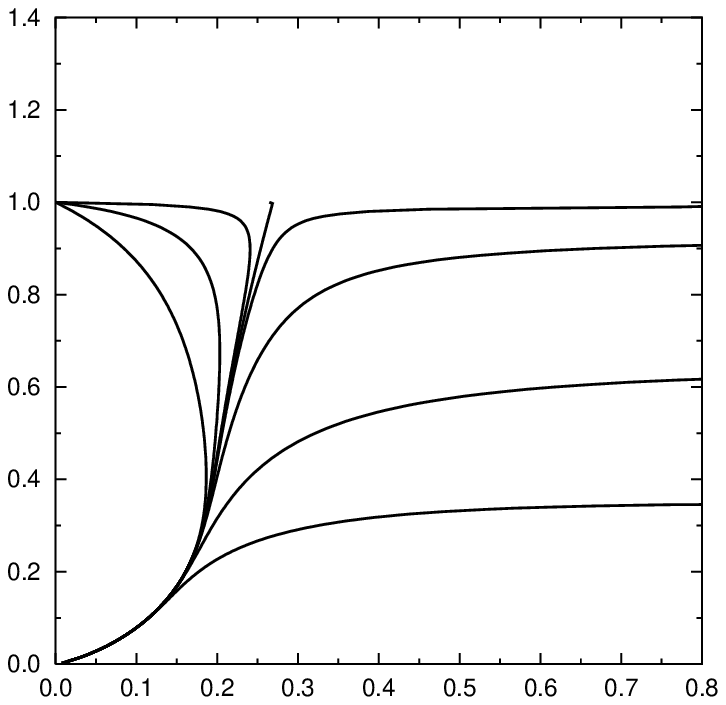}}
\put(45,-2){$J_s/J_p$} \put(6,34){$T/J_p$}
\put(80,47){\small $n\!=\!1.99$}
\put(80,42){\small $n\!=\!1.9$}
\put(80,31){\small $n\!=\!1.5$}
\put(80,20){\small $n\!=\!1$}
\put(40,50){\small $n\!=\!2$}
\put(25,48){\small $n\!=\!2.01$}
\put(24,38){\small $n\!=\!2.1$}
\put(23,28){\small $n\!=\!2.5$}
\put(60,55){HS}
\end{picture}
\vspace*{2mm} \caption{Continuous bifurcations from swollen ($m=0$, HS) to collapsed ($m\neq 0$) states, for several
$n$ values around $n=2$,
 for the case where
$J_p>J_g$ and $\nu>0$ (homogeneous polarity sequences, unlike proteins). The corresponding curve for $n=\infty$ is shown in figure \ref{fig:phasediag_ninfty_coex}.
We see that, if there were no discontinuous transitions, reentrance would occur upon lowering $T$ for $n>2$, where for $n<2$
the continuous transition temperature is monotonic in $J_s/J_p$. This suggests strongly that there is a discontinuous bifurcation to HM
phase for $n>2$, but not for $n<2$.
 } \label{fig:phasediag_nfinite_coex}
\end{figure}

 It turns out that the (anti-symmetric) functional
perturbation $\Delta \Psi_A(x)$ that solves equation
(\ref{eq:PsiA}) can be expressed in terms of
$\Delta m$. We show this by substituting  for $\lambda\neq 1$ the
ansatz
\begin{eqnarray}
\Delta\Psi_A(x)&=&\frac{\lambda J_pk}{\lambda\!-\!1}
~\delta^\prime(x) \Delta m \label{eq:PsiAansatz}
 \end{eqnarray}
 into the leading orders
of (\ref{eq:PsiA}). Using integration by parts and the properties
$\partial_x B(x,y)|_{x=0}=0$ and $\partial_x
A(x,y)|_{x=0}=\tanh(\beta y)$ this is found to give
\begin{eqnarray}
\hspace*{-20mm}
\lambda \delta^\prime(x) &=& -\frac{\int\!dx^\prime
\delta^\prime(x^\prime) \int\!d\eta~w(\eta) \delta[x-
A(x^\prime\!,\eta J_s)]
 e^{n\beta [B(x^\prime\!,\eta
J_s)- \nu\eta]}} { \int\!d\eta~w(\eta) e^{n\beta [B(0,\eta J_s)-
\nu\eta]}} \nonumber
\\
\hspace*{-20mm}
&=& - \frac{\int\!dx^\prime \delta(x^\prime)
\int\!d\eta~w(\eta)\Big\{n\beta \delta[x\!-\! A(x^\prime\!,\eta
J_s)] \frac{\partial}{\partial x^\prime}B(x^\prime\!,\eta
J_s)
\Big\}
 e^{n\beta [B(x^\prime\!,\eta
J_s)- \nu\eta]}} { \int\!d\eta~w(\eta) e^{n\beta [B(0,\eta J_s)-
\nu\eta]}} \nonumber
\\
\hspace*{-20mm}
&&
+ \frac{\int\!dx^\prime \delta(x^\prime)
\int\!d\eta~w(\eta)\Big\{\delta^\prime[x\!-\! A(x^\prime\!,\eta
J_s)]\frac{\partial}{\partial x^\prime}A(x^\prime\!,\eta J_s)
\Big\}
 e^{n\beta [B(x^\prime\!,\eta
J_s)- \nu\eta]}} { \int\!d\eta~w(\eta) e^{n\beta [B(0,\eta J_s)-
\nu\eta]}} \nonumber
\\
\hspace*{-20mm}
&=&  \delta^\prime(x)~ \frac{ \int\!d\eta~w(\eta) \tanh(\beta \eta
J_s) e^{n\beta [B(0,\eta J_s)- \nu\eta]}} { \int\!d\eta~w(\eta)
e^{n\beta [B(0,\eta J_s)- \nu\eta]}}
\end{eqnarray}
 This confirms that (\ref{eq:PsiAansatz}) indeed
solves our bifurcation equation, with
\begin{eqnarray}
\lambda&=& \frac{ \int\!d\eta~w(\eta) \tanh(\beta \eta J_s)
e^{n\beta [B(0,\eta J_s)- \nu\eta]}} { \int\!d\eta~w(\eta)
e^{n\beta [B(0,\eta J_s)- \nu\eta]}} \label{eq:lambda}
\end{eqnarray}
This result allows us to compactify our bifurcation conditions
further. Upon substituting (\ref{eq:PsiAansatz}) into
(\ref{eq:general_bif_m}) and carrying out the remaining integral,
we obtain the following simple set of bifurcation conditions:
\begin{eqnarray}
&\Delta m\neq 0:&~~~~ 1= \beta J_p\Big[
\int\!d\xi~\xi^2p(\xi)-\frac{2\lambda k^2}{\lambda\!-\!1}\Big]
\label{eq:final_bif_m}
 \\
&\Delta k\neq 0:&~~~~ 1= n\beta
(J_p\!-\!J_g)\Big[\!\int\!d\xi~\xi^2p(\xi)\!-\!k^2\Big]
\label{eq:final_bif_k}
\end{eqnarray}
where
\begin{eqnarray}
p(\xi)&=& \frac{w(\xi)e^{n\beta \xi(J_p-J_g)(k-k_0)}}
 {\int\!d\xi^\prime w(\xi^\prime)e^{n\beta
 \xi^\prime(J_p-J_g)(k-k_0)}}
 \end{eqnarray}
 For $\beta=0$, infinite temperature, the right-hand sides of
 (\ref{eq:final_bif_m}) and (\ref{eq:final_bif_k}) are zero.
 Hence the physical transitions occur at the highest temperature
 for which the right-hand sides have increased to the value 1. If the first transition to take place is
(\ref{eq:final_bif_k}), then $m$ will remain zero and equation
(\ref{eq:final_bif_m}) will still apply to predict a further
$m\neq 0$ transition. If (\ref{eq:final_bif_m})  is the first transition to
occur, then (\ref{eq:final_bif_k}) will no longer apply. \vsp

 As a simple but nontrivial test we can recover from
(\ref{eq:final_bif_m},\ref{eq:final_bif_k}) our earlier
predictions for the limit $n\to\infty$. Taking
$n\to\infty$ in (\ref{eq:lambda}) gives the simple result
$\lim_{n\to\infty}\lambda = -\sgn(\nu)\tanh(\beta J_s)$. In the
HS, HC and HM phases we have $J_p>J_g$ and $k=\pm 1$, so
$\lim_{n\to\infty}p(\xi)=\delta(\xi-k)$ and therefore
$\lim_{n\to\infty}\int\!d\xi~\xi^2p(\xi)=1$.
 This simplifies condition (\ref{eq:final_bif_m}) for the continuous bifurcation of $m\neq 0$  in the $k=\pm 1$ phases
 to the expression found earlier in analyzing the $n\to\infty$
  equations, as it should:
\begin{eqnarray}
\beta J_s&=&
 \frac{1}{2}\sgn(\nu)\log(\beta J_p)
\end{eqnarray}
For the $k=k_0$ states the $m\neq 0$ bifurcation is
discontinuous, involving a jump in the polarity statistics as
measured by $\pi(\xi)$; so there equations
(\ref{eq:final_bif_m},\ref{eq:final_bif_k}) do not apply.

As an application of (\ref{eq:final_bif_m},\ref{eq:final_bif_k}) we have solved these equations
numerically for $J_g/J_p=\nu=\frac{1}{2}$ and $k_0=0$, to investigate the  effect of genetic noise on the
phase diagram in figure \ref{fig:phasediag_ninfty_coex} (although this is the biologically less relevant case of polymers with homogeneous polarity,
it has the more interesting phase diagram). The result is shown in figure \ref{fig:phasediag_nfinite_coex}. Although based on equations that only apply to continuous transitions, the figure allows us to predict  on topological grounds that discontinuous
transitions will occur for $n\geq 2$. This is a remarkable result:
the critical value $n=2$ for the onset of first order transitions was found persistently in earlier coupled dynamics models \cite{Coolen1993,Penney1993,Jongen1998,Jongen2001},
 but since these did not involve short range forces, its re-appearance in the present model strongly suggests an unexpected universality
 which at present we do not understand.

\section{Numerical results}

\subsection{Numerical solution of order parameter equations via population dynamics}

 The goal of this
section is to verify numerically the phases predicted in
previous sections, and to provide phase diagrams for those cases where
solutions of equations (\ref{eq:normalized_eve_psi},\ref{eq:final_m_with_fields},\ref{eq:final_k_with_fields},\ref{eq:effective_fields})
 for the observables $m$, $k$ and
$\Psi(x)$ can not be found analytically.
To limit the number of control parameters to be varied  we choose $J_s=0.1$, $J_p=1$, $\mu=J_pk^\star$ (so $k_0=k^\star$), and $k^\star=0.7$  throughout, since this still allows us to probe all
the phases
in figures \ref{fig:phasediag_ninfty_nocoex} and \ref{fig:phasediag_ninfty_coex}.
We followed
the mathematically related studies
\cite{Theo,Theo2,HSN2005,HSBB2006} and solved the functional equation
(\ref{eq:normalized_eve_psi}) using a so-called population dynamics
algorithm (with a population of size $10^4$), which exploits the interpretation of such equations as fixed-point conditions for a suitably chosen stochastic process for the local fields.

\begin{figure}[t]
\vspace*{-5mm} \hspace*{-3mm}\setlength{\unitlength}{0.93mm}
\begin{picture}(80,70)

\put(5,5){\epsfysize=60\unitlength\epsfbox{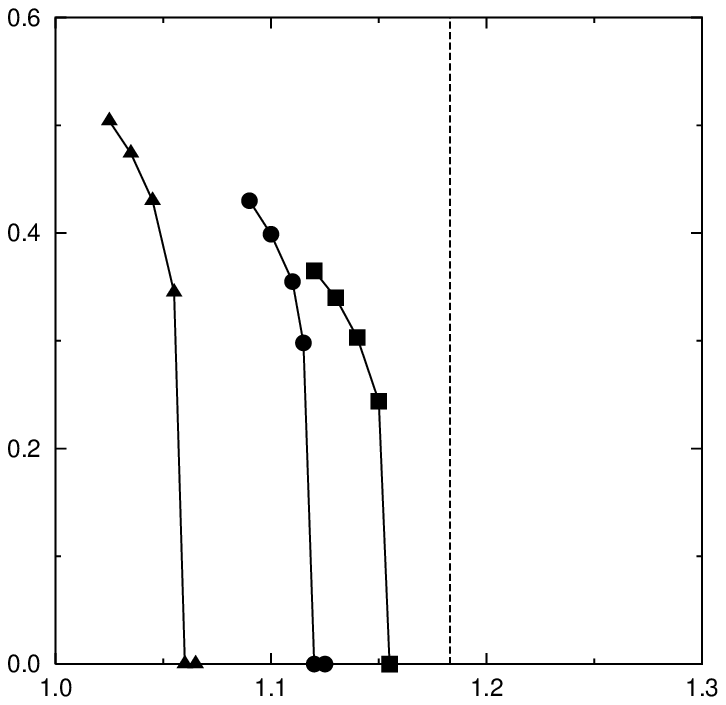}}
\put(37,-2){$T$} \put(1,34){$m$} \put(17,57){\small $n=100$}
\put(30,50){\small $n=200$} \put(37,43){\small $n=400$}

\put(80,5){\epsfysize=60\unitlength\epsfbox{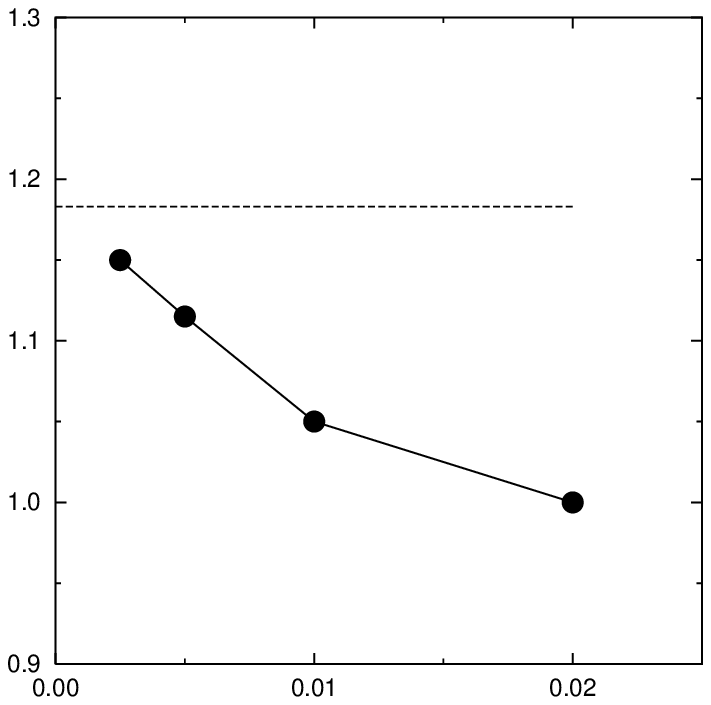}}
\put(107,-2){$1/n$} \put(76,34){$T_c$}
\end{picture}
\vspace*{3mm}
\caption{Left: dependence of order parameter $m$ on the folding temperature $T$, obtained by numerical solution of the order
 parameter equations, for control parameters $(J_s,J_p,J_g)=(0.1,1,2)$, $k_0=0.7$, $\mu=0.2$, $\nu=0.5$. The relative genetic noise levels $n=\tilde{T}/T$ were $n=100$ (connected triangle), $n=200$ (connected circles) and $n=400$ (connected squares).
 According to our earlier analysis, for $n\to\infty$
 the phases should be those shown in figure \ref{fig:phasediag_ninfty_coex}. For the present values of control parameters this predicts
 for $n\to\infty$ a continuous transition from $m\neq 0$ (IS phase) to $m=0$ (IC phase) at $\lim_{n\to\infty}T_c=1.183$ (shown as a vertical dashed line).
 Right: the IS$\to$IC transition temperatures $T_c$ shown versus $1/n$, for the same values of the remaining control parameters. The data are perfectly consistent with  the analytically determined
 value $\lim_{n\to\infty}T_c=1.183$ (dashed).
 }
 \label{fig:numerics_ninfty}
\end{figure}

\begin{figure}[t]
\vspace*{-5mm} \hspace*{-3mm}\setlength{\unitlength}{0.93mm}
\begin{picture}(80,70)

\put(5,5){\epsfysize=60\unitlength\epsfbox{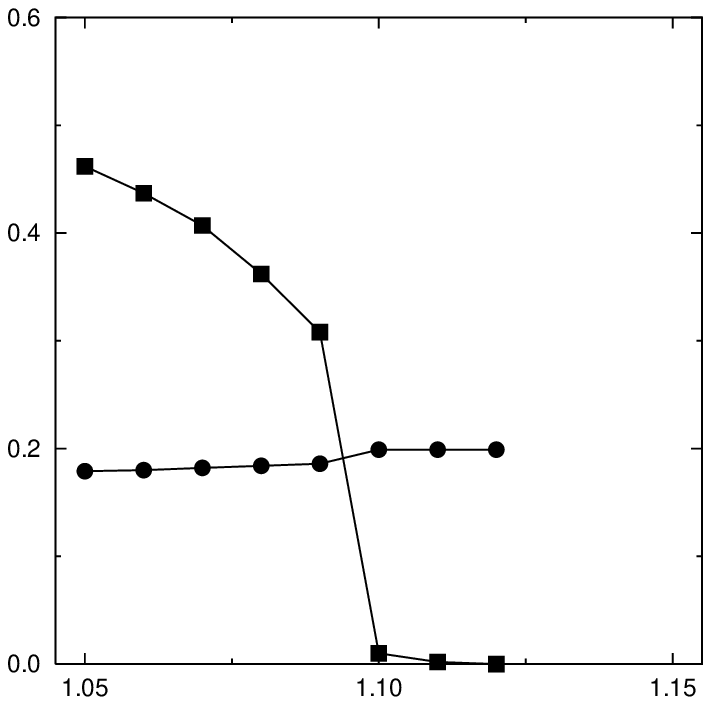}}
\put(37,-2){$T$} \put(-2,34){$m,k$} \put(50,57){$\nu=0.5$}

\put(80,5){\epsfysize=60\unitlength\epsfbox{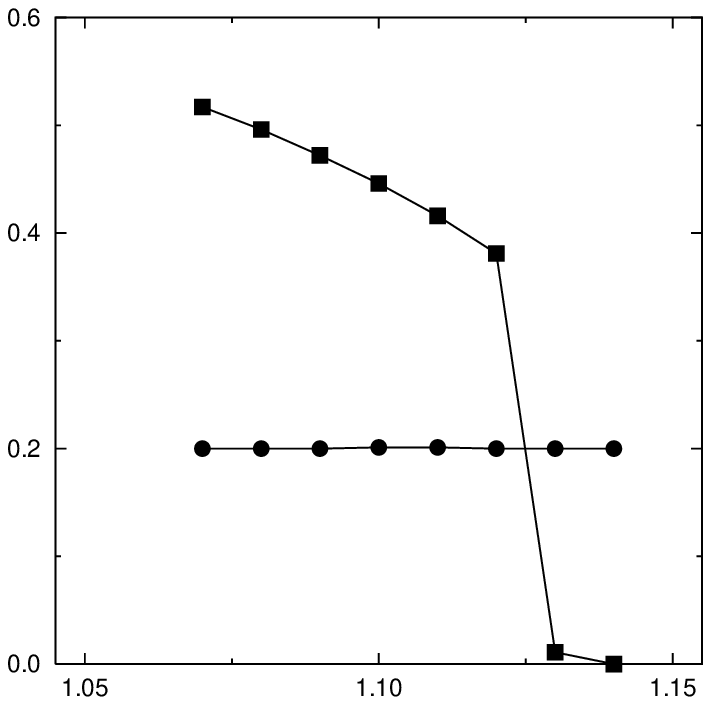}}
\put(112,-2){$T$}\put(73,34){$m,k$} \put(123,57){$\nu=-0.5$}
\end{picture}
\vspace*{3mm}
\caption{Dependence of order parameter $m$ (connected squares) and $k$ (connected circles) on the folding temperature $T$, obtained by numerical solution of the order
 parameter equations, for control parameters $(J_s,J_p,J_g)=(0.1,1,2)$, $k_0=0.2$, and $\mu=0.7$.
In both graphs the relative genetic noise level is $n=\tilde{T}/T=200$. Left graph: $\nu=0.5$ (promoting different orientations for adjacent amino-acids). Right graph: $\nu=-0.5$ (promoting identical orientations).
The large $n$ theory of the previous section predicted that the sub-leading order in $n$ for the $k=k_0$ solution (as shown here) is $\order(n^{-1})$
when $\nu<0$, but $\order(n^{-1/2})$ when $\nu>0$. The numerical data shown here are consistent with this prediction.}
 \label{fig:numerics_n100}
\end{figure}

We turn first to the most important and realistic  case of (near-)deterministic sequence selection, where for
$n\to\infty$ we expect to recover the behaviour shown in the phase diagrams of figures
\ref{fig:phasediag_ninfty_nocoex} and \ref{fig:phasediag_ninfty_coex}. Here we face the practical
problem that in our equations $n$ appears usually in exponents, which limits our numerical analysis
to values $n\leq 400$. It turns out that to observe the $n\to\infty$ predictions one needs values of $n$ that are
significantly larger than this; furthermore, for large but finite $n$ the limiting values of transition temperatures
and the nature of the various transitions can vary significantly from one phase to another.
In figure \ref{fig:numerics_ninfty} we present numerical results for
positive $\nu$, where the steric forces make it energetically favourable for adjacent amino-acids to
have different side chain orientations.
 We plot the order parameter $m$ versus temperature (left panel) to locate the IS$\to$IC phase transition, which for $n\to\infty$ was predicted to be continuous, and which for the present control parameters should occur at $T_c=1.183$.
It turns out that for large but finite $n$ the transition is in fact
{\em discontinuous} and at a lower temperature than the $n\to\infty$ one. However, a study of the asymptotic
scaling with $n$ of the transition temperature, within the numerically accessible regime, confirms that for $n\to\infty$ the correct value is found, see figure
\ref{fig:numerics_ninfty} (right).
The observed strong dependence on $n$ of the exact location of the transition is remarkable;  the
system appears to be very sensitive to the ratio of temperatures of the two coupled processes,
and the deterministic regime is achieved only
asymptotically. For $n=100$ the location of the transition point differs by more than $10\%$ from its $n\to\infty$ value. If one carries out a scaling analysis of the magnitude of the jump in $m$ found at the transition temperature
for large but finite $n$, one finds that for $n\to\infty$ this jump will indeed vanish, in agreement with our previous asymptotic analysis.

\begin{figure}[t]
\vspace*{-5mm} \hspace*{-3mm}\setlength{\unitlength}{0.93mm}
\begin{picture}(80,70)

\put(5,5){\epsfysize=60\unitlength\epsfbox{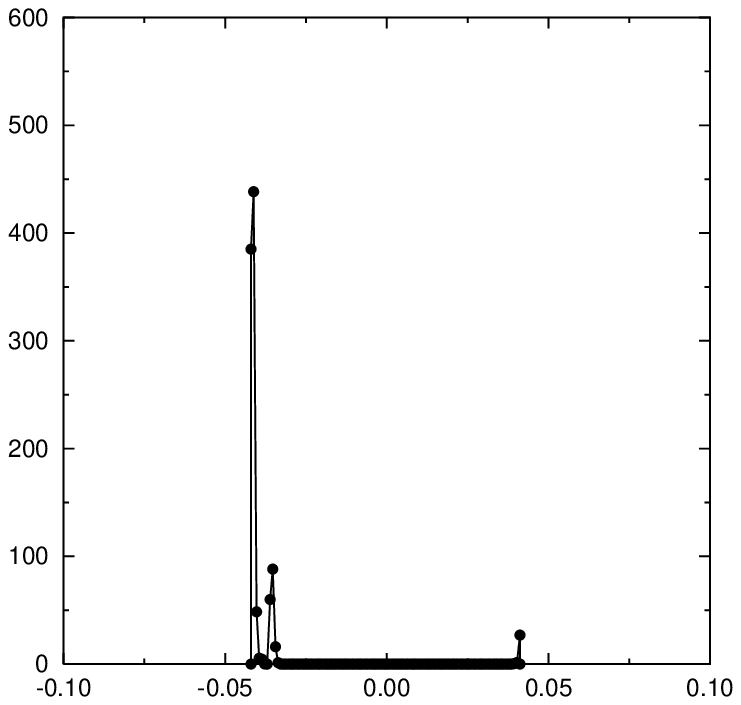}}
\put(38,-2){$x$} \put(-2,36){$\Psi(x)$}

\put(80,5){\epsfysize=60\unitlength\epsfbox{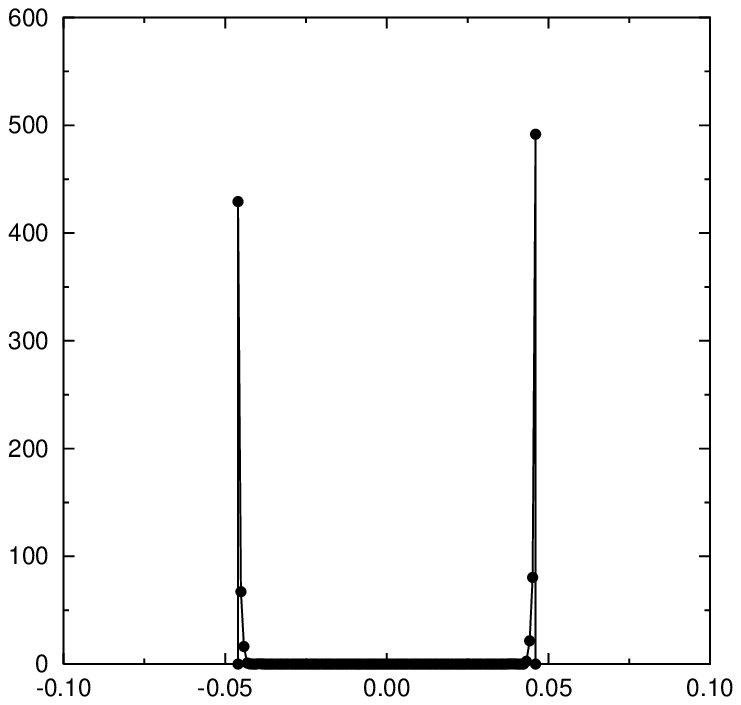}}
\put(113,-2){$x$} \put(73,36){$\Psi(x)$}
\end{picture}
\vspace*{4mm}
\caption{
Distribution $\Psi(x)$ of the short-range contributions to the local effective fields, for
$(J_s,J_p,J_g)=(0.1,1,2)$,
$n=200$, $k_0=0.2$, $\mu=0.7$, and $T=1.07$, as obtained via a population dynamics algorithm.
Left:  $\nu=0.5$. Right: $\nu=-0.5$. Since for $n\to\infty$ the function $\Psi(x)$ is symmetric,
so these results confirm that finite-$n$ effects are more profound for $\nu>0$ (where they are predicted to be $\order(n^{-1/2})$)
than for $\nu<0$ (where they should be $\order(n^{-1})$).
}
 \label{fig:Psix}
\end{figure}

Upon carrying out a similar analysis for  negative values of $\nu$, where steric forces are such that adjacent amino-acids
prefer identical chain orientations, the resulting graphs and the physical picture are similar to those of $\nu>0$.
For large but finite $n$ the phase transition is again
discontinuous,  and a scaling analysis shows once more good
agreement with the theory in the limit $n\to\infty$.
However, there is an important difference between the cases $\nu>0$ and $\nu<0$
which concerns the sub-leading orders in $n^{-1}$ for the state $k=k_0$, as
$n\rightarrow \infty$, which is reflected in both the field distribution
$\Psi(x)$ and in the order parameter $k$ close the transition.
This is an important success of the population dynamics algorithm,
which allows us to evaluate in a simple and straightforward way the
distribution $\Psi(x)$ of the short-range contributions to the local effective fields. In addition, it is a crucial test to verify
the scaling of the sub-leading orders in $n^{-1}$ predicted by our
theory. Figure \ref{fig:numerics_n100}  shows how $m$ and $k$
behave close to the transition. One notes that the (discontinuous) behaviour of $m$ is qualitatively similar in both cases, whereas
the polarity
$k$ behaves in a very different way: in contrast to $\nu<0$, for $\nu>0$ there
is a noticeable (small) jump in $k$ at the transition. This can
be explained if we assume that for $\nu>0$ the
solution scales in a different way with $n^{-1}$. Close inspection of the jump
shows that this jump for $\nu>0$ is indeed of order $1/\sqrt{n}$, again in perfect
agreement with the theory.
The difference between the regimes $\nu<0$ and $\nu>0$ is also observed in the
distribution $\Psi(x)$ of the short-range contributions to the local effective fields; see Figure \ref{fig:Psix}.

\begin{figure}[t]
\vspace*{-5mm} \hspace*{45mm}\setlength{\unitlength}{1.0mm}
\begin{picture}(80,70)
\put(5,5){\epsfysize=60\unitlength\epsfbox{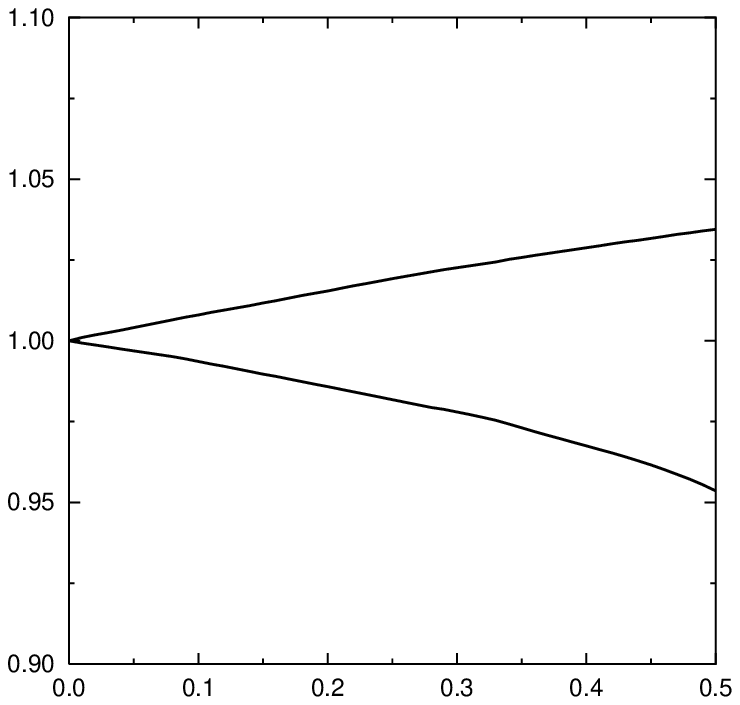}}
\put(35,-2){$J_s/J_p$}
\put(-3,34){$T/J_p$}
\put(20,20){IC/HC}\put(20,50){IS/HS}

\put(55,38){$\nu=-1$}\put(56,19){$\nu=1$}
\end{picture}
\vspace*{2mm} \caption{Phase diagram cross-sections for
$n=1$ (strongly noisy sequence selection),
$\mu=0.7$, and $J_g/J_p=2$.
Top curve: $\nu=-1$ (promoting identical orientations of adjacent amino-acids). Bottom curve: $\nu=1$.
(promoting opposite orientations).
  Solid line: transition marking the continuous
 bifurcation of $m\neq 0$ states.
 Phases are defined and described in the main text.
 } \label{fig:phasediag_n1}
\end{figure}

Although less relevant from a biological point of view, it is interesting to compare the phase diagrams of $n\to\infty$
(or at least large), describing (near-) deterministic sequence selection, to those one would have found for
very noisy sequence selection. An example is shown in Figure
\ref{fig:phasediag_n1}, for $n=1$. Compared
to the phase diagram of Figure \ref{fig:phasediag_ninfty_nocoex}, we see that in the presence of high  genetic noise
the impact of the short range forces, as measured by $J_s$ is reduced drastically (note the different vertical scales),
with as expected a corresponding reduction of sequence selection specificity.

\subsection{Numerical simulations}

\begin{figure}[t]
\vspace*{-5mm} \hspace*{-3mm}\setlength{\unitlength}{0.93mm}
\begin{picture}(80,70)

\put(5,5){\epsfysize=60\unitlength\epsfbox{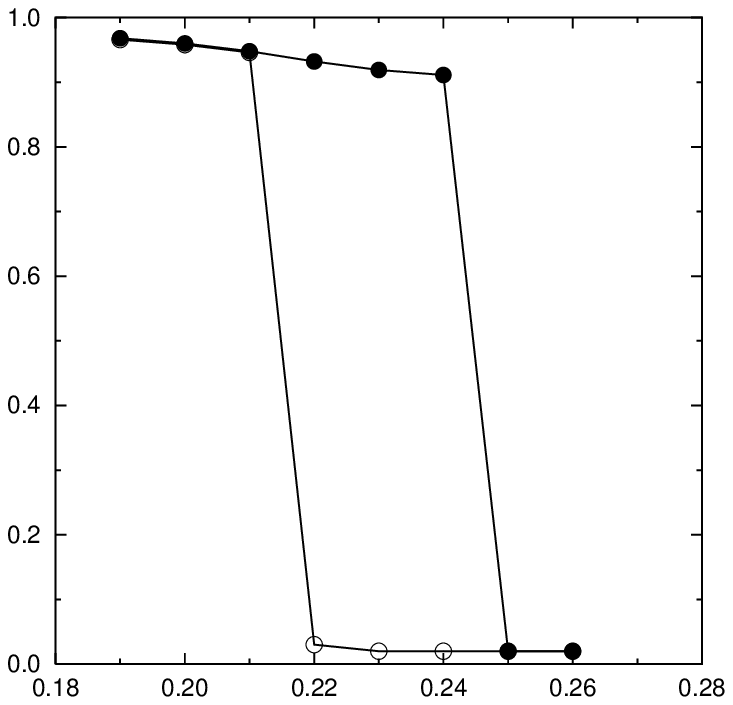}}
\put(35,-2){$J_s/J_p$} \put(-2,36){$m$}

\put(80,5){\epsfysize=60\unitlength\epsfbox{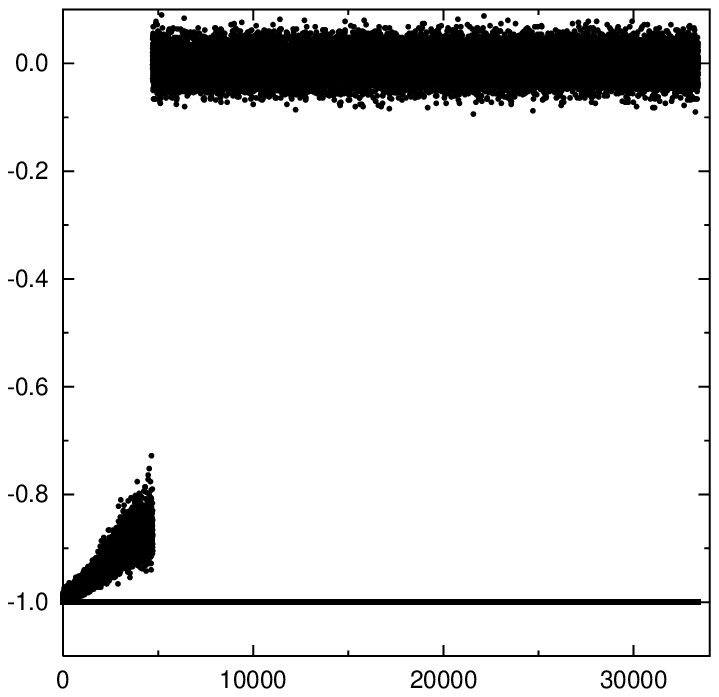}}
\put(113,-2){$t$} \put(73,36){$m,k$}
\end{picture}
\vspace*{4mm}
\caption{Results of numerical simulations of the coupled stochastic processes of (fast) folding and (slow)
primary sequence selection, for $N=1000$, at $T=0.3$ and $n=200$. Further system parameters:
$\nu=J_g=\frac{1}{2}$, $J_p=1$, and and $k^\star=0.7$.
 For $n\to\infty$ one expects to find the phenomenology of Figure \ref{fig:phasediag_ninfty_coex}, with $k=-1$ and with a
 region where $m=0$ (swollen) and $m\neq 0$ (folded) states are simultaneously stable; for $n=200$ one expects this to remain true but with
 shifted values of $J_s/J_p$.
 The left picture shows the equilibrated values of $m$, found upon increasing $J_s$ in stages from below (full circles) and alternatively upon decreasing $J_s$ in stages from above (open circles).
 It confirms that there is a coexistence region at the predicted range of values
 for $J_s/J_p$.
 The right picture, measured at $J_s/J_p=0.25$, shows the evolution in time of $m$ (upper) and $k$ (lower), upon initializing the system in
 the folded state that is stable for lower values of $J_s$. It suggests that the chosen duration of $5.10^4$ iterations per monomer suffices
 in the present parameter regime to achieve equilibration.
}
 \label{fig:simulations}
\end{figure}

The theory presented in this manuscript makes a large number of
predictions about the cooperative long-time behaviour of the
polymeric chain. In some asymptotic limits it is possible to work
out the expressions for the relevant order-parameters of the system
and find simplified algebraic equations which allow to plot phase
diagrams. In other cases we had to rely on population dynamics algorithms to solve
our functional order parameter equations and detect the relevant transition lines.

In order to have independent tests of our formulae we have also performed Monte-Carlo simulations
of the stochastic processes that would lead to equilibration with the Hamiltonians (\ref{eq:H}) and (\ref{eq:Heff}).
 This is of course the cleanest way to check the theory. However, due
to the special character of the coupled dynamics which requires nested equilibrations of two complex processes at
widely separated time-scales,  these simulations
are highly nontrivial and extremely time consuming, and one is severely limited in both the number and the precision of
simulation experiments
that can be completed reliably. A systematic scan of all possible parameter regimes is certainly ruled out.
Instead we focused on the regime $n\to\infty$ (genetic evolution of sequences at low noise levels).
This is not only the most relevant one biologically, but is also
the regime where our predictions take their most explicit form, as here we could go beyond population dynamics analyses.
In particular, we chose to invest our computing resources
in verifying the existence and location of the predicted coexistence region in the phase diagram
shown in Figure \ref{fig:phasediag_ninfty_coex}.

 We simulated the coupled Monte-Carlo dynamics associated with (\ref{eq:H}) and (\ref{eq:Heff}) for $n=200$, with $\nu=J_g=\frac{1}{2}$, $J_p=1$ (so Figure \ref{fig:phasediag_ninfty_coex} is predicted to apply at least in the limit $n\to\infty$)
 and $k^\star=0.7$, for a system of $N=1000$ monomers at folding temperature $T=0.3$.
 We employed careful on-line tests to ensure equilibration of folding angles before carrying out monomer substitutions (i.e. genetic updates),
 and we allowed for $5.10^4$ iterations per  monomer.
 For these parameter choices our
$n\to\infty$ theory predicts that always $k=-1$, and that there are two critical values for $J_s$: one should find
$m\neq 0$ (a folded state) for $J_s< 0.181$, $m=0$ (a swollen state) for $J_s>0.209$,
with coexistence of the $m=0$ and $m\neq 0$ states for $0.181<J_s<0.209$.  For large but finite $n$ (here: $n=200$) one should expect on the basis
of the earlier data in e.g. Figure \ref{fig:numerics_ninfty} to observe a shift of about 10\% in these critical values relative to those of $n\to\infty$.
 The results of the simulation experiments are shown in figure \ref{fig:simulations}. Each individual
  point in the left figure represents the outcome of
 a simulation where both the fast and the slow process have equilibrated. The figures confirm the existence of a coexistence region, with
 critical $J_s$ values compatible with the predicted 10\% shift relative to those calculated for $n\to\infty$. The associated value of the order parameter $k$ is indeed $k=-1$. The graph showing the evolution in time of the scalar observables illustrates for $J_s/J_p=0.25$ how the $m\neq 0$ destabilizes if $J_s$ has become too large, and supports the claim that in the present parameter regime our
 simulations have equilibrated sufficiently. The remaining fluctuations in $m$ are finite size effects, of the expected order $\Delta m\sim N^{-1/2}$.

\section{Discussion}

In this paper we have studied the coupled stochastic dynamics of primary and secondary structure formation (i.e. slow
genetic sequence selection and fast folding) in the context of a solvable microscopic model that includes both short-range steric
forces and and long-range polarity-driven forces. The rationale behind our approach is that it allows us to circumvent the basic obstacle
in the application of disordered systems techniques to protein folding, which is the need to specify in a mathematical formula
the statistics of the disorder, i.e. the statistics of the amino-acid sequences. Here this is not necessary,
the sequences are themselves allowed to evolve in time, albeit slowly (to model genetic selection) and in a manner that takes account of the
folding properties of the associated chain,
and the statistics of sequences are now an implicit {\em output} of the model rather than an {\em input}.
Our solution is based on exploiting recent mathematical progress \cite{Theo,Theo2} in the diagonalization of replicated
 transfer matrices, and leads in the thermodynamic limit
 to explicit predictions regarding phase transitions and phase diagrams at genetic equilibrium.

In order to apply the methodology of replicated transfer matrices (which require a formulation in the form of a
pseudo-one-dimensional system)
we limited ourselves to effective Hamiltonians of a type
that represents the physical feasibility and energetic gain of three-dimensional folds indirectly, as in e.g. \cite{Skan}.
Even then, in order to keep the remaining mathematics manageable, we chose to limit ourselves further by retaining only polarity forces and steric forces, we reduced the orientation degrees of freedom of individual monomers, and we made the simplest
statistical assumptions regarding polarity and steric properties of amino-acids. However, in contrast to the above limitation
to pseudo-one-dimensional models, these latter restrictions and choices are not strictly required  and can in principle be lifted
if one is willing to accept the inevitable associated quantitative increase in mathematical complexity.
Even in its reduced form, our model and its solution still have a large number of control parameters to be varied, and a full exploration
of its phase phenomenology would have required more than double the present page numbers. Instead we have largely
focused on the regime which we believe to be the most relevant one biologically: the large $n$ regime, where the genetic noise is low.
We have tried to explain the phases observed and their transitions, and understand these qualitatively.

Our model was found to exhibit a parameter regime where protein-like behaviour is observed, i.e. where the genetic selection results in inhomogeneous polarity sequences, and where the folding process describes transitions between swollen and collapsed phases. There was also a parameter regime where the genetic dynamics leads to polymers which are homogeneous in polarity. However, this un-biological
behaviour requires unphysical values of the control parameters. There is a simple argument to see this. The reason for the energetic
advantage of homogeneous polarity sequences  is the mean-field contribution $-(J_p/N)\sum_{ij}\xi(\lambda_i)
\xi(\lambda_j)~\delta_{\phi_i,\phi_j}$ to (\ref{eq:H}), which even for completely random angles $\{\phi_i\}$, where $\bra \delta_{\phi_i,\phi_j}\ket=q^{-1}$, retains on average a value $-J_pN (N^{-1}\sum_{i}\xi(\lambda_i))^2$. In random hereropolymer models
with frozen sequences this term is irrelevant, but here the sequences $\{\lambda_i\}$ evolve, so the system can reduce its energy by increasing $|N^{-1}\sum_i\xi(\lambda_i)|$. A rational alternative definition would be to replace the
mean field term in (\ref{eq:H}) by $-(J_p/N)\sum_{ij}\xi(\lambda_i)
\xi(\lambda_j)[\delta_{\phi_i,\phi_j}-q^{-1}]$, expressing energy gain via folding in terms of {\em correlation} between side-chain orientations and
polarity, rather than {\em covariance}. This would generate a term similar to the polarity balance energy, and
result in the replacement $J_g v(k-k^\star)\to J_g v(k-k^\star)+J_pk^2/q$. For the simple choices $q=2$, $v(x)=\frac{1}{2}x^2$, and $k^\star=0$, in particular, the change would translate into the simple parameter re-scaling $J_g\to J_g+J_p$. The natural parameter regime is apparently $J_g>J_p$, the one with inhomogeneous polarities.

There is certainly significant scope for improvement and expansion of this study.
All our simplifying choices, made for the sake of mathematical convenience,
should however be judged in the light of the complexity of
the resulting equations even for the presently studied simplified model.
The obvious directions to move into next are clear. First there is
the search for more realistic Hamiltonians describing
the fast process, by improving the energetic description of the effects of 3D folding (possibly via a formulation involving contact maps, which would replace the long range all-to-all forces by a sparse connectivity version), and by including hydrogen bonds.
Second, we would like to work out our formulae for the case where the monomers' mechanical degrees of freedom consist of two angles, that furthermore can each take more than just two values (preferably a continuum, which would replace the replicated transfer matrices by replicated kernels).
Thirdly, one would like to
find more realistic alternatives for the sequence selection Hamiltonian, that is more precise in terms of quantifying
a sequence's biological functionality, and
that employ a better proxy for the unique foldability of a sequence than just its folding free energy.

We see this paper as a proof of principle, demonstrating that it is in principle possible to construct solvable microscopic models of
primary and secondary structure formation in heteropolymers, with both long- and short-range forces, in which
there is no need to assume (and average over) random amino-acid sequences or to find a formula for
suitably non-random sequence statistics. This study represents a small step, but we believe it to be a step
in a promising  direction.

\section*{Acknowledgements}

It is our pleasure to thank Isaac Perez-Castillo, Nikos Skantzos and Jort van Mourik
for valuable discussions. One of the the authors (CJPV) acknowledges financial
support from project FIS2006-13321-C02-01 and grant PR2006-0458.

\clearpage
\appendix
\section{Identification of observables}
\label{app:meaning}

\subsection{`Slow' free energy as generator of observables}

\noindent In the stationary state, where both the fast degrees of
freedom ($\bphi$, giving the secondary structure) and the slow
degrees of freedom ($\blambda$, giving the primary structure) have
equilibrated, expectation values of observables are given by two
nested Boltzmann averages. Using definition (\ref{eq:Heff}) and
$\tilde{\beta}=n\beta$ the result can be written as
\begin{eqnarray}
\hspace*{-10mm}
\bra\bra G(\bphi,\blambda)\ket_{\rm fast}\ket_{\rm slow}&=&
\frac{\sum_{\blambda}e^{-\tilde{\beta}H_{\rm eff}(\blambda)} \bra
G(\bphi,\blambda)\ket_{\rm
fast}}{\sum_{\blambda}e^{-\tilde{\beta}H_{\rm eff}(\blambda)}}
\nonumber
\\
\hspace*{-10mm}
&=& e^{\tilde{\beta}Nf_N}
\sum_{\blambda}e^{-\tilde{\beta}H_{\rm eff}(\blambda)}\left\{
\frac{\sum_{\bphi}e^{-\beta H_{\rm f}(\bphi|\blambda)}
G(\bphi,\blambda)}{\sum_{\bphi}e^{-\beta H_{\rm
f}(\bphi|\blambda)}}\right\} \nonumber
\\
\hspace*{-10mm}
&=& e^{\tilde{\beta}Nf_N}
\sum_{\blambda}\frac{e^{-\tilde{\beta}[U(\blambda)+V(\blambda)]}}{\Z^{1-n}_{\rm
f}(\blambda)}\sum_{\bphi}G(\bphi,\blambda)e^{-\beta H_{\rm
f}(\bphi|\blambda)} \nonumber
\\
\hspace*{-10mm}
&=& e^{\tilde{\beta}N \!f_N}
\!\sum_{\blambda}\!\sum_{\bphi^1\!\ldots\!
\bphi^n}\!\!G(\bphi^1\!,\blambda)
 e^{-\beta \sum_{\alpha} \left[H_{\rm
f}(\bphi^\alpha\!|\blambda)+U(\blambda)+V\!(\blambda)\right]}
\nonumber \\[-2mm]&& \label{eq:expectation_values}
\end{eqnarray}
with $\alpha=1\ldots n$. This latter expression is also obtained
as the derivative of the `slow' free energy $f_N$, provided we add
a suitable generating term to the `fast' Hamiltonian $H_{\rm
f}(\bphi|\blambda)$. To be precise, upon replacing
\begin{equation}
H_{\rm f}(\bphi|\blambda)\to H_{\rm f}(\bphi|\blambda)+\chi N
G(\bphi,\blambda) \label{eq:extraterm}\end{equation} one obtains
\begin{equation}
\bra\bra G(\bphi,\blambda)\ket_{\rm fast}\ket_{\rm
slow}=\lim_{\chi\to 0}\frac{\partial}{\partial \chi} f_N
\label{eq:generator}\end{equation} The validity of
(\ref{eq:generator}), which allows us to use the free energy as a
generating function for expectation values, follows immediately
upon substituting (\ref{eq:extraterm}) into (\ref{eq:overallf}):
\begin{eqnarray}
\hspace*{-25mm}
\lim_{\chi\to 0}\frac{\partial}{\partial\chi}f_N&=&
 -\lim_{\chi\to 0}\frac{1}{nN\beta }\frac{\partial}{\partial\chi}\log \sum_{\blambda}\sum_{\bphi^1\ldots
\bphi^n}
 e^{-\beta \sum_{\alpha=1}^n \left[\chi NG(\bphi^\alpha\!,\blambda)+H_{\rm
f}(\bphi^\alpha|\blambda)+U(\blambda)+V(\blambda)\right]}
\nonumber
\\
\hspace*{-25mm}
&=&  \lim_{\chi\to 0}\frac{1}{n}\sum_{\gamma=1}^n\frac{
\sum_{\blambda}\sum_{\bphi^1\ldots \bphi^n}
G(\bphi^\gamma\!,\blambda)
 e^{ -\beta \sum_{\alpha=1}^n \left[\chi G(\bphi^\alpha\!,\blambda)+ H_{\rm
f}(\bphi^\alpha|\blambda)+U(\blambda)+V(\blambda)\right]} } {
\sum_{\blambda}\sum_{\bphi^1\ldots \bphi^n}
 e^{-\beta \sum_{\alpha=1}^n \left[\chi G(\bphi^\alpha\!,\blambda)+  H_{\rm
f}(\bphi^\alpha|\blambda)+U(\blambda+V(\blambda))\right]}}\nonumber
\hspace*{-10mm}
\\
\hspace*{-25mm}
&=&  \frac{ \sum_{\blambda}\sum_{\bphi^1\ldots \bphi^n}
G(\bphi^1\!,\blambda)
 e^{ -\beta \sum_{\alpha=1}^n \left[H_{\rm
f}(\bphi^\alpha|\blambda)+U(\blambda)+V(\blambda)\right]} } {
\sum_{\blambda}\sum_{\bphi^1\!\ldots\! \bphi^n}
 e^{-\beta \sum_{\alpha=1}^n \left[ H_{\rm
f}(\bphi^\alpha|\blambda)+U(\blambda)+V(\blambda)\right]}}
\nonumber
\\
\hspace*{-25mm}
&=& e^{\tilde{\beta}Nf_N} \!\sum_{\blambda}\sum_{\bphi^1\!\ldots\!
\bphi^n}\!G(\bphi^1\!,\blambda) e^{-\beta \sum_{\alpha=1}^n
\left[H_{\rm
f}(\bphi^\alpha|\blambda)+U(\blambda)+V(\blambda)\right]}\nonumber
\\
\hspace*{-25mm}
&=& \bra\bra
G(\bphi,\blambda)\ket_{\rm fast}\ket_{\rm slow}
\end{eqnarray}

\subsection{Identification of order parameters for $q=2$}

\noindent We next apply the general relations
(\ref{eq:extraterm},\ref{eq:generator}) for $q=2$ to
 observables of the form
$G(\bsigma,\blambda)=N^{-1}\sum_i g(\sigma_i,\xi_i,\eta_i)$. Here
equations (\ref{eq:extraterm},\ref{eq:generator}) translate into
\begin{eqnarray}
&& H_{\rm f}(\bsigma|\blambda)\to H_{\rm f}(\bsigma|\blambda)+\chi
\sum_i g(\sigma_i,\xi_i,\eta_i) \label{eq:generate1}
\\
&&\frac{1}{N}\sum_i\bra\bra g(\sigma_i,\xi_i,\eta_i)\ket_{\rm
fast}\ket_{\rm slow}=\lim_{\chi\to 0}\frac{\partial}{\partial
\chi}f_N~~~~~ \label{eq:generate2}
\end{eqnarray}
We repeat our previous derivation of the free energy per
amino-acid  (\ref{eq:q2phi_RS}) but now with the new contribution
$\chi \sum_i g(\sigma_i,\xi_i,\eta_i)$ included in the fast
Hamiltonian $H_{\rm f}(\bsigma)$, in leading order in $\chi$. The
new term changes (\ref{eq:q2M}) into
\begin{eqnarray}
\hspace*{-10mm}
 M[\bsigma_{i-1},\bsigma_i,\bsigma_{i+1}|\mb,\bk]&=&\big\bra\big\bra
 e^{\beta\xi[ J_p
\sum_\alpha(k_\alpha+m_\alpha\sigma_i^\alpha)-n\mu]} \nonumber
\\
\hspace*{-10mm}
&& \hspace*{-35mm}
\times e^{-\beta \xi n J_g v^\prime(\frac{1}{n}\sum_\alpha
k_\alpha-k^\star)+
\beta\eta[J_s\bsigma_{i+1}\cdot\bsigma_{i-1}-n\nu]-\beta \chi \sum_\alpha g(\sigma_i^\alpha,\xi,\eta)}
\big\ket\big\ket_{\xi,\eta}~~~~~
\end{eqnarray}
 From this we can immediately recover the
identifications (\ref{eq:meaning_m},\ref{eq:meaning_k}). For
instance, choosing $g(\sigma,\xi,\eta)=\sigma \xi$ gives
\begin{eqnarray}
\hspace*{-10mm}
 M[\bsigma_{i-1},\bsigma_i,\bsigma_{i+1}|\mb,\bk]&=&
M[\bsigma_{i-1},\bsigma_i,\bsigma_{i+1}|\mb-\frac{\chi}{
J_p}(1,\ldots,1),\bk]~~~~
\end{eqnarray}
From this we extract, due to
$M[\bsigma_{i-1},\bsigma_{i},\bsigma_{i+1}]$ only affecting the
transfer matrix eigenvalue $\lambda(\bm,\bk)$, within the replica
symmetric ansatz:
\begin{eqnarray}
\hspace*{-10mm} \lim_{N\to\infty}\frac{1}{N}\sum_i \bra\bra
\xi_i\sigma_i\ket_{\rm fast}\ket_{\rm slow}&=& \lim_{\chi\to
0}\frac{\partial}{\partial \chi}f_N= -\frac{1}{\beta n}\lim_{\chi\to
0}\frac{\partial}{\partial \chi}\log\lambda_{\rm max}^{\rm
RS}(m\!-\!\frac{\chi}{J_p},k) \nonumber
\\
\hspace*{-10mm}
&&\hspace*{-30mm}= \frac{1}{\beta n J_p}\frac{\partial}{\partial
m}\log\lambda_{\rm max}^{\rm RS}(m,k)|_{\chi=0}~=~m
\end{eqnarray}
according to (\ref{eq:eqn_for_m}). Similarly, making the
alternative choice $g(\sigma,\xi,\eta)=\xi$ gives in leading order
in $\chi$:
\begin{eqnarray}
 M[\bsigma_{i-1},\bsigma_i,\bsigma_{i+1}|\mb,\bk]&=&
 \\
&&\hspace*{-20mm}
M[\bsigma_{i-1},\bsigma_i,\bsigma_{i+1}|\mb,\bk\!-\!\frac{\chi}{
J_p}(1,\ldots,1)]_{v^\prime(.)\to v^\prime(.)-\chi
v^{\prime\prime\!}(.)/J_p} \nonumber
\end{eqnarray}
From this we extract, within the replica symmetric ansatz:
\begin{eqnarray} \lim_{N\to\infty}\frac{1}{N}\sum_i \bra\bra
\xi_i\ket_{\rm fast}\ket_{\rm slow}&=& \lim_{\chi\to
0}\frac{\partial}{\partial \chi}f_N\nonumber
\\
&&\hspace*{-30mm}= -\frac{1}{\beta n}\lim_{\chi\to
0}\frac{\partial}{\partial \chi}\log\lambda_{\rm max}^{\rm
RS}(m,k\!-\!\frac{\chi}{J_p})|_{v^\prime(.)\to v^\prime(.)-\chi
v^{\prime\prime\!}(.)/J_p} \nonumber
\\
&&\hspace*{-30mm}= -\frac{1}{\beta n}\lim_{\chi\to
0}\frac{\partial}{\partial \chi}\log\lambda_{\rm max}^{\rm
RS}(m,k\!-\!\frac{\chi}{J_p}\!+\!\frac{\chi}{J_p}[\frac{J_g}{J_p}v^{\prime\prime\!}(k\!-\!k^\star)])\nonumber
\\
&&\hspace*{-30mm}= \frac{1}{\beta n J_p}\Big[
1\!-\!\frac{J_g}{J_p}v^{\prime\prime\!}(k\!-\!k^\star)\Big]
\frac{\partial}{\partial k}\log\lambda_{\rm max}^{\rm
RS}(m,k)|_{\chi=0} = ~k
\end{eqnarray}
according to (\ref{eq:eqn_for_k}). The above identification of the
scalar order parameters $m$ and $k$ was relatively easy since we
could absorb the extra generating terms into those already
present. This will generally not be the case.

\subsection{Joint distribution of primary structure variables}

\noindent We next turn to the calculation of the equilibrium
amino-acid statistics as measured by
$\pi(\hat{\xi},\hat{\eta})=\lim_{N\to\infty}N^{-1}\sum_i\bra\bra
\delta(\hat{\xi}-\xi_i)\delta(\hat{\eta}-\eta_i)\ket_{\rm
fast}\ket_{\rm slow}$. This distribution follows from
(\ref{eq:generate1},\ref{eq:generate2}) upon making the choice
$g(\sigma,\xi,\eta)=\delta(\xi-\hat{\xi})\delta(\eta-\hat{\eta})$:
\begin{eqnarray}
&& H_{\rm f}(\bsigma|\blambda)\to H_{\rm f}(\bsigma|\blambda)+\chi
\sum_i\delta(\xi_i-\hat{\xi})\delta(\eta_i-\hat{\eta})
\\
&&\pi(\hat{\xi},\hat{\eta})=\lim_{N\to\infty}\lim_{\chi\to
0}\frac{\partial}{\partial \chi}f_N~~~~~
\end{eqnarray}
The calculation is now complicated by the fact that
the convenient decomposition identity (\ref{eq:decomposition}) no
longer holds. Instead we now find, in replica-symmetric ansatz:
\begin{eqnarray}
\hspace*{-20mm}
 M[\bsigma_{i-1},\bsigma_i,\bsigma_{i+1}|\mb,\bk]&=&
 M[\bsigma_{i-1},\bsigma_i,\bsigma_{i+1}|\mb,\bk]|_{\chi=0}
 \nonumber
 \\[1mm]
 \hspace*{-20mm}
 && \hspace*{-45mm} -n\beta \chi \big\bra \delta(\xi\!-\!\hat{\xi})e^{\beta
 \xi[J_p\sum_\alpha(k+m\sigma_i^\alpha)-n\mu- n J_g v^\prime(k-k^\star)]}\big\ket_{\xi} \big\bra \delta(\eta\!-\!\hat{\eta})
e^{\beta
\eta[J_s\bsigma_{i+1}\cdot\bsigma_{i-1}-n\nu]}\big\ket_{\eta}\nonumber
\\
\hspace*{-20mm} && \hspace*{50mm}+\order(\chi^2)
\end{eqnarray}
so that
\begin{eqnarray}
\hspace*{-20mm}
\prod_i M[\bsigma_{i-1},\bsigma_i,\bsigma_{i+1}|\mb,\bk]&=&
\prod_i
\Gamma_{\bsigma_{i-1},\bsigma_{i+1}}(\mb,\bk)~+\order(\chi^2)
 \nonumber
 \\
 \hspace*{-20mm}
 &&\hspace*{-50mm} -n\beta \chi \sum_j \Big\{\big\bra \delta(\xi\!-\!\hat{\xi})e^{\beta
 \xi[J_p\sum_\alpha(k+m\sigma_i^\alpha)-n\mu- n J_g v^{\prime\!}(k-k^\star)]}\big\ket_{\xi} \big\bra \delta(\eta\!-\!\hat{\eta}) e^{\beta
\eta[J_s\bsigma_{j+1}\cdot\bsigma_{j-1}-n\nu]}\big\ket_{\eta}
\nonumber
\\
\hspace*{-20mm}
&&\hspace*{20mm}\times\prod_{i\neq
j} M[\bsigma_{i-1},\bsigma_i,\bsigma_{i+1}|\mb,\bk]\Big\}
 \nonumber
\\
\hspace*{-20mm}
&&\hspace*{-45mm}=\Big[\prod_i
\Gamma_{\bsigma_{i-1},\bsigma_{i+1}}(\mb,\bk)\Big]\Big[1\!-\!n\beta
\chi \sum_j
V_{\bsigma_j}(m,k)W_{\bsigma_{j-1}\bsigma_{j+1}}(m,k)\!+\!\order(\chi^2)\Big]
\end{eqnarray}
with
\begin{eqnarray}&&\hspace*{-5mm}
V_{\bsigma}(m,\!k)=\frac{ \big\bra
\delta(\xi\!-\!\hat{\xi})e^{n\beta
 \xi[\frac{J_p}{n}\sum_\alpha(k+m\sigma_i^\alpha)-\mu- J_g v^\prime(k-k^\star)]}\big\ket_{\xi}} { \big\bra e^{n\beta
 \xi[\frac{J_p}{n}\sum_\alpha(k+m\sigma_i^\alpha)-\mu- J_g v^\prime(k-k^\star)]}\big\ket_{\xi}}~~~~~~
\\
&& \hspace*{-5mm}  W_{\bsigma\bsigma^\prime}(m,\!k)=\frac{
\big\bra \delta(\eta\!-\!\hat{\eta}) e^{n\beta
\eta[\frac{J_s}{n}\bsigma\cdot\bsigma^\prime-\nu]}\big\ket_{\eta}}
{ \big\bra e^{n\beta
\eta[\frac{J_s}{n}\bsigma\cdot\bsigma^\prime-\nu]}\big\ket_{\eta}}
\end{eqnarray}
This leads us to
\begin{eqnarray}
&&\hspace*{-20mm} \lim_{N\to\infty}\frac{1}{N}\log
\sum_{\bsigma_1\ldots \bsigma_N} \prod _i
 M[\ldots|\ldots]~=~ \log \lambda^{\rm RS}_{\rm
 max}(m,k)|_{\chi=0}
 \\
 &&\hspace*{-20mm}
-n\beta \chi \lim_{N\to\infty}\frac{1}{N}\sum_j
\frac{\sum_{\bsigma_1\ldots \bsigma_N} \big[\prod _i
 \Gamma_{\bsigma_{i-1}\bsigma_{i+1}}(\mb,\bk)\big]
 V_{\bsigma_j}(m,k)W_{\bsigma_{j-1}\bsigma_{j+1}}(m,k)} {\sum_{\bsigma_1\ldots \bsigma_N} \big[\prod _i
 \Gamma_{\bsigma_{i-1}\bsigma_{i+1}}(\mb,\bk)\big]}
 \nonumber
 \\
 &&\hspace*{80mm}
+\order(\chi^2)
 \end{eqnarray}
 and hence, with $\bGamma(m,k)$ denoting the replica-symmetric
 version (\ref{eq:GammaRS}) of the transfer matrix $\bGamma(\mb,\bk)$, with $\lambda^{\rm RS}_{\rm max}(m,k)$ denoting the largest eigenvalue of
 $\bGamma(m,k)$, and using the periodicity of the
 chain:
  \begin{eqnarray}
  \hspace*{-20mm}
\lim_{N\to\infty}  f_N&=& {\rm extr}_{m,k}\Big\{
\frac{1}{2}J_p(m^2\!+k^2) + J_g
\big[v(k\!-\!k^\star)\!-\!kv^\prime(k\!-\!k^\star)\big]
\nonumber
\\
\hspace*{-20mm}
&& +\chi\lim_{N\to\infty} \frac{\sum_{\bsigma_1\ldots \bsigma_N}
\big[\prod _i
 \Gamma_{\bsigma_{i-1}\bsigma_{i+1}}(\mb,\bk)\big]
 V_{\bsigma_1}(m,k)W_{\bsigma_{N}\bsigma_{2}}(m,k)} {\big({\rm
 Tr}[\bGamma^{N/2}(m,k)]\big)^2}
 \nonumber
 \\
 \hspace*{-20mm}&&\hspace*{30mm}
-\frac{1}{\beta n}\log\Lambda
-\frac{1}{\beta n} \log \lambda^{\rm
RS}_{\rm max}(m,k)+\order(\chi^2)\Big\}
\\
\hspace*{-20mm}
\pi(\hat{\xi},\hat{\eta})&=& \left[\lim_{N\to\infty}
\frac{\sum_{\bsigma_1\bsigma_3} \bGamma^{N/2-1}_{\bsigma_3
\bsigma_1}(m,k)\Gamma_{\bsigma_{1}\bsigma_{3}}(m,k)
 V_{\bsigma_1}(m,k)}  {{\rm
 Tr}[\bGamma^{N/2}(m,k)]}\right]
\nonumber
\\
\hspace*{-20mm}
&& \times
 \left[\lim_{N\to\infty}
 \frac{\sum_{\bsigma_2\bsigma_N} \bGamma^{N/2-1}_{\bsigma_{2}\bsigma_{N}}(m,k)
\Gamma_{\bsigma_{N}\bsigma_{2}}(m,k)
W_{\bsigma_{N}\bsigma_{2}}(m,k)}  {{\rm
 Tr}[\bGamma^{N/2}(m,k)]}\right]
\end{eqnarray}
  In the latter expression one must substitute for $(m,k)$ the
solution of the original $\chi=0$ saddle-point problem. We find
once more a convenient effective decoupling of the odd sites from
the even sites, as well as statistical independence of the
single-site polarity and steric angle statistics, giving
$\pi(\hat{\xi},\hat{\eta})= \pi(\hat{\xi}) \pi(\hat{\eta})$ with
the individual distributions
\begin{eqnarray}
\hspace*{-25mm}
\pi(\hat{\xi})&=&
\\
\hspace*{-25mm}&&\hspace*{-15mm} \lim_{N\to\infty}
\frac{\sum_{\bsigma\bsigma^\prime} \bGamma^{N/2-1}_{\bsigma^\prime
\bsigma}(m,k) \big\bra \delta(\xi\!-\!\hat{\xi})e^{\beta
 \xi[J_p(nk+m\sum_\alpha\sigma_i^\alpha)-n\mu- n J_g v^{\prime\!}(k-k^\star)]}\big\ket_{\xi}
  \big\bra
e^{\beta\eta[J_s\bsigma\cdot\bsigma^\prime-n\nu]} \big\ket_{\eta}
 }  {{\rm
 Tr}[\bGamma^{N/2}(m,k)]}
 \nonumber
\\
\hspace*{-25mm}
\pi(\hat{\eta})&=&\\
\hspace*{-25mm}&& \hspace*{-15mm}
 \lim_{N\to\infty}
 \frac{\sum_{\bsigma\bsigma^\prime} \bGamma^{N/2-1}_{\bsigma^\prime\bsigma}(m,k)
\big\bra e^{\beta
 \xi[J_p(nk+m\sum_\alpha\sigma_i^\alpha)-n\mu- n J_g v^{\prime\!}(k-k^\star)]}\big\ket_{\xi}
 \big\bra \delta(\eta\!-\!\hat{\eta}) e^{\beta
\eta[J_s\bsigma\cdot\bsigma^\prime-n\nu]}\big\ket_{\eta}}  {{\rm
 Tr}[\bGamma^{N/2}(m,k)]}
  \nonumber
\end{eqnarray}
 The limit $N\to\infty$ can now be taken upon using
the fact that for $N\to\infty$ one may write in leading order
$\bGamma^{N}_{\bsigma\bsigma^\prime}(m,k)\to
\lambda^{N}(m,k)u_{\bsigma}^{\rm R}u_{\bsigma^\prime}^{\rm
L}/\sum_{\bsigma^\pprime}u_{\bsigma^\pprime}^{\rm
L}u_{\bsigma^\pprime}^{\rm R}$, where $\{u^{\rm L}_{\bsigma}\}$
and  $\{u^{\rm R}_{\bsigma}\}$ denote the left- and
right-eigenvectors of $\bGamma(m,k)$ associated with the largest
eigenvalue. In the result we can then substitute our expression
(\ref{eq:lambdaPsi}) for the largest eigenvalue and the
replica-symmetric forms (\ref{eq:right_ev},\ref{eq:left_ev}) for
the eigenvectors. For the polarity distribution $\pi(\hat{\xi})$
this gives, after some further manipulations and with help of the
definitions (\ref{eq:p_xi},\ref{eq:effective_fields}):
\begin{eqnarray}
\hspace*{-20mm}
\pi(\hat{\xi})&=&  \frac{\sum_{\bsigma\bsigma^\prime}
u_{\bsigma}^{\rm L}  u_{\bsigma^\prime}^{\rm R}\big\bra
\delta(\xi\!-\!\hat{\xi})e^{\beta
 \xi[J_p(nk+m\sum_\alpha\sigma_i^\alpha)-n\mu- n J_g v^{\prime\!}(k-k^\star)]}\big\ket_{\xi}
  \big\bra
e^{\beta\eta[J_s\bsigma\cdot\bsigma^\prime-n\nu]} \big\ket_{\eta}
 }  {\lambda(m,k) \sum_{\bsigma}u_{\bsigma}^{\rm
L}u_{\bsigma}^{\rm R}} \nonumber
\\
\hspace*{-20mm}
&=&\frac{p(\hat{\xi})\int\!dh~\cosh^n(\beta h)\int\!dx~ \Psi(x)
\Psi(h-x-\hat{\xi} J_pm) \ket_{\xi}
 }  {\int\!d\xi~p(\xi)\int\!dh~\cosh^n(\beta h)
 \int\!dx~
\Psi(h-x-\xi J_pm)\Psi(x) }\nonumber \\
\hspace*{-20mm}&=&
\int\!dh~W(\hat{\xi},h)
\end{eqnarray}
For the steric angle distribution $\pi(\hat{\eta})$ one finds an
expression with a similar structure:
{\small
\begin{eqnarray}
\hspace*{-20mm}
\pi(\hat{\eta})&=&
 \frac{\sum_{\bsigma\bsigma^\prime}u_{\bsigma}^{\rm
L} u_{\bsigma^\prime}^{\rm R} \big\bra e^{\beta \xi[ J_p (
nk+m\sum_\alpha\!\sigma_\alpha)-n\mu- n J_g
v^{\prime\!}(k-k^\star)]}\big\ket_{\xi}
 \big\bra \delta(\eta\!-\!\hat{\eta}) e^{\beta
\eta[J_s\bsigma\cdot\bsigma^\prime-n\nu]}\big\ket_{\eta}}
{\lambda(m,k)\sum_{\bsigma}u_{\bsigma}^{\rm L}u_{\bsigma}^{\rm R}}
\nonumber
\\
\hspace*{-20mm}
&&\hspace*{-15mm} =\frac{ \int\!dx dx^\prime \Phi(x^\prime)\Psi(x)
\Big\bra\!\Big\bra
 \delta(\eta\!-\!\hat{\eta}) e^{n\beta[B(x^\prime\!,\eta J_s)+\xi(J_p
k-\mu-J_g v^{\prime\!}(k-k^\star))-\eta\nu]}\cosh^n[\beta (x\!+\!\xi
J_p m\!+\!A(x^\prime\!,\eta J_s))]\Big\ket\!\Big\ket_{\xi,\eta}}
{\int\!dx~\Phi(x)\Big\bra\!\!\Big\bra
 e^{n\beta [B(x,\eta
J_s)+\xi( J_p k  -\mu-J_g v^{\prime\!}(k-k^\star))- \nu\eta]}
\Big\ket\!\!\Big\ket_{\!\xi,\eta} \int\!dxdx^\prime
\Phi(x^\prime)\Psi(x)\cosh^n[\beta (x+ x^\prime)] } \nonumber
\\
\hspace*{-20mm}
&=&
 \frac{\int\!dh~\cosh^n(\beta h) \int\!dx~\Phi(x)\Phi(h-A(x,\hat{\eta} J_s))\bra
 \delta(\eta\!-\!\hat{\eta}) e^{n\beta[B(x,\eta J_s)-\eta\nu]}\ket_{\eta}}
{\int\!dh~\cosh^n(\beta h) \int\!dx~ \Phi(x)\Phi(h-A(x,\hat{\eta}
J_s))\bra
  e^{n\beta[B(x,\eta J_s)-\eta\nu]}\ket_{\eta} }
\end{eqnarray}}
Both in the limit $n\to 0$ (fully random sequence selection) and
in the limit $\beta\to 0$ one sees both equilibrated distributions
reducing to the prior statistics $w(\hat{\xi})$ and
$w(\hat{\eta})$, as it should. In general, however, one will find
non-trivial distributions $\pi(\hat{\xi})$ and $\pi(\hat{\eta})$,
which reflect the complicated interplay between secondary and
primary structure generation. Finally we observe that
$\pi(\xi)\neq p(\xi)$, except when $J_pm=0$; this suggests that,
rather than the polarity distribution in the equilibrated system,
the physical interpretation of $p(\xi)$ is that of a prior
distribution which would have been found in the absence of
secondary structure formation.

\section{Saddle-point treatment of order parameter equations in the limit $n\to\infty$}
\label{app:saddlepoint_deterministic}

\subsection{Saddle-point treatment of the equation for $\Psi(x)$}

\noindent Since  we know that $\Psi(x)= 0$
 for $|x|>J_s$, we may write without loss of generality $\Psi(x)=e^{n\beta \psi(x)}$
for $x\in\Omega\subseteq[-J_s,J_s]$ and $\Psi(x)=0$ for
$x\notin\Omega$,
 where $\int_{\Omega}\!dx~e^{n\beta \psi(x)}=1$.
We also define for $|x|\leq y$ and $y>0$
 the function
 \begin{eqnarray}
C(x,y)&=&\frac{1}{\beta}\tanh^{-1}[\tanh(\beta x)/\tanh(\beta
y)]~~~~
\end{eqnarray}
 It is the inverse of the function $A(x,y)$ with respect to
the variable $x$, since $C(A(x,y),y)=x$ for all $|x|<|y|$. We note
that $C(0,y)=0$ and $\sgn[C(x,y)]=\sgn(x)$. We can now  insert our
expression for $p(\xi)$ and the definition
$\Psi(x)=e^{n\beta\psi(x)}$ (for $x\in\Omega$, with $\Psi(x)=0$
elsewhere) into our equation for $\Psi(x)$, and use the function
$C(x,y)$ to subsequently transform variables inside the
$\delta$-distribution in the right-hand side. Since the Jacobian
of this transformation will not be exponential in $n$ as
$n\to\infty$, as a result of these manipulations we find for all
$x\in\Omega$ an equation for $\psi(x)$ that is for $n\to\infty$
evaluated by steepest descent:
\begin{eqnarray}
\hspace*{-25mm}
\lim_{n\to\infty} \psi(x) &=& \lim_{n\to\infty}\frac{1}{\beta
n}\log \left\{\bigroom
\right.
\\
\hspace*{-25mm}&&\hspace*{-22mm}
\left.\frac{\int_{\Omega}\! dy\int\! d\xi d\eta~
w(\eta)w(\xi) \delta\big[C(x,\eta J_s)\!-\! y\!-\!J_pm\xi\big]
 e^{n\beta[\psi(y)+ \xi(J_p-J_g)(k-k_0)+B(y+J_pm\xi,\eta
J_s)- \nu\eta]}} { \int_{\Omega}\!dy\int\! d\xi d\eta ~w(\eta)
w(\xi)
 e^{n\beta[\psi(y)+ \xi(J_p-J_g)(k-k_0)+B(y+J_pm\xi,\eta
J_s)- \nu\eta]}} \right\}\hspace*{-10mm}
 \nonumber
 \\[1mm]
 \hspace*{-25mm}
 &&\hspace*{-16mm}= \max_{y\in\Omega,~y=C(x,\eta J_s)-J_pm\xi,~|\xi|\leq 1,~|\eta|\leq 1}\Big\{
 \psi(y)\!+\! \xi(J_p\!-\!J_g)(k\!-\!k_0)\!+\!B(y\!+\!J_pm\xi,\eta J_s)\!-\! \nu\eta\Big\}\nonumber
 \\
 \hspace*{-25mm}
 &&
 \hspace*{0mm}
 -\max_{y\in\Omega,~|\xi|\leq 1,~|\eta|\leq 1}\Big\{
 \psi(y)\!+\! \xi(J_p\!-\!J_g)(k\!-\!k_0)\!+\!B(y\!+\!J_pm\xi,\eta J_s)\!-\! \nu\eta\Big\} ~~~~ \label{eq:ninftyproblem}
\end{eqnarray}
 Solving the optimization problem
(\ref{eq:ninftyproblem}) means calculating both the set
$\Omega\subseteq[-J_s,J_s]$ and the function $\lim_{n\to\infty}
\psi(x)$ for $x\in\Omega$. Let us inspect some properties of this
optimization problem in more detail. Since the maximization in the
first line of (\ref{eq:ninftyproblem}) is over a {\em subset} of
the set in the second line (instead of allowing for all
$y\in\Omega$, in the first line we impose $y=C(x,\eta
J_s)-J_pm\xi$), it is inevitable that $\lim_{n\to\infty}
\psi(x)\leq 0$ for all $x\in\Omega$. We now know that $\psi_{\rm
max}=\lim_{n\to\infty}\max_{x\in\Omega}\psi(x)\leq 0$. This leaves
two options: $\psi_{\rm max}<0$ versus $\psi_{\rm max}=0$. In the
first case, however, we would get $\lim_{n\to\infty}
\Psi(x)=\lim_{n\to\infty}e^{n\beta\psi(x)}\leq
\lim_{n\to\infty}e^{n\beta\psi_{\rm max}}=0$ for all $x\in\Omega$;
this function can never be normalized. We conclude that $\psi_{\rm
max}=0$.

Let us turn to those values of $x$ for which one has
$\lim_{n\to\infty} \psi(x)=\psi_{\rm max}=0$. We call the set of
those values $\Omega^\star\subseteq \Omega$:
\begin{eqnarray}
\hspace*{-25mm}&&
x\in\Omega^\star:~ \max_{y \in\Omega,~|\xi|,|\eta|\leq
1,~x=A(y+J_pm\xi,\eta J_s)}\Big\{
 \psi(y)\!+\! \xi(J_p\!-\!J_g)(k\!-\!k_0)\!+\!B(y\!+\!J_pm\xi,\eta J_s)\!-\! \nu\eta\Big\}
 \hspace*{-10mm}\nonumber
 \\
 \hspace*{-25mm}
 &&
 \hspace*{30mm}=
 \max_{y\in\Omega,~|\xi|,|\eta|\leq 1}\Big\{
 \psi(y)\!+\! \xi(J_p\!-\!J_g)(k\!-\!k_0)\!+\!B(y\!+\!J_pm\xi,\eta J_s)\!-\! \nu\eta\Big\}
 \nonumber
 \\
 \hspace*{-25mm}&&
 \label{eq:Omegastar}
 \end{eqnarray}
 We see that with every combination $(y,\xi,\eta)$ that gives the
 maximum value in the second line there corresponds a value of
 $x\in\Omega^\star$. If the maximum is obtained for a {\em unique}
 combination $(y^\star,\xi^\star,\eta^\star)$, which apart from symmetries one must expect to be the generic case,
  then the set $\Omega^\star$ contains
 just one element $x^\star=A(y^\star+J_pm\xi^\star,\eta^\star
 J_s)$. It follows that one must generally anticipate
 $\lim_{n\to\infty}\Psi(x)$ to be a sum of a small number of
 $\delta$-peaks.

We can finally also use saddle-point arguments to express the
limit $n\to\infty$ of the free energy per monomer
(\ref{eq:final_fRS}) in terms of the function $\psi(x)$, the
scalar order parameters $(k,m)$, and the set $\Omega$:
\begin{eqnarray}
\hspace*{-10mm}
\lim_{n\to\infty}\varphi \!&=&\!  \frac{1}{2}J_p(m^2\!\!+\!k^2)-
\frac{1}{2}J_g(k^2\!\!-\!k^{\star 2})- |J_p\!-\!J_g||k\!-\!k_0|
 \nonumber
\\
\hspace*{-10mm}
&&\hspace*{-15mm} - \max_{x\in\Omega,~\xi,\eta\in[-1,1]}
 \Big\{\psi(x)+\xi(J_p\!-\!J_g)(k\!-\!k_0)
  +
 B(x\!+\!J_pm\xi,\eta J_s)- \nu\eta
 \Big\}
 \label{eq:free_energy_ninfty}
\end{eqnarray}

In the remainder of this section we will not attempt to solve the
problem (\ref{eq:ninftyproblem}) in its full generality, but
rather construct two qualitatively different specific  solutions
of (\ref{eq:ninftyproblem}), for which indeed $\Psi(x)$ is found
to reduce to either one or two $\delta$-peaks, and which both
reduce exactly to the unique solutions that we established earlier
in the two limits $J_s\to 0$ or $J_pm\to 0$.

\subsection{Homogeneous polarity states $k=\pm 1$}

\noindent Here we construct solutions of (\ref{eq:ninftyproblem})
where $\Omega=\{x^\star\}$, and show that these represent the
continuation to arbitrary $J_s>0$ and $J_pm\neq 0$ of the
homogeneous polarity states $k=\pm 1$.  Now we must have
$\Omega^\star=\Omega$ and $\lim_{n\to\infty}
\psi(x^\star)=\psi_{\rm max}=0$, and (\ref{eq:Omegastar}) becomes
\begin{eqnarray}
&&
\max_{\xi,\eta\in[-1,1],~x^\star=A(x^\star+J_pm\xi,\eta
J_s)}L(\xi,\eta)
 =
 \max_{\xi,\eta\in[-1,1]} L(\xi,\eta)
 \label{eq:Omega_Onepeak}
 \\
 && L(\xi,\eta)=\xi(J_p\!-\!J_g)(k\!-\!k_0)+B(x^\star\!+\!J_pm\xi,\eta J_s)- \nu\eta
 \end{eqnarray}
In both sides of (\ref{eq:Omega_Onepeak}) we maximize exactly the
same object, but in the left-hand side we have the additional
constraint that the values $(\xi,\eta)$ for which the maximum is
found {\em must} allow the equation
$x^\star=A(x^\star+J_pm\xi,\eta J_s)$ to have a solution
$x^\star\in[-J_s,J_s]$. If the maximum in the (less constrained)
right-hand side is obtained for an $(\xi,\eta)$ such that the
equation $x^\star=A(x^\star+J_pm\xi,\eta J_s)$ has {\em no}
solution $x^\star\in[-J_s,J_s]$, then the extra constraint
apparently interferes with the maximization and the two sides {\em
cannot} be the same, so no solution with $\Omega=\{x^\star\}$ can
exist. We conclude that the present type of solution exists if and
only if both sides of (\ref{eq:Omega_Onepeak}) find their maximum
at the same value $(\hat{\xi},\hat{\eta})$ (values that will
depend on $x^\star$, since $x^\star$ appears in the function to be
maximized), with the value of $x^\star$ subsequently following
from solution of the nonlinear equation
$x^\star=A(x^\star+J_pm\hat{\xi},\hat{\eta} J_s)$:
\begin{eqnarray}
(\hat{\xi},\hat{\eta})&=&{\rm argmax}_{\xi,\eta\in[-1,1]}
L(\xi,\eta)
\\
 x^\star&=& A(x^\star+J_pm\hat{\xi},\hat{\eta} J_s)
\end{eqnarray}
Since $B(x,y)=B(|x|,|y|)$, and is monotonically increasing with
both $|x|$ and $|y|$  we can immediately maximize with respect to
$\eta\in[-1,1]$, giving $\hat{\eta}=-\sgn(\nu)$. This simplifies
our remaining problem to solving
\begin{eqnarray}
 x^\star\!&=& -\sgn(\nu) A(x^\star+J_pm\hat{\xi},J_s)
 \label{eq:homopol1}
\\
 \hat{\xi}&=&\!{\rm argmax}_{\xi\in[-1,1]}L(\xi)
 \\
 L(\xi)&=&
  \xi(J_p\!-\!J_g)(k\!-\!k_0)+B(x^\star\!\!+\!J_pm\xi,J_s)~~~~
\end{eqnarray}
To resolve the remaining extremization we inspect the properties
of $B(x,y)$, in particular its second partial derivative in $x$.
We find that the function $L(\xi)$ is convex:
\begin{eqnarray}
\frac{\partial^2 L(\xi)}{\partial \xi^2}&=&
J^2_pm^2\Big\{1-\frac{1}{2}\tanh^2[\beta(x^\star\!\!+\!J_pm\xi+J_s)]~~~~
\nonumber \\ &&
-\frac{1}{2}\tanh^2[\beta(x^\star\!\!+\!J_pm\xi-J_s)]\Big\}\geq 0
\end{eqnarray}
$L(\xi)$ can therefore only be maximal at the boundaries
$\xi\in\{-1,1\}$. Next we can rule out states with $x^\star=0$,
since substitution
 into (\ref{eq:homopol1}) shows that it would be incompatible with
$\hat{\xi}=\pm 1$. Due to $J_pm\neq 0$, $x^\star\neq 0$, and the
monotonicity and symmetry of $B(x,y)$, the function $L(\xi)$ is
not symmetric in $\xi$, hence its maximum is unique:
\begin{eqnarray}
\hspace*{-10mm}
\hat{\xi}&=& \sgn\Big\{(J_p\!-\!J_g)(k\!-\!k_0) +\frac{1}{2}[B(x^\star\!\!+\!J_pm,J_s)
-B(x^\star\!\!-\!J_pm,J_s)] \Big\}~~~~ \label{eq:homopol2}
\end{eqnarray}

Having solved the extremization problem for solutions with
$\Omega=\{x^\star\}$, resulting in the two coupled equations
(\ref{eq:homopol1},\ref{eq:homopol2}) we turn to the $n\to\infty$
limit of  the order parameter equations
(\ref{eq:ninfty_m},\ref{eq:ninfty_k}) for $m$ and $k$. We define
\begin{eqnarray}
R(\xi)&=&\xi(J_p\!-\!J_g)(k\!-\!k_0)+\frac{1}{\beta}\log
\cosh[\beta(J_p m\xi\!+\!2x^\star)]
\end{eqnarray}
This is again a convex function, which is asymmetric in $\xi$ (due
to $x^\star\neq 0$), and therefore takes is maximal value on the
interval $[-1,1]$ at the boundary
\begin{eqnarray}
\bar{\xi}&=&
 \sgn\Big\{(J_p\!-\!J_g)(k\!-\!k_0)  +\frac{1}{2\beta}\log \Big[\frac{\cosh[\beta(J_p
m\!+\!2x^\star)]}{\cosh[\beta(J_p m\!-\!2x^\star)]}\Big]
 \Big\}~~~~
 \end{eqnarray}
Our equations for $m$ and $k$ can now be written as
\begin{eqnarray}
m &=& \lim_{n\to\infty} \frac{\int\!d\xi~w(\xi)\xi \tanh[\beta(J_p
m\xi+2x^\star)]e^{n\beta R(\xi)}
  } {\int\!d\xi~w(\xi)e^{n\beta R(\xi)}
 }
 \nonumber
\\
&=& \tanh[\beta(J_p m+2x^\star\bar{\xi})]
\\
 k&=& \lim_{n\to\infty} \frac{\int\!d\xi~w(\xi)\xi e^{n\beta R(\xi)}
  } {\int\!d\xi~w(\xi)e^{n\beta R(\xi)}
 }~=~\bar{\xi}
 \end{eqnarray}
 We have now confirmed that the present family of solutions
 with $\Omega=\{x^\star\}$ are indeed the generalization to arbitrary values of $J_s$ and $J_p m$
 of the solutions $k=\pm 1$ with homogeneous polarity, as claimed.
 Putting all our final equations together, replacing $\bar{\xi}$ by $k\in\{-1,1\}$ and using the full definition of $B(x,y)$,
  gives the new set
\begin{eqnarray}
 x^\star\!&=& -\sgn(\nu) A(x^\star+J_pm\hat{\xi},J_s)
 \label{eq:Homopol1}
 \\
m &=& \tanh[\beta(J_p m+2x^\star k)] \label{eq:Homopol2}
\\
k&=&
 \sgn\Big\{(J_p\!-\!J_g)(k\!-\!k_0) +\frac{1}{2\beta}\log \Big[\frac{\cosh[\beta(2x^\star+J_p
m)]}{\cosh[\beta(2x^\star-J_p m)]}\Big]
 \Big\}~~~~
 \label{eq:Homopol3}
\\
\hat{\xi}&=& \sgn\Big\{(J_p\!-\!J_g)(k\!-\!k_0) \nonumber
\\
&&  +\frac{1}{4\beta}\log
\Big[\frac{\cosh[\beta(x^\star\!\!+\!J_pm\!+\!J_s)]\cosh[\beta(x^\star\!\!+\!J_pm\!-\!J_s)]}
{\cosh[\beta(x^\star\!\!-\!J_pm\!+\!J_s)]\cosh[\beta(x^\star\!\!-\!J_pm\!-\!J_s)]}
 \Big\} \label{eq:Homopol4}
\end{eqnarray}
In both of the limits $J_s\to 0$ and $J_pm\to 0$ we recover
correctly the equations of the $k=\pm 1$ states as derived earlier
for these special cases, viz. $x^\star= 0$, $m = \tanh(\beta J_p
m)$, and $k=\hat{\xi}=
 \sgn[(J_p\!-\!J_g)(k\!-\!k_0)]$.
 \vsp

Finally we try to compactify and simplify our equations. We first
solve $x^\star$ from (\ref{eq:Homopol2}), which gives
\begin{eqnarray}
x^\star&=& k[ \frac{1}{2\beta}\tanh^{-1}(m)-\frac{1}{2}J_pm]
\label{eq:xstar}
\end{eqnarray}
Subsequent insertion into (\ref{eq:Homopol1})  leaves us with
\begin{eqnarray}&&
 \frac{\tanh[\frac{1}{2}{\rm
arctanh}(m)-\frac{1}{2}\beta J_pm]}{\tanh[\frac{1}{2}{\rm
arctanh}(m)-\frac{1}{2}\beta J_pm(1-2
k\hat{\xi})]}
=-\sgn(\nu)\tan(\beta J_s)
 \label{eq:Homopolnearlyb1}
\end{eqnarray}
Furthermore, we notice that with $k\hat{\xi}\in\{-1,1\}$ only the
choice $k=\hat{\xi}$ will allow the above equations to reduce to
the equations for $m\to 0$ that were found earlier, and that the
alternative $k=-\hat{\xi}$ would make it extremely difficult to
satisfy both (\ref{eq:Homopol3}) and (\ref{eq:Homopol4})
simultaneously. Upon choosing  $\hat{\xi}=k$ and after additional
rearranging and manipulation we can reduce our set of equations
further to
\begin{eqnarray}
\hspace*{-10mm}
\sgn(\nu)\tan(\beta J_s) &\!=\!& \frac{\tanh[\frac{1}{2}\beta
J_p|m|\!-\!\frac{1}{2}\tanh^{-1}|m|]}{\tanh[\frac{1}{2}\beta
J_p|m|\!+\!\frac{1}{2}\tanh^{-1}|m|]}
 \label{eq:Homopolb1}
 \\
 \hspace*{-10mm}
(J_p\!-\!J_g)(1\!-\!k_0k) &\!>\!&  \frac{1}{2\beta}\log
\Big[\frac{\cosh[\tanh^{-1}|m|\!-\!2\beta J_p|m|]}{\cosh[{\rm
arctanh}|m|]}\Big] \nonumber
\\
\hspace*{-10mm}
&&
 \label{eq:Homopolb2}
 \\
 \hspace*{-10mm}
(J_p\!-\!J_g)(1\!-\!k_0k) &\!>\!& \frac{1}{4\beta}\log \Big[\frac{\cosh[{\rm
arctanh}|m|\!-\!3\beta J_p|m|]\!+\!\cosh(2\beta J_s)}{\cosh[{\rm
arctanh}|m|\!+\!\beta J_p|m|]\!+\!\cosh(2\beta J_s)}\Big]
\label{eq:Homopolb3}
\end{eqnarray}

The joint distribution $W(h,\xi)$ of effective fields and
polarities for the present solution is very simple:
\begin{eqnarray}
W(h,\xi)&=& \delta[h-k\beta^{-1}\tanh^{-1}(m)]\delta(\xi-k)
\end{eqnarray}
Working out the free energy per monomer
(\ref{eq:free_energy_ninfty}) for the above solution gives, using
$k_0\in(-1,1)$ and equation (\ref{eq:xstar}) to eliminate
$x^\star$:
\begin{eqnarray}
\hspace*{-10mm}
\lim_{n\to\infty}\varphi \!&=&\!  \frac{1}{2}J_p(m^2\!\!+\!1)-
\frac{1}{2}J_g(1\!\!-\!k^{\star 2})- |J_p\!-\!J_g|(1\!-\!kk_0)
 - (J_p\!-\!J_g)(1\!-\!kk_0) \nonumber \\
 \hspace*{-10mm} &&
 -| \nu|
 -B(\frac{1}{2\beta}\tanh^{-1}(m)+\frac{1}{2}J_pm, J_s)
\label{eq:f_homopol}
\end{eqnarray}

 Equation (\ref{eq:Homopolb1}) gives a single transparent law
 from which to solve our order parameter $m$. Equations
 (\ref{eq:Homopol2},\ref{eq:Homopol3}) give conditions for the
 solution of (\ref{eq:Homopolb1}) to be acceptable; they are
 guaranteed to be satisfied for small $m$ if $J_p>J_g$ (due to $|k_0|<1$), whereas for larger
 $m$ their validity needs to be checked explicitly.
 Equations  (\ref{eq:Homopol2},\ref{eq:Homopol3}) also suggest
 that, as was found explicitly in the simple cases $J_s=0$ and $J_pm=0$, the most
 stable solution (and hence the thermodynamic state) will generally be the
 one with $k=-\sgn(k_0)$.
This completes our analysis of solutions with
$\Omega=\{x^\star\}$. We always find $k=\pm 1$, viz.
sequences with homogeneous polarity, provided $J_p>J_g$.

\subsection{Inhomogeneous polarity states $k=k_0$}

\newcommand{\ul}{u_{\ell}}
\newcommand{\ur}{u_{r}}

 \noindent In the same manner we now construct the continuation to arbitrary values of $J_s$ and $J_pm$
 of the inhomogenous polarity states,  where $k=k_0$. For this case, where $\Omega$ no longer contains just one point, our equation
 (\ref{eq:ninftyproblem}) from which to
solve $\lim_{n\to\infty}\psi(x)$ takes the following form
\begin{eqnarray}
\hspace*{-15mm}
 \psi(x) &=&  \max_{\xi,\eta\in[-1,1],~y\in\Omega,~y=C(x,\eta
J_s)-J_pm\xi}L(\xi,\eta,y)
-\max_{\xi,\eta\in[-1,1],~y\in\Omega}L(\xi,\eta,y)
\nonumber
\\
\hspace*{-15mm}&&
\label{eq:ninftyproblemsymmetries}
\\
\hspace*{-15mm}
L(\xi,\eta,y)&=&
 \psi(y)+B(y+J_pm\xi,\eta J_s)- \nu\eta
\end{eqnarray}
(provided $x\in\Omega$). In contrast to the $k\neq k_0$ case, this
equation has symmetries that can be exploited: it allows for will
solutions with $\psi(-x)=\psi(x)$ for all $x\in\Omega$, with
$\Omega$ symmetric around the origin. This is easily confirmed by
working out the right-hand side of
(\ref{eq:ninftyproblemsymmetries}) under the assumption of this
symmetry (via transformations like $y\to -y$ and $\xi\to -\xi$,
which are allowed by the constraints) upon making the replacement
$x\to -x$ in the left-hand side and using
$L(-\xi,\eta,-y)=L(\xi,\eta,y)$ and $C(-x,y)=C(x,y)$:
\begin{eqnarray}
\psi(\!-\!x)\!-\!\psi(x)&\!=\!&
\max_{\xi,\eta\in[-1,1],~y\in\Omega,~y=C(-x,\eta
J_s)-J_pm\xi}L(\xi,\eta,y)\nonumber\\ && \hspace*{-2mm} -
\max_{\xi,\eta\in[-1,1],~y\in\Omega,~y=C(x,\eta
J_s)-J_pm\xi}L(\xi,\eta,y) \nonumber
\\
&&\hspace*{-7mm} =\max_{\xi,\eta\in[-1,1],~y\in\Omega,~y=C(x,\eta
J_s)-J_pm\xi}L(-\xi,\eta,-y)\nonumber\\ && \hspace*{-2mm}-
\max_{\xi,\eta\in[-1,1],~y\in\Omega,~y=C(x,\eta
J_s)-J_pm\xi}L(\xi,\eta,y)~=~0
\end{eqnarray}
We will now construct solutions for $k=k_0$ with this reflection
symmetry. Inside (\ref{eq:ninftyproblemsymmetries}) it allows us
to  transform without punishment $y\to y\sgn(\eta)$ and $\xi\to
\xi\sgn(\eta)$, which gives a new expression that shows (using
$C(x,-y)=-C(x,y)$ and $B(x,y)=B(|x|,|y|)$) that both terms are
maximized for $\sgn(\eta)=-\sgn(\nu)$, and the second term more
specifically for the value $\eta=-\sgn(\nu)$. Upon abbreviating
$\Omega(x,\xi,\eta)=\{y\in\Omega|~y=C(x,\eta J_s)-J_pm\xi\}$:
\begin{eqnarray*}
\hspace*{-23mm}
\max_{\xi,\eta\in[-1,1],~y\in\Omega(x,\xi,\eta)}L(\xi,\eta,y)
&=& \max_{\xi,\eta\in[-1,1],~y\in\Omega(x,\xi,\eta)} \Big\{
 \psi(y)\!+\!B(y\!+\!J_pm\xi,|\eta| J_s)\!-\! \nu\eta
\Big\} \nonumber\\ &=& \max_{|\xi|,|\eta|\leq 1
,~y\in\Omega(x,\xi,|\eta|)} \Big\{
 \psi(y)\!+\! B(y\!+\!J_pm\xi,|\eta| J_s)\!+\! |\nu\eta|
\Big\}
\end{eqnarray*}
and
\begin{eqnarray*}
\hspace*{-15mm}
\max_{\xi,\eta\in[-1,1],~y\in\Omega}L(\xi,\eta,y)
&=& \max_{\xi,\eta\in[-1,1],~y\in\Omega} \Big\{
 \psi(y)+B(y\!+\!J_pm\xi,|\eta| J_s)- \nu\eta
\Big\} \nonumber\\ &=& \max_{|\xi|\leq 1 ,~y\in\Omega} \Big\{
 \psi(y)+B(y\!+\!J_pm\xi, J_s)\Big\}+ |\nu|
\end{eqnarray*}
 We observe the potential consistency of assuming
$\psi(x)$ to incease monotonically for $x\geq 0$. An increase in
$x$ leads via the constraint $y\in \Omega(x,\xi,|\eta|)$ to an
increase of $y$ inside the first maximization, so that $\psi(y)$
will increase. The term with $B(.,.)$ will also increase if the
sign of $\xi$ is chosen right. So we make the ansatz that
$\psi(x)$ is differentiable, and that $\psi^\prime(x)\geq 0$ on
$x\geq 0$. This implies that $\Omega=[-u,u]$, with
$\max_{x\in\Omega}\psi(x)=\psi(u)=0$. The second maximization in
(\ref{eq:ninftyproblemsymmetries}) now reduces to
\begin{eqnarray}
\hspace*{-10mm} \max_{y\in\Omega,~|\xi|\leq 1}\Big\{
 \psi(y)\!+\!B(y\!+\!J_pm\xi,J_s)\Big\}+|\nu|
 &=&~ \psi(u)+B(u+J_p|m|,J_s)+|\nu|\nonumber
 \\
 &=&~ B(u+J_p|m|,J_s)+|\nu|
\end{eqnarray}
This simplifies our equation (\ref{eq:ninftyproblemsymmetries})
for $\psi(x)$. For all $x\in[0,u]$ we now have
\begin{eqnarray}
\hspace*{-25mm}
\psi(x) &=&
  \max_{|y|\leq u,~y=C(x,|\eta|
J_s)-J_pm\xi,~|\xi|,|\eta|\leq 1}\Big\{
 \psi(y)\!+\!B(C(x,|\eta|J_s),|\eta| J_s)\!+\! |\nu|(|\eta|\!-\!1)\Big\}
 \nonumber\\
 \hspace*{-25mm}&&\hspace*{70mm}
 - B(u\!+\! J_p|m|,J_s)~~~~\nonumber
 \\
 \hspace*{-25mm}
  &=&
  \max_{|y|\leq u,~|y-C(x,|\eta|
J_s)|\leq J_p|m|,~|\eta|\leq 1}\Big\{
 \psi(y)\!+\!B(C(x,|\eta|J_s),|\eta| J_s)\!+\! |\nu|(|\eta|\!-\!1)\Big\}
 \nonumber
 \\
 \hspace*{-25mm}&&\hspace*{70mm}
 - B(u\!+\!J_p|m|,J_s)~~~~\nonumber
 \\
 \hspace*{-25mm}
  &&\hspace*{-10mm}=\max_{|\eta|\leq 1}~
  \max_{y\in[-u,u]\cap [C(x,|\eta|
J_s)-J_p|m|,C(x,|\eta|J_s)+J_p|m|]}\Big\{
 \psi(y)\!+\!B(C(x,|\eta|J_s),|\eta| J_s)\!+\! |\nu|(|\eta|\!-\!1)\Big\}
 \hspace*{-10mm}\nonumber
 \\
 \hspace*{-25mm}&& \hspace*{70mm}
  \!-\! B(u\!+\!J_p|m|,J_s)
  \label{eq:inhom_psi2}
\end{eqnarray}
 Since $\psi(y)$ is monotonic in $|y|$,
 we need $|y|$ to be as large as possible
for any given $|\eta|$. Since the intersection interval (if it
exists) is always biased to the right, we must find the largest
allowed value $y$ in the intersection interval. The intersection
is seen to be empty if $C(x,|\eta|J_s)>u+J_p|m|$, whereas  the
remaining possible scenarios are
\begin{eqnarray*}
& u\!-\!J_p|m|<C(x,|\eta|J_s)<u\!+\!J_p|m|:&~~~ y_{\rm max}=u
\\
& C(x,|\eta|J_s)<u\!-\!J_p|m| :&~~~ y_{\rm
max}=C(x,|\eta|J_s)\!+\!J_p|m|
\end{eqnarray*}
Consistency with the premise $x\in[0,u]$ demands that we must
identify the point where $x$ becomes so large that the
intersection interval is empty for {\em any} value of $|\eta|$
should be the boundary $x=u$. This, together with
$\min_{|\eta|\leq 1}C(x,|\eta|J_s)=C(x,J_s)$,  immediately gives
us an equation for $u$: $C(u,J_s)=u+J_p|m|$, or equivalently
\begin{eqnarray}
u=A(u+J_p|m|,J_s)
\end{eqnarray}
Graphical inspection shows that this equation always has one
unique non-negative solution $u$. Within our present construction
 we can always achieve a non-empty intersection set in  (\ref{eq:inhom_psi2}) for suitable
$(\xi,\eta)$, and we may proceed with maximization over $|\eta|$.
For each $x\in\Omega$ we now have
{\small
\begin{eqnarray}
\hspace*{-25mm}
\psi(x) &=&\nonumber
\\
\hspace*{-25mm}&&\hspace*{-20mm}
 \max_{|\eta|\leq 1,~C(x,|\eta|J_s)\leq
 u+J_p|m|}\left\{\begin{array}{lll}
 \psi(C(x,|\eta|J_s)\!+\!J_p|m|)\!+\!B(C(x,|\eta|J_s),|\eta| J_s)\!+\! |\nu||\eta|
&& {\rm if}~~C(x,|\eta|J_s)<u\!-\!J_p|m|
\\
 B(u+J_p|m|,|\eta| J_s)\!+\! |\nu||\eta|
&& {\rm if}~~C(x,|\eta|J_s)> u\!-\!J_p|m|
\end{array}
\right. \nonumber
\\[1mm]
\hspace*{-25mm}
&&\hspace*{20mm} -~ B(u\!+\!J_p|m|,J_s)-|\nu| \nonumber
\\[1mm]
\hspace*{-25mm}
&&\hspace*{-10mm}=
 \max_{z\in[0,J_s],~C(x,z)\leq
 u+J_p|m|}\left\{\begin{array}{lll}
 \psi(C(x,z)\!+\!J_p|m|)\!+\!B(C(x,z),z)\!+\! |\nu|z/J_s
&& {\rm if}~~C(x,z)<u\!-\!J_p|m|
\\
 B(u+J_p|m|,z)\!+\! |\nu|z/J_s
&& {\rm if}~~C(x,z)> u\!-\!J_p|m|
\end{array}
\right. \nonumber
\\[1mm]
\hspace*{-25mm}
&&\hspace*{20mm} -~ B(u\!+\!J_p|m|,J_s)-|\nu| \nonumber
\\[1mm]
\hspace*{-25mm}
&&\hspace*{-10mm}=
 \max_{z\in[C(x,u+J_p|m|),J_s]}\left\{\begin{array}{lll}
 \psi(C(x,z)\!+\!J_p|m|)\!+\!B(C(x,z),z)\!+\! |\nu|z/J_s
&& {\rm if}~~z>C(x,u-J_p|m|)
\\
 B(u+J_p|m|,z)\!+\! |\nu|z/J_s
&& {\rm if}~~z<C(x,u-J_p|m|)
\end{array}
\right. \nonumber
\\[1mm]
\hspace*{-25mm}
&&\hspace*{20mm} -~ B(u\!+\!J_p|m|,J_s)-|\nu|
\label{eq:inhom_psi3}
\end{eqnarray}
}
Since both $C(x,z)$ and $B(C(x,z),z)$ decrease
monotonically with increasing $z$ (see \ref{app:functions}) we are sure that for sufficiently small
values of $\nu$ we always find the maximum in
(\ref{eq:inhom_psi3}) by substituting the smallest allowed value
for $z$. We now proceed by assuming this property to hold for {\em
any} value of $\nu$. If indeed we always need the smallest $z$,
viz. $z=C(x,u+J_p|m|)$, we obtain for all $x\in[0,u]$:
\begin{eqnarray}
\psi(x)&=&
 B(u\!+\!J_p|m|,C(x,u\!+\!J_p|m|))- B(u\!+\!J_p|m|,J_s)\nonumber
 \\
 && + \frac{|\nu|C(x,u\!+\!J_p|m|)}{J_s}-|\eta|
 \label{eq:final_psix}
\end{eqnarray}
This expression meets our requirements: it increases
monotonically on $[0,u]$, and (using the general identity
$C(x,C(x,y))=y$ in combination with our previously established
relation $C(u,J_s)=u+J_p|m|$) one verifies that $\psi(u)=0$. We
take this as sufficient support for our ans\"{a}tze; in addition
we will find that for the purpose of evaluating the scalar order
parameters $(m,k)$ and the phase diagrams we do not need the
full shape of $\psi(x)$ but only the property that
$\psi(-u)=\psi(u)=\max_{x\in\Omega}\psi(x)$ with
$u=A(u+J_p|m|,J_s)$. \vsp

What remains in our present analysis is to work out the order
parameter equations for $m$ and $k$, and confirm that these
support the premise $\lim_{n\to\infty} k=k_0$. For large but
finite $n$ one would expect to have $k=k_0+k_1/n+\order(n^{-2})$
for $n\to\infty$, which implies that $n\beta
\xi(J_p-J_g)(k-k_0)=\beta \xi(J_p-J_g)k_1+\order(n^{-1})$.
Similarly one would expect for large but finite $n$ that
$\log\Psi(x)=n\beta\psi(x)+\psi_1(x)+\order(n^{-1})$, with
$\psi(x)$ as given by (\ref{eq:final_psix}).
 Insertion of
 these forms into (\ref{eq:ninfty_m},\ref{eq:ninfty_k}) gives
 integrals over $(x,y)$ that can be evaluated by steepest descent, with the
 relevant saddle-point obtained for $x=y=u\sgn(m\xi)$:
 {\small
 \begin{eqnarray}
 \hspace*{-25mm}
m &=& \lim_{n\to\infty}\nonumber
\\
\hspace*{-25mm}
&&\hspace*{-8mm} \frac{\int\!d\xi~w(\xi)\xi
\int_{-u}^u\!dxdy~ \tanh[\beta(J_p
m\xi\!+\!x\!+\!y)]e^{n[\log\cosh[\beta(J_p
m\xi+x+y)]+\beta\psi(x)+\beta\psi(y)]+\psi_1(x)+\psi_1(y)+\beta
\xi(J_p-J_g)k_1}
  } {\int\!d\xi~w(\xi) \int_{-u}^u\!dxdy~
e^{n[\log\cosh[\beta(J_p
m\xi+x+y)]+\beta\psi(x)+\beta\psi(y)]+\psi_1(x)+\psi_1(y)+\beta
\xi(J_p-J_g)k_1}
 }
 \nonumber
 \\
 \hspace*{-25mm}
&=& \lim_{n\to\infty} \frac{\int\!d\xi~w(\xi)|\xi|\sgn(m)
\tanh[\beta(J_p |m\xi|+2u)]e^{n\log\cosh[\beta(J_p
|m\xi|+2u)]+2\psi_1(u\sgn(m\xi))+\beta \xi(J_p-J_g)k_1}
  } {\int\!d\xi~w(\xi)
e^{n\log\cosh[\beta(J_p |m\xi|+2u)]+2\psi_1(u\sgn(m\xi))+\beta
\xi(J_p-J_g)k_1}
 }
 \nonumber
 \\
 \hspace*{-25mm}
 &&
\end{eqnarray}
}
so
\begin{eqnarray}
|m| &=& \tanh[\beta(J_p |m|+2u)]
 \label{eq:inhom_m}
\end{eqnarray}
Similarly we must solve
\begin{eqnarray}
\hspace*{-23mm}
 k_0&=& \lim_{n\to\infty}
\frac{\int\!d\xi~w(\xi)\xi
\int_{-u}^{u}\!dxdy~e^{n[\log\cosh[\beta(J_p
m\xi+x+y)]+\beta\psi(x)+\beta\psi(y)]+\psi_1(x)+\psi_1(y)+\beta
\xi(J_p-J_g)k_1}
  } {\int\!d\xi~w(\xi) \int_{-u}^u\!dxdy~
e^{n[\log\cosh[\beta(J_p
m\xi+x+y)]+\beta\psi(x)+\beta\psi(y)]+\psi_1(x)+\psi_1(y)+\beta
\xi(J_p-J_g)k_1}} \nonumber
\\
\hspace*{-23mm}
&=& \lim_{n\to\infty} \frac{\int\!d\xi~w(\xi)\xi
e^{n\log\cosh[\beta(J_p |m\xi|+2u)]+2\psi_1(u\sgn(m\xi))+\beta
\xi(J_p-J_g)k_1}
  } {\int\!d\xi~w(\xi)
e^{n\log\cosh[\beta(J_p |m\xi|+2u)]+2\psi_1(u\sgn(m\xi))+\beta
\xi(J_p-J_g)k_1}}
 \nonumber
\\
\hspace*{-23mm}
&=&  \frac{e^{2\psi_1(u\sgn(m))+\beta(J_p-J_g)k_1} -
e^{2\psi_1(-u\sgn(m))-\beta(J_p-J_g)k_1}
  } {
e^{2\psi_1(u\sgn(m))+\beta(J_p-J_g)k_1} +
e^{2\psi_1(-u\sgn(m))-\beta(J_p-J_g)k_1}}
 \label{eq:inhom_k}
 \end{eqnarray}
 As soon as a solution for the non-leading order
$\psi_1(x)$ exists, there will be a value of $k_1$ that give the
desired value $k=k_0$. However, careful inspection of the
sub-leading orders in the functional saddle-point equation for
$\Psi(x)$ reveals that the above construction works for $\nu<0$,
but no finite solution $\psi_1(x)$ exists when $\nu>0$. In the
latter case it turns out that the solution of the problem scales
with $n$ as $\log\Psi(x)=\beta
n\psi(x)+\psi_1(x)\sqrt{n}+\order(n^0)$ and
$k=k_0+k_1/\sqrt{n}+\ldots$. For a detailed analysis of the
different sub-leading orders see \ref{app:subleading}.  \vsp

The final result is that $k=k_0$ solutions always exist (although
they will be locally stable only for $J_g>J_p$), and that the
associate value of the order parameter $m$ is to be solved from
the two coupled equations
\begin{eqnarray}
 |m|&=& \tanh[\beta (2u\!+\!J_p|m|)]
 \\
 \tanh(\beta u)&=&\tanh[\beta(u+J_p|m|)\tanh(\beta J_s)
\end{eqnarray}
The sign of $m$ is arbitrary, both solutions $m=\pm |m|$ are
allowed and equally likely. We solve the first equation for $u$,
giving $u=\frac{1}{2}\beta^{-1}\tanh^{-1}(|m|)-\frac{1}{2}J_p|m|$,
and obtain an equation involving $|m|$ only:
\begin{eqnarray}
\frac{\tanh[\frac{1}{2}\tanh^{-1}(|m|)-\frac{1}{2}\beta
J_p|m|]}{\tanh[\frac{1}{2}\tanh^{-1}(|m|)+\frac{1}{2}\beta
J_p|m|]}=\tanh(\beta J_s)
\end{eqnarray}
The joint distribution $W(h,\xi)$ of effective fields and
polarities for the present solution, where $J_pm\neq 0$,  is found
to be
\begin{eqnarray}
W(h,\xi)&=&
\frac{1}{2}(1\!+\!k_0)\delta(\xi\!-\!1)\delta[h\!-\!J_pm\!-\!2u\sgn(m)]
\nonumber
\\
&&
+\frac{1}{2}(1\!-\!k_0)\delta(\xi\!+\!1)\delta[h\!+\!J_pm\!+\!2u\sgn(m)]
~~~~~~
\end{eqnarray}
The free energy per monomer (\ref{eq:free_energy_ninfty}) for the
present type of solution is found to reduce to
\begin{eqnarray}
\lim_{n\to\infty}\varphi &=&  \frac{1}{2}J_p(m^2\!\!+\!k_0^2)-
\frac{1}{2}J_g(k_0^2\!\!-\!k^{\star 2})- |\nu|
\nonumber
\\
&&
 - B(\frac{1}{2}\beta^{-1}\tanh^{-1}(|m|)+\frac{1}{2}J_p|m|, J_s)
~~~~ \label{eq:f_inhomopol}
\end{eqnarray}

\section{Properties of the functions $C(x,y)$ and
$B(C(x,y),y)$} \label{app:functions}

\noindent The functions $B(x,y)$ and $C(x,y)$ are defined as
 \begin{eqnarray}
B(x,y)&=&\frac{1}{2\beta}\log[4\cosh[\beta(x\!+\!y)]\cosh[\beta(x\!-\!y)]]
~~~~
\\
C(x,y)&=&\beta^{-1}\tanh^{-1}[\tanh(\beta x)/\tanh(\beta y)]
\end{eqnarray}
 We are only interested in the regime where $y\geq 0$ and $|x|<y$.
 The function $C(x,y)$ is monotonic and anti-symmetric in $x$, and
obeys $\sgn[C(x,y)]=\sgn(xy)$ and $|C(x,y)|\geq |x|$. It is the
$x$-inverse of $A(x,y)$, since
\begin{eqnarray*}
A(C(x,y),y)&=&\beta^{-1}\tanh^{-1}[\tanh(\beta C(x,y)) \tanh(\beta
y)]\nonumber \\ &=& \beta^{-1}\tanh^{-1}[\tanh(\beta x)]=x
\\
C(A(x,y),y)&=&\beta^{-1}\tanh^{-1}[\tanh(\beta A(x,y))/\tanh(\beta
y)]\nonumber \\ &=&\beta^{-1}\tanh^{-1}[\tanh(\beta x)]=x
\end{eqnarray*}
Furthermore $C(x,y)$ obeys the general identity
\begin{eqnarray}
C(x,C(x,y))&=& \beta^{-1}\tanh^{-1}\Big[\frac{\tanh(\beta
x)}{\tanh(\beta x)/\tanh(\beta y)}\Big]\nonumber
\\
&=& \beta^{-1}\tanh^{-1}[\tanh(\beta y)]=y
\end{eqnarray}
The function $B(x,y)$ is symmetric in $x$; thus also the function
$B(C(x,y),y)$ is symmetric in $x$. The partial derivatives of
$C(x,y)$ are
\begin{eqnarray}
\frac{\partial}{\partial x}C(x,y)&=&\frac{\tanh(\beta
y)~[1\!-\!\tanh^2(\beta x)]}{\tanh^2(\beta y)-\tanh^2(\beta x)}
\\
\frac{\partial}{\partial y}C(x,y)&=&-\frac{\tanh(\beta
x)~[1\!-\!\tanh^2(\beta y)]}{\tanh^2(\beta y)-\tanh^2(\beta x)}
\end{eqnarray}
Next we work out and simplify the quantity $B(C(x,y),y)$ with the
help of identities such as
\begin{eqnarray*}
&& 2\cosh[\tanh^{-1}(m)+\beta y]= e^{\beta
y}\Big(\frac{1+m}{1-m}\Big)^{\frac{1}{2}}+e^{-\beta y}
\Big(\frac{1+m}{1-m}\Big)^{-\frac{1}{2}}
\\
&& 2\cosh\Big[\tanh^{-1}\Big(\frac{\tanh(\beta x)}{\tanh(\beta
y)}\Big)+\beta y\Big]\cosh\Big[\tanh^{-1}\Big(\frac{\tanh(\beta
x)}{\tanh(\beta y)}\Big)-\beta y\Big]\nonumber
\\
&&\hspace*{40mm}
= \frac{\tanh^2(\beta y)+\tanh^2(\beta x)}{\tanh^2(\beta y) -\tanh^2(\beta x)}
+\cosh(2\beta y)
\end{eqnarray*}
This results in
\begin{eqnarray}
\hspace*{-25mm}
B(C(x,y),y)&=& \frac{1}{2\beta}\log\left\{4\cosh\Big(
\tanh^{-1}\Big[\frac{\tanh(\beta x)}{\tanh(\beta y)}\Big]\!+\!\beta
y\Big)\cosh\Big( \tanh^{-1}[\frac{\tanh(\beta x)}{\tanh(\beta
y)}]\!-\!\beta y\Big)\right\} \nonumber
\\
\hspace*{-25mm}
 &=&
 \frac{1}{\beta}\log[2 \cosh(\beta y)]-\frac{1}{\beta}\log \cosh(\beta x)
 -\frac{1}{2\beta}\log\Big[1-\frac{\tanh^2(\beta x)}{\tanh^2(\beta y)}\Big]
\end{eqnarray}
  Hence we have
\begin{eqnarray}
\frac{\partial}{\partial x}B(C(x,y),y)&=& \frac{\tanh(\beta
x)~[1\!-\!\tanh^2(\beta y)]} {\tanh^2(\beta y)-\tanh^2(\beta
x)}~~~~
\end{eqnarray}
Thus, in the region $|x|<|y|$ we have $\frac{\partial}{\partial x}
B(C(x,y),y)<0$ for $x<0$ and $\frac{\partial}{\partial x}
B(C(x,y),y)>0$ for $x>0$. The function $B(C(x,y),y)$ is symmetric
in $x$, diverges at $x=\pm y$, and has a unique
minimum $B(C(0,y),y)=\beta^{-1}\log[ 2\cosh(\beta y)]$ at $x=0$.

\section{Analysis of sub-leading orders for the state $k=k_0$ as $n\to\infty$}
\label{app:subleading}

\noindent Here  we analyze in more detail the sub-leading terms in
$n$ of the nontrivial solution of our equations
(\ref{eq:ninfty_psi},\ref{eq:ninfty_m},\ref{eq:ninfty_k})
 for the case where $J_g>J_p$, i.e.  where $m\neq 0$ and $k=k_0$, as $n\to\infty$. Given the exponential
scaling with $n$ of the kernel in (\ref{eq:ninfty_psi}), we may
without loss of generality  for $n\to\to\infty$
 always write $\Psi(x)$ in one of the following two forms:
\begin{eqnarray}
{\rm either:}~~~& \Psi(x)=& e^{n\psi(x)+\psi_1(x)+ \order(n^{-1})}
\label{eq:scaling1}
\\ {\rm
or:}~~~& \Psi(x)=& e^{n\psi(x)+\sqrt{n}\psi_1(x)+ \order(n^0)}
\label{eq:scaling2}
\end{eqnarray}
Since $\psi(x)$ was found to be maximal at $x=\pm u$ (where
$u>0$), we find in both cases
\begin{equation}
\lim_{n\to\infty}\Psi(x)=\alpha \delta(x-u)+(1-\alpha)\delta(x+u)
\label{eq:limit_of_Psi}
\end{equation}
where
\begin{eqnarray}
{\rm scaling~(\ref{eq:scaling1}):}&~~~~&
 \alpha = \frac{e^{\psi_1(u)}}{e^{\psi_1(u)}\!+\!
e^{\psi_1(-u)}} \label{eq:scaling1alpha}\\
 {\rm scaling~(\ref{eq:scaling2}):}&~~~~&
\alpha = \theta[\psi_1(u)-\psi_1(-u)] \label{eq:scaling2alpha}
\end{eqnarray}
We will show below that for $\nu<0$ the solution is of the form
(\ref{eq:scaling1}), with $k=k_0+k_1/n+\ldots$,
\begin{eqnarray}
&&
k_1=0,~~~~~~
\alpha=\frac{\sqrt{1+\sgn(m)k_0}~\big(\sqrt{1+
|k_0|}\!-\!\sqrt{1-|k_0|}\big)}{2|k_0|}~~~
\end{eqnarray}
and with $\lim_{n\to\infty}p(\xi) =w(\xi)$, whereas for $\nu>0$
the solution is of the form (\ref{eq:scaling2}), with
$k=k_0+k_1/\sqrt{n}+\ldots$,
\begin{eqnarray}
&& k_1 =
\frac{\psi_1(\minus u)\!-\!\psi_1(u)}{\beta \sgn(m)( J_p\!-\!J_g)
},~~~~~~
\alpha=\theta[\psi_1(u)\!-\!\psi_1(\minus u)]
\end{eqnarray}
and with $\lim_{n\to\infty}p(\xi)= \delta[\xi\!+\!\sgn(k_1)]$.

\subsection{First scaling ansatz: $\order(n^0)$ sub-leading terms}

\noindent If we simply substitute (\ref{eq:limit_of_Psi})
 and $k=k_0+k_1/n+\ldots$ into equation
 (\ref{eq:ninfty_psi}), we find
\begin{eqnarray}
\lim_{n\to\infty}p(\xi) &=& \frac{w(\xi)
 e^{\beta \xi( J_p-J_g) k_1 }} { \int\!d\xi^\prime
~w(\xi^\prime)e^{\beta \xi^\prime( J_p-J_g)k_1 }  }
\end{eqnarray}
and
\begin{eqnarray}
\hspace*{-20mm}
\alpha\delta(x-u)+(1\minus\alpha)\delta(x+u) &=&
\\
\hspace*{-20mm}
&&
\hspace*{-40mm}
\lim_{n\to\infty}  \frac{\alpha\int\!d\xi
d\eta~p(\xi)w(\eta)\delta\big[x\minus A(J_pm\xi\plus
u,\eta J_s)\big]
 e^{n\beta [B(J_pm\xi+u\!,\eta
J_s)- \nu\eta]}} {\int\!d\xi d\eta~p(\xi)w(\eta)\Big\{
\alpha e^{n\beta [B(J_pm\xi+u\!,\eta J_s)-
\nu\eta]}+(1\minus\alpha)
 e^{n\beta [B(J_pm\xi-u\!,\eta
J_s)- \nu\eta]}\Big\}}
\nonumber
\\
\hspace*{-20mm}
&& \hspace*{-40mm} +
\lim_{n\to\infty}
\frac{(1\minus \alpha)\int\!d\xi
d\eta~p(\xi)w(\eta)\delta\big[x\minus A(J_pm\xi\minus
u,\eta J_s)\big]
 e^{n\beta [B(J_pm\xi-u\!,\eta
J_s)- \nu\eta]}} {\int\!d\xi d\eta~p(\xi)w(\eta)\Big\{
\alpha e^{n\beta [B(J_pm\xi+u\!,\eta J_s)-
\nu\eta]}+(1\minus\alpha)
 e^{n\beta [B(J_pm\xi-u\!,\eta
J_s)- \nu\eta]}\Big\}}
\nonumber
\end{eqnarray}
 Since $\eta\in[-1,1]$ and $B(.,.)$ is symmetric
and monotonically increasing in both arguments, the leading
exponentials are maximal for $\eta=-\sgn(\nu)$ and $\xi=\pm
\sgn(m)$, so
\begin{eqnarray}
\hspace*{-20mm}
\alpha\delta(x-u)+(1\minus\alpha)\delta(x+u) &=&
\\[1mm]
\hspace*{-20mm}
&&\hspace*{-50mm} \frac{\alpha
p(\sgn(m))\delta\big[x\plus\sgn(\nu) A(J_p|m|\plus u,J_s)\big]
+(1\minus \alpha)p(-\sgn(m))\delta\big[x\minus
\sgn(\nu)A(J_p|m|\plus u,J_s)\big] } {\alpha
p(\sgn(m))+(1\minus\alpha)
 p(-\sgn(m))}\nonumber
\end{eqnarray}
There are two possibilities for solution,
dependent on how we match the two $\delta$-peaks on either side of
this equation. One always ends up with $u$ to be solved from
\begin{eqnarray}
u= A(J_p|m|+ u,J_s) \label{eq:recover_ueqn}
\end{eqnarray}
but, since $u>0$, the specific matching depends on $\nu$. For
$\nu<0$ one is forced to choose
\begin{eqnarray}
\alpha= \frac{\alpha e^{\beta\sgn(m)( J_p-J_g) k_1} } {\alpha
e^{\beta \sgn(m)( J_p-J_g) k_1 }\!+\!(1\minus\alpha)
 e^{-\beta \sgn(m)( J_p-J_g) k_1 }}
 \label{eq:alpha1}
 \end{eqnarray}
 whereas for $\nu>0$ the only option is
 \begin{eqnarray}
\alpha= \frac{(1\minus \alpha)e^{-\beta\sgn(m)( J_p-J_g) k_1} }
{\alpha e^{\beta \sgn(m)( J_p-J_g) k_1} \!+\!(1\minus\alpha)
 e^{-\beta \sgn(m)( J_p-J_g) k_1 }}
 \label{eq:alpha2}
\end{eqnarray}
To proceed with equations (\ref{eq:ninfty_m},\ref{eq:ninfty_k})
for $m$ and $k$ we first calculate
\begin{eqnarray}
\hspace*{-20mm}
\int\!d\xi~W(h,\xi)\xi f(h)&=&
\\
\hspace*{-20mm}&& \hspace*{-30mm}\lim_{n\to\infty} \frac{\int\!d\xi
dxdy~w(\xi)
 e^{\beta \xi( J_p-J_g) k_1 } \Psi(x)
\Psi(y) f(x+y+J_pm\xi)\xi e^{n\log \cosh[\beta (x+y+J_pm\xi)]}
  } {\int\!d\xi dxdy~w(\xi)
 e^{\beta \xi( J_p-J_g) k_1 } \Psi(x)
\Psi(y) e^{n\log \cosh[\beta (x+y+J_pm\xi)]}} \nonumber
\\
\hspace*{-20mm}
&&\hspace*{-35mm}= \sgn(m) \frac{ \alpha^2 f(2u\plus J_p|m|)
 e^{\beta
\sgn(m)( J_p-J_g) k_1 }\!-\!(1\minus\alpha)^2 f(\minus 2u\minus
J_p|m|)
 e^{-\beta
\sgn(m)( J_p-J_g) k_1 }} { \alpha^2
 e^{\beta
\sgn(m)( J_p-J_g) k_1 }+(1\minus\alpha)^2
 e^{-\beta
\sgn(m)( J_p-J_g) k_1 }}\nonumber
\end{eqnarray}
Application of this formula to $f(h)=\tanh(\beta h)$ and $f(h)=1$
gives
\begin{eqnarray}
|m|&=& \tanh[\beta( 2u+ J_p|m|)] \label{eq:recover_meqn}
\\
k_0&=&  \sgn(m)~ \frac{ \alpha^2
 e^{\beta
\sgn(m)( J_p-J_g) k_1 }-(1\minus\alpha)^2
 e^{-\beta
\sgn(m)( J_p-J_g) k_1 }} { \alpha^2
 e^{\beta
\sgn(m)( J_p-J_g) k_1 }+(1\minus\alpha)^2
 e^{-\beta
\sgn(m)( J_p-J_g) k_1 }} \label{eq:k0eqn}
\end{eqnarray}
 So far we have successfully recovered the
equations for $u$ and $m$ are as derived earlier; the next
question is whether we can find a corresponding solution for $k_1$
and $\alpha$.

 Both (\ref{eq:alpha1}) and (\ref{eq:alpha2}) are
quadratic equations for $\alpha$, so we expect at most two
solutions. In fact for $\nu>0$ only one of these is in the
interval $[0,1]$:
\begin{eqnarray}
\nu<0:&~~~~&\alpha\in\{0,1\}
\\
\nu>0:&~~~~&\alpha= \frac{1}{1+e^{\beta \sgn(m)( J_p-J_g) k_1}}
\end{eqnarray}
For $\nu<0$,  combination with
(\ref{eq:scaling1alpha},\ref{eq:k0eqn}) subsequently gives
\begin{eqnarray}
&&
k_1=0,~~~~~~
\alpha= \frac{\sqrt{1\plus\sgn(m)k_0}~\big(\sqrt{1\plus
|k_0|}-\sqrt{1\minus |k_0|}\big)}{2|k_0|}
\end{eqnarray}
For $\nu>0$, on the other hand, the solution breaks down.  Upon
writing $k_1$ in terms of $\alpha$ and substituting the result
into (\ref{eq:k0eqn}), we find the trivial $k_0=0$. Thus, only for
the degenerate special case $k_0=0$ is the solution of our
equations for $\nu>0$ of the form (\ref{eq:scaling1}). We conclude
that the generic solution for $\nu>0$ scales differently with $n$.

\subsection{Second scaling ansatz: $\order(\sqrt{n})$ sub-leading terms}

\noindent If we substitute (\ref{eq:limit_of_Psi})
 and $k=k_0+k_1/\sqrt{n}+\ldots$ into equation
 (\ref{eq:ninfty_psi}) (where $k_1\neq 0$, since otherwise
 we return to the previous scaling case)  we get
\begin{eqnarray}
 \lim_{n\to\infty}p(\xi) &=&
\delta[\xi+\sgn(k_1)]
\end{eqnarray}
and
\begin{eqnarray}
\hspace*{-20mm}
\alpha\delta(x-u)+(1\minus\alpha)\delta(x+u) &=&
\\
\hspace*{-20mm}
&&\hspace*{-40mm}
\lim_{n\to\infty} \frac{\alpha\int\!d\xi
d\eta~p(\xi)w(\eta)\delta\big[x\minus A(J_pm\xi\plus
u,\eta J_s)\big]
 e^{n\beta [B(J_pm\xi+u\!,\eta
J_s)- \nu\eta]}} {\int\!d\xi d\eta~p(\xi)w(\eta)\Big\{
\alpha e^{n\beta [B(J_pm\xi+u\!,\eta J_s)-
\nu\eta]}+(1\minus\alpha)
 e^{n\beta [B(J_pm\xi-u\!,\eta
J_s)- \nu\eta]}\Big\}}\nonumber
\\
\hspace*{-20mm}
&&\hspace*{-40mm}
+\lim_{n\to\infty} \frac{(1\minus \alpha)\int\!d\xi
d\eta~p(\xi)w(\eta)\delta\big[x\minus A(J_pm\xi\minus
u,\eta J_s)\big]
 e^{n\beta [B(J_pm\xi-u\!,\eta
J_s)- \nu\eta]}} {\int\!d\xi d\eta~p(\xi)w(\eta)\Big\{
\alpha e^{n\beta [B(J_pm\xi+u\!,\eta J_s)-
\nu\eta]}+(1\minus\alpha)
 e^{n\beta [B(J_pm\xi-u\!,\eta
J_s)- \nu\eta]}\Big\}}\nonumber
\end{eqnarray}
Once more the dominant exponent is maximal when $\eta=-\sgn(\nu)$
and $\xi=\pm \sgn(m)$, so
\begin{eqnarray}
\hspace*{-20mm}\alpha\delta(x-u)+(1\minus\alpha)\delta(x+u)
&=&\nonumber
\\
\hspace*{-20mm} &&\hspace*{-40mm}\lim_{n\to\infty}
\frac{e^{\sqrt{n}[ \psi_1(u)+\beta \sgn(m)( J_p-J_g) k_1]}~\delta\big[x\!+\! \sgn(\nu)A(J_p|m|\!+\!u,J_s)\big]}
{e^{\sqrt{n}[\psi_1(u)+\beta \sgn(m)( J_p-J_g) k_1 ]}
+e^{\sqrt{n}[\psi_1(-u)-\beta \sgn(m)( J_p-J_g) k_1 ]}}
 \nonumber
\\
\hspace*{-20mm} &&\hspace*{-40mm} +\lim_{n\to\infty}
\frac{e^{\sqrt{n}[\psi_1(-u)-\beta \sgn(m)( J_p-J_g) k_1 ]}~\delta\big[x\!-\!\sgn(\nu) A(J_p|m|\!+\! u, J_s)\big]}
{e^{\sqrt{n}[\psi_1(u)+\beta \sgn(m)( J_p-J_g) k_1 ]}
+e^{\sqrt{n}[\psi_1(-u)-\beta \sgn(m)( J_p-J_g) k_1 ]}}
\end{eqnarray}
Again we have to match the two $\delta$-peaks on
both sides. Since we know that the equation $u= -A(J_p|m|+ u,J_s)$
has no non-negative solutions $u$ (for $J_pm\neq 0$), we are
forced to match $\delta(x \pm u)$ to $\delta\big[x\pm
A(J_p|m|\plus u,J_s)\big]$. From this we recover equation
(\ref{eq:recover_ueqn}), as required,  but now with
\begin{eqnarray}
\hspace*{-20mm}
\nu<0:&~~~&\alpha=\lim_{n\to\infty} \frac{e^{\sqrt{n}[
\psi_1(u)+\beta \sgn(m)( J_p-J_g) k_1]}}
{e^{\sqrt{n}[\psi_1(u)+\beta \sgn(m)( J_p-J_g) k_1 ]}
+e^{\sqrt{n}[\psi_1(-u)-\beta \sgn(m)( J_p-J_g) k_1 ]}}
\nonumber
\\
\hspace*{-20mm}&&
\label{eq:alpha1b}
\\
\hspace*{-20mm}
\nu>0:&~~~&\alpha=\lim_{n\to\infty}
\frac{e^{\sqrt{n}[\psi_1(-u)-\beta \sgn(m)( J_p-J_g) k_1 ]}}
{e^{\sqrt{n}[\psi_1(u)+\beta \sgn(m)( J_p-J_g) k_1 ]}
+e^{\sqrt{n}[\psi_1(-u)-\beta \sgn(m)( J_p-J_g) k_1 ]}}
\nonumber
\\
\hspace*{-20mm}&&
\label{eq:alpha2b}
\end{eqnarray}
 Our present
equations can be obtained from those of the previous scaling
regime upon substituting $k_1\to \sqrt{n}k_1$ and
$\psi_1(x)\to\sqrt{n}\psi_1(x)$. This allows us to take over the
previous evaluation of the order parameter equations for $m$ and
$k$, provided we make the appropriate substitutions. For $m$ we
then recover equation (\ref{eq:recover_ueqn}) (as required),
whereas the equation for $k$ gives
\begin{eqnarray}
\hspace*{-18mm}
k_0~\sgn(m)&=& \lim_{n\to\infty}  \frac{
e^{\sqrt{n}[2\psi_1(u)+\beta \sgn(m)( J_p-J_g) k_1
]}-e^{\sqrt{n}[2\psi_1(-u) -\beta \sgn(m)( J_p-J_g) k_1 ]}} {
e^{\sqrt{n}[2\psi_1(u)+\beta \sgn(m)( J_p-J_g) k_1
]}+e^{\sqrt{n}[2\psi_1(-u)-\beta \sgn(m)( J_p-J_g) k_1 ]}}
\nonumber\\
\hspace*{-18mm}&&
\label{eq:last_k0_eqn}
\end{eqnarray}
 We have now successfully recovered the expressions for
$u$ and $m$ derived earlier; the remaining question is whether we
can find a corresponding solution for $k_1$ and $\alpha$ from the
coupled equations
(\ref{eq:scaling2alpha},\ref{eq:alpha1b},\ref{eq:alpha2b},\ref{eq:last_k0_eqn}).
 Since $|k_0|<1$
we conclude from (\ref{eq:last_k0_eqn}) that the following must be
true, so that the $\order(\sqrt{n})$ terms cancel and the
$\order(n^0)$ terms can indeed give us $|k_0|<1$:
\begin{equation}
k_1= \frac{\psi_1(-u)-\psi_1(u)}{\beta \sgn(m)( J_p-J_g) }
\end{equation}
This solution for $k_1$ we can insert into our previous equations
for $\alpha$, which gives
\begin{eqnarray}
\nu<0:&~~~&\alpha=\lim_{n\to\infty} \frac{e^{\sqrt{n} \psi_1(-u)}}
{e^{\sqrt{n}\psi_1(-u)} +e^{\sqrt{n}\psi_1(u) }}=1-\alpha
\\
\nu>0:&~~~&\alpha=\lim_{n\to\infty} \frac{e^{\sqrt{n}\psi_1(u) }}
{e^{\sqrt{n}\psi_1(-u)} +e^{\sqrt{n}\psi_1(u)}}=\alpha
\end{eqnarray}
Apparently, for $\nu>0$ the present scaling ansatz gives
self-consistent solutions. For $\nu<0$ we find
$\alpha=\frac{1}{2}$, and hence $k_1=0$ which is forbidden since
it effectively brings us back to the previous scaling regime. We
conclude that, apart from degenerate limits, the two scaling
ans\"{a}tze (\ref{eq:scaling1},\ref{eq:scaling2}) are
complementary: for $\nu<0$ the system is in a state  of the type
(\ref{eq:scaling1}), whereas for $\nu>0$ it is in a state of the
type (\ref{eq:scaling2}).

\end{document}